\documentclass[aps,prd,showpacs,superscriptaddress,preprintnumbers,nofootinbib,notitlepage]{revtex4-1}
\pdfoutput=1
\usepackage[T1]{fontenc}
\usepackage{bm,amsmath,amssymb,color,bbm,float,multirow} 
\usepackage[pdftex]{graphicx}
\usepackage[colorlinks=true, pdfstartview=FitH, linkcolor=blue, citecolor=blue, urlcolor=blue]{hyperref}

\graphicspath{{./1_Figures/}}

\DeclareMathOperator\diag{diag}
\DeclareMathOperator\Tr{Tr}
\DeclareMathOperator\rme{\mathrm{e}}

\newcommand{\nn}{\nonumber}
\newcommand{\eref}[1]{(\ref{#1})}
\renewcommand{\Im}{\mathrm{Im}}
\renewcommand{\Re}{\mathrm{Re}}
\newcommand{\der}{\partial}
\renewcommand{\bar}[1]{\overline{#1}}
\newcommand{\calO}{\mathcal{O}}
\newcommand{\bep}{\begin{pmatrix}} 
\newcommand{\eep}{\end{pmatrix}}
\newcommand{\USp}{\text{USp}}
\newcommand{\SU}{\text{SU}}
\newcommand{\SO}{\text{SO}}
\renewcommand{\O}{\text{O}}
\newcommand{\U}{\text{U}}
\newcommand{\1}{\mathbbm{1}}
\newcommand{\RR}{\mathbb{R}}

\newcommand{\ZZ}{\mathbb{Z}}
\renewcommand{\epsilon}{\varepsilon}
\newcommand{\rmd}{\mathrm{d}}
\newcommand{\be}{\begin{eqnarray}}
\newcommand{\ee}{\end{eqnarray}}
\newcommand{\ea}{\end{align}}
\newcommand{\ms}{\hspace{-10pt}}
\newcommand{\KK}{\mathcal{K}}
\DeclareMathOperator\Pf{\mathrm{Pf}}

\newcommand{\sign}{{\rm sign}}

\def\ba#1\ea{\begin{align}#1\end{align}}
\def\mkakko#1{\left( #1 \right)}
\def\ckakko#1{\left\{ #1 \right\}}
\def\kkakko#1{\left[ #1 \right]}
\def\wt#1{\widetilde{#1}}
\newcommand{\calZ}{\mathcal{Z}}

\begin{document}
\title{\boldmath Random matrix approach to three-dimensional QCD with a Chern-Simons term}

\author{Takuya Kanazawa}
\affiliation{Research and Development Group, Hitachi, Ltd., Kokubunji, Tokyo 185-8601, Japan}
\author{Mario Kieburg}
\affiliation{Department of Physics, Bielefeld University, Postfach 100131, D-33501 Bielefeld, Germany}
\author{Jacobus J.~M.~Verbaarschot}
\affiliation{Department of Physics and Astronomy, Stony Brook University, Stony Brook, NY 11794, USA}
\date{\today}
\allowdisplaybreaks

\begin{abstract}
  We propose a random matrix theory for QCD in three dimensions with a
  Chern-Simons term at level $k$ which spontaneously breaks the flavor
  symmetry according to U($2N_{\rm f}$) $\to $ U($N_{\rm f}+k$)$\times$U($N_{\rm f}-k$).
  This random matrix model is obtained by adding a complex part to
  the action for the $k=0$ random matrix model.  We derive the
  pattern of spontaneous symmetry breaking from the analytical solution of the model.
  Additionally, we obtain explicit analytical results for the spectral density
  and the spectral correlation functions for the Dirac operator
  at finite matrix dimension, that become complex. In the
  microscopic domain where the matrix size tends to infinity,
  they are expected to be universal, and give an
  exact analytical prediction to the spectral properties of the Dirac
  operator in the presence of a Chern-Simons term. Here, we calculate the
  microscopic spectral density. It shows exponentially large (complex) oscillations
  which cancel the phase of the $k=0$ theory.

\end{abstract}
\maketitle
%\tableofcontents

\section{Introduction}

It is believed that in the QCD vacuum the strong interactions of gluons and quarks induce spontaneous breaking of 
chiral symmetry $\SU(N_{\rm f})_R\times\SU(N_{\rm f})_L\to \SU(N_{\rm f})_V$ when 
the number of massless Dirac fermions $N_{\rm f}$ is below the conformal window. 
The quantum fluctuations of the gauge fields in the broken phase 
manifest themselves in characteristic spectral fluctuations 
of the Dirac operator in the microscopic domain \cite{Leutwyler:1992yt}, which can be exactly reproduced 
by a zero-dimensional matrix model with the same global symmetries as QCD known as  
chiral random matrix theory \cite{Shuryak:1992pi,Verbaarschot:1993pm,Verbaarschot:1994qf}. 
We refer to \cite{Verbaarschot:1997bf,Verbaarschot:2000dy,
Verbaarschot:2005rj,Kanazawa:2012zzr,Akemann:2016keq} for reviews. 

In three-dimensional spacetime, whether the symmetry of fermions is dynamically  
broken or not has remained a matter of debate for decades 
\cite{Pisarski:1984dj,Appelquist:1986fd,Appelquist:1986qw}.  
Three-dimensional gauge theories are of distinguished importance in various contexts, ranging from  
domain walls and surface (boundary) states in four dimensions to quantum Hall effects, 
graphene, spin liquids and high-temperature superconductivity 
\cite{Lee2006doping,Nayak2008,Balents2010,QiZhang2011,Hansson2017}. Part of this 
rich physics stems from the existence of a Chern-Simons term. 
In three-dimensional QED (QED$_3$) the interplay of a Chern-Simons term and 
fermionic symmetry breaking was investigated in  
\cite{Hong:1992ww,Kondo:1994bt,Hong:1997ws,Itoh:1998xk,
Matsuyama:1999br,Liu:2002yc,Hofmann:2010zy}. A consensus from these 
studies is that dynamical symmetry breaking is generally suppressed by a Chern-Simons 
term, because photons acquire a gauge-invariant mass term which in turn 
quenches quantum fluctuations. However, recent lattice simulations 
\cite{Karthik:2015sgq,Karthik:2016ppr,Karthik:2017hol} report that dynamical 
mass generation of fermions does not occur in QED$_3$
even in the absence of a Chern-Simons term (see also \cite{Roscher:2016wox} 
for a RG study with the same conclusion). 

In contrast, in three-dimensional QCD (QCD$_3$) with an even number $2N_{\rm f}$ of 
massless two-component flavors and \emph{without} a Chern-Simons term, 
dynamical symmetry breaking%
\footnote{A Vafa-Witten-type argument \cite{Vafa:1983tf} 
shows that the $\U(N_{\rm f})\times\U(N_{\rm f})$ symmetry is unbroken. Here 
we assume that fermions are in a complex representation of the gauge 
group.}
\ba
	\U(2N_{\rm f})\to\U(N_{\rm f})\times\U(N_{\rm f})
	\label{eq:UUU}
\ea
is believed to take place through fermion bilinear condensation 
when $N_{\rm f}$ is below a certain threshold 
\cite{Appelquist:1989tc,Ferretti:1992fga,Diamantini:1993vn}. 
The fermion condensate has been observed in quenched lattice simulations 
\cite{Damgaard:1998yv,Karthik:2016bmf}. A non-chiral matrix model 
corresponding to \eqref{eq:UUU} is also known \cite{Verbaarschot:1994ip}; 
see \cite{NagaoSlevin93,Damgaard:1997pw,Akemann:1998ta,
Christiansen:1998mu,Magnea:1999iv,Magnea:1999ey,Hilmoine:2000ca,
Nagao:2000um,Szabo:2000qq,Szabo:2005gi} for further developments. 
However, not much is known about QCD$_3$ at nonzero Chern-Simons 
level $k$.%
\footnote{As is well known, non-Abelian gauge invariance 
forces the Chern-Simons coefficient to be quantized \cite{Deser:1981wh}. 
Here we label it as $k\in\ZZ$.} 
This is partly due to the sign problem that makes 
a direct lattice simulation prohibitively hard. 
Recently, it was argued \cite{Komargodski:2017keh} that there is a finite 
window of $N_{\rm f}$ in which a novel symmetry breaking  
\ba
	\label{eq:ssbnew}
	\U(2N_{\rm f}) \to \U(N_{\rm f}+k)\times\U(N_{\rm f}-k) \qquad 
	\text{for}\quad |k|<N_{\rm f}
\ea  
occurs. New boson-fermion dualities describing the transition region 
of \eqref{eq:ssbnew} were also proposed \cite{Komargodski:2017keh}. 
While a proof is not available yet, this conjecture passes 
nontrivial tests such as the matching of symmetries and anomalies, and 
consistency under mass deformations. Related discussions can be 
found in \cite{Gaiotto:2017tne,Armoni:2017jkl}. 

In this paper, we propose a new random matrix model that realizes the 
symmetry breaking scenario \eqref{eq:ssbnew}.%
\footnote{To avoid confusion we note that 
the approach of the present paper is unrelated to  
Chern-Simons matrix models in \cite{Marino:2002fk,Aganagic:2002wv,Tierz:2002jj,
Kapustin:2009kz,Drukker:2010nc}, where the dynamics of fermions 
was not the main focus.} This is made possible 
through a judicious choice of a non-Gaussian weight for matrix elements 
in which $k$ enters as a parameter. 
We show that in the large-$N$ limit with $N$ the matrix size, 
the model reduces to a sigma model with a target space of 
the complex Grassmannian $\U(2N_{\rm f})/[\U(N_{\rm f}+k)\times\U(N_{\rm f}-k)]$. 
When $k$ is varied there occurs a sequence of first order phase transitions 
that separate phases with different complex Grassmannians. 
By solving the model we delineate the structure of the Dirac operator 
spectrum that underlies the exotic 
symmetry breaking \eqref{eq:ssbnew}. Under the assumption that \eqref{eq:ssbnew} 
indeed characterizes the vacuum of QCD$_3$ with a Chern-Simons term, 
our approach offers an entirely new way to probe the interplay of strongly coupled 
fermion dynamics and a topological term within a tractable framework. 

This paper is organized as follows. 
In Section~\ref{sc:rmm} we define the model and derive key properties. 
The partition function of the model is computed and the phase structure 
as a function of $k$ and the fermion mass is investigated.  
Section~\ref{sc:specf} is devoted to the spectral functions that are first derived at finite matrix dimension $N$. 
As $k$ grows the spectral density evolves from a smooth semicircle 
to a distorted complex oscillatory form. In addition, we compute the large-$N$ microscopic limit of the spectral 
density in the quenched and unquenched ensemble. The details of the calculations to get these results are given in Appendix~\ref{sec:app}.  
Concluding remarks are made in section~\ref{sc:concl}.

\section{\label{sc:rmm}Random matrix model}

We introduce three new random matrix models 
labeled by the Dyson index $\beta$ in Subsection~\ref{sec:model}. They are associated with 
QCD$_3$ in the presence of a Chern-Simons term
with fermions transforming 
in a complex/pseudoreal/real ($\beta=2/1/4$) 
representation of the gauge group, respectively. 
The class $\beta=2$ comprises quarks in the fundamental representation 
of $\SU(N_c)$ with $N_c\geq 3$; $\beta=1$ includes quarks 
in the fundamental representation of $\USp(N_c)$ (here, $N_c$ must be even)\footnote{Our convention is such that $\USp(2)=\SU(2)$.}; and 
$\beta=4$ corresponds to quarks in the adjoint representation of $\SU(N_c)$ 
and in the vector representation of $\SO(N_c)$. In Subsection~\ref{sec:eta} we give a discussion on the renormalization of the Chern-Simons term due to the dynamical quarks that are related to the $\eta$-invariant.

Each of the above three random matrix models produces a universal non-linear sigma model that is derived in detail for $\beta=2$ in Subsection~\ref{sec:sigma}. As in the case of
three dimensional QCD, the model experiences a phase transition from one to another sigma model due to the Chern-Simons-like term.  In Subsection~\ref{sec:phase} we show how the mechanism works in general, and in Subsection~\ref{sec:part} we illustrate our findings by studying the two-flavor case for $\beta=2$. Therein, we also present finite results for  the partition function at finite matrix dimension.

\subsection{Partition function of our model}\label{sec:model}

The partition functions of our model are defined by
\be
Z^{\beta=2} & =& \int \rmd A \;
\exp\kkakko{\frac{\alpha_2}{2}  (\Tr A-2ik)^2 - \frac{N}{2}\Tr A^2}
	\prod_{f=1}^{2N_{\rm f}}\det \mkakko{iA+ m_f \mathbbm{1}_{N}} \,,
	\label{eq:Zbeta2}
	\\
	Z^{\beta=1} & =& \int 
	\rmd A' \; \exp\kkakko{ {\alpha_1} 
	( \Tr A' -2 i k)^2	- N \Tr {A'}^2}
	\prod_{f=1}^{2N_{\rm f}}\det \mkakko{i A'+ m_f \mathbbm{1}_{N}} \,,
	\label{eq:Zbeta1}
	\\
	Z^{\beta=4} & =& \int \rmd A'' \;  \exp\kkakko{
		\frac {\alpha_4}4 (\Tr A''-2 ik)^2 - \frac{N}{2}\Tr {A''}^2
	}  
	\prod_{f=1}^{2N_{\rm f}} \sqrt{\det \mkakko{i A''+ m_f \mathbbm{1}_{2N}}} \,,
	\label{eq:Zbeta4}
\ee
where $A$ is a complex hermitian $N\times N$ matrix,  
$A'$ is a real symmetric $N\times N$ matrix, and $A''$ 
is a self-dual $N\times N$ matrix whose elements are real
quaternions\footnote{The matrix size $N$ is not equal to the number of colors 
  in QCD$_3$. It replaces the dimension of the Hilbert
  space that is the product of the space-time volume, number of colors
  and spinor dimension (dimension of the gauge group representation).}.
The measures $\rmd A$, $\rmd A'$ and $\rmd A''$ are the
corresponding Lebesgue  measures, in particular the products
of the differentials of the real independent matrix entries. 
The square root of the determinants for $\beta=4$, see~\eqref{eq:Zbeta4},
is exact and may be implemented as a Pfaffian  determinant, 
\begin{equation}
\sqrt{\det \mkakko{iA''+m_f \1_{2N}}}=\Pf[(-i\sigma_2\otimes\1_N)\mkakko{iA''+m_f \1_{2N}}].
\end{equation}
The masses are gathered in the diagonal matrix $M=\diag(m_1,\ldots,m_{2N_{\rm f}})$.

  The positive constant $\alpha_\beta$ determines the effective strength of the
  ``Chern-Simons'' coupling. Starting with section~\ref{sec:part} it will
  be chosen
  \be
  \alpha_\beta = \frac N{N+4 N_{\rm f}/\beta + 1}<1.
\label{alfbet}
  \ee
  In sections~\ref{sec:sigma} and~\ref{sec:phase} the detailed form of
  $\alpha_\beta$ is irrelevant apart from the convergence requirements
  of the integrals and, thence, remain unspecified therein.
  Indeed the integrability of the variable $\Tr A$ is guaranteed when
  $\alpha_1<1$, $\alpha_2<1$ and $\alpha_4<2$, which is satisfied by Eq.~\eqref{alfbet}. 
    
In Eq.~\eqref{eq:Zbeta4} $A''$ is regarded as a complex $2N\times 2N$ matrix,  
using $(\1_2, i\sigma_a)$ as the quaternion basis.  
Note that all matrices are square, reflecting the absence of 
topological zero modes in $2+1$ dimensions. 
The real parameter $k$ corresponds to the Chern-Simons level, 
and $2N_{\rm f}$ represents the number of two-component Dirac fermions 
for $\beta=1,2$ and of two-component Majorana fermions for $\beta=4$. 
The case of an odd number of flavors will not be considered in this paper. We expect that the mechanism described below should work similar to the even number of flavors case though their is a significant difference; the Goldstone manifold is disconnected for an odd number of flavors~\cite{Verbaarschot:1994ip,Magnea:1999iv,Magnea:1999ey}.

The models~\eqref{eq:Zbeta2},~\eqref{eq:Zbeta1} and~\eqref{eq:Zbeta4} 
differ from the conventional random matrix models for QCD$_3$ \cite{Verbaarschot:1994ip,
Magnea:1999iv,Magnea:1999ey} by the presence of the squared trace term 
in the exponent.%
\footnote{A squared trace term was first also introduced in matrix models
in~\cite{Das:1989fq} with application to 2D quantum gravity, see also~\cite{Cicuta:1990uc,Ueda:1991xa,Sawada:1991uh,Korchemsky:1992tt,David:1996vp}.
Additionally, they appear in random matrix theories for the Wilson Dirac operator~\cite{Kieburg:2013xta}.
} 
At $k\ne 0$, the latter makes the statistical weight complex-valued, just 
as the Chern-Simons term does in Euclidean QCD$_3$ causing the 
infamous ``sign problem''. It is not problematic for us because we 
can still solve the matrix models exactly 
without recourse to numerical simulations. 
Our motivation to include a squared trace term is that 
this deformation changes the pattern of flavor symmetry breaking. 
The microscopic large-$N$ limit~\cite{Shuryak:1992pi,Verbaarschot:1993pm}
makes this more lucid. For this purpose, we take
$N\to\infty$ and $m_f\to 0$ with $\widehat{m}_f=Nm_f$ and $k$ fixed.%
\footnote{This limit should not be confused with the large-$N_c$ limit 
in gauge theory. We keep the number of colors $N_c$ in QCD$_3$ finite
 throughout our work.}  
If $N$ is identified with the volume of space-time, this limit is equivalent to the 
leading order of the $\varepsilon$-expansion in chiral perturbation
theory ~\cite{Gasser:1987ah,Leutwyler:1992yt}, 
in which the partition function reduces to a non-linear sigma model 
of static Nambu-Goldstone modes. If $k\in\ZZ$ with $|k|\le N_{\rm f}$, 
one can show for the partition functions~\eqref{eq:Zbeta2},~\eqref{eq:Zbeta1}
and~\eqref{eq:Zbeta4}  in the microscopic limit reduce to
\ba
	Z^{\beta=2} &\sim  
	\int\limits_{\U(2N_{\rm f})} \ms~ \rmd \mu(U)\; 
	\exp\big[ \Tr U^\dagger \Lambda_k U \widehat{M}  \big]\,,
	\label{eq:zN2}
	\\
	Z^{\beta=1} &\sim  \int\limits_{\USp(4N_{\rm f})} \ms \rmd \mu(U) ~
	\exp \big[ \Tr  
	U^{\dagger} \diag(\Lambda_k,-\Lambda_k) U
		\diag(\widehat{M},-\widehat{M})
	 \big] \,, 
	%\label{eq:234s}
	\label{eq:zN1}
	\\
	Z^{\beta=4} &\sim  
	\int\limits_{\O(2N_{\rm f})}\ms \;\rmd \mu(O)\; 
	\exp \big[
		 \Tr O^{\rm T} \Lambda_k O\widehat{M} 
	\big]\,,
	\label{eq:zN4}
\ea
where $\widehat{M}\equiv \diag(\widehat{m}_1,\ldots,\widehat{m}_{2N_{\rm f}})$ and 
\ba
	\Lambda_k\equiv \diag \big(
		\1_{N_{\rm f}+k}\,,-\1_{N_{\rm f}-k}
	\big) \,.
	\label{eq:Lpdef}
\ea  
The Haar measure of the respective groups are denoted by $\rmd\mu$. This result is derived in the next subsection.

Effectively we do not integrate over the whole group but a coset. These cosets are the Goldstone manifolds and reflect the patterns of flavor symmetry breaking in the ``chiral'' limit  given by
\ba
	\label{eq:ssblist}
	\begin{array}{lrcl}
	\beta=2:& \quad \U(2N_{\rm f})&~~\to~~&\U(N_{\rm f}+k)\times\U(N_{\rm f}-k),
	\\[5pt]
	\beta=1:& \quad \USp(4N_{\rm f})&~~\to~~&
	\USp\big(2(N_{\rm f}+k)\big)\times\USp\big(2(N_{\rm f}-k)\big),
	\\[5pt]
	\beta=4:& \quad \O(2N_{\rm f})&~~\to~~&\O(N_{\rm f}+k)\times\O(N_{\rm f}-k)
	\end{array}
\ea
yielding $2(N_{\rm f}^2-k^2)$, $4(N_{\rm f}^2-k^2)$ and 
$N_{\rm f}^2-k^2$ Nambu-Goldstone modes, respectively. 
When $k=0$ they recover the usual symmetry breaking  
patterns proposed for parity-invariant  
QED$_3$ and QCD$_3$ with no Chern-Simons term 
\cite{Pisarski:1984dj,Appelquist:1986qw,Appelquist:1989tc,
Verbaarschot:1994ip,Magnea:1999iv,Magnea:1999ey,
Nagao:2000um,Szabo:2005gi}. 
This agreement is nontrivial because the partition functions 
\eqref{eq:Zbeta2}, \eqref{eq:Zbeta1} and \eqref{eq:Zbeta4} 
are different from  those in \cite{Verbaarschot:1994ip,Magnea:1999iv,Magnea:1999ey} 
even at $k=0$ due to the squared trace term. It highlights the universality of the large-$N$ limit.  
When $k\ne 0$, the symmetry breaking schemes \eqref{eq:ssblist}
coincide with the generalizations proposed 
recently \cite{Komargodski:2017keh} 
for QCD$_3$ with a Chern-Simons term at level $k$.  

In \eqref{eq:zN2}, \eqref{eq:zN1} and \eqref{eq:zN4} 
we omitted overall multiplicative factors, which are all 
proportional to $(-1)^{Nk}$. Therefore choosing even $N$ is mandatory 
to ensure positivity of the partition function, although 
the overall normalization of $Z$ does not affect physical expectation values.

%%%%%%%%%%%%%%%%%%%%%%%%%%%%%%%
\begin{figure}[t!]
	\centering
	\includegraphics[width=.4\textwidth]{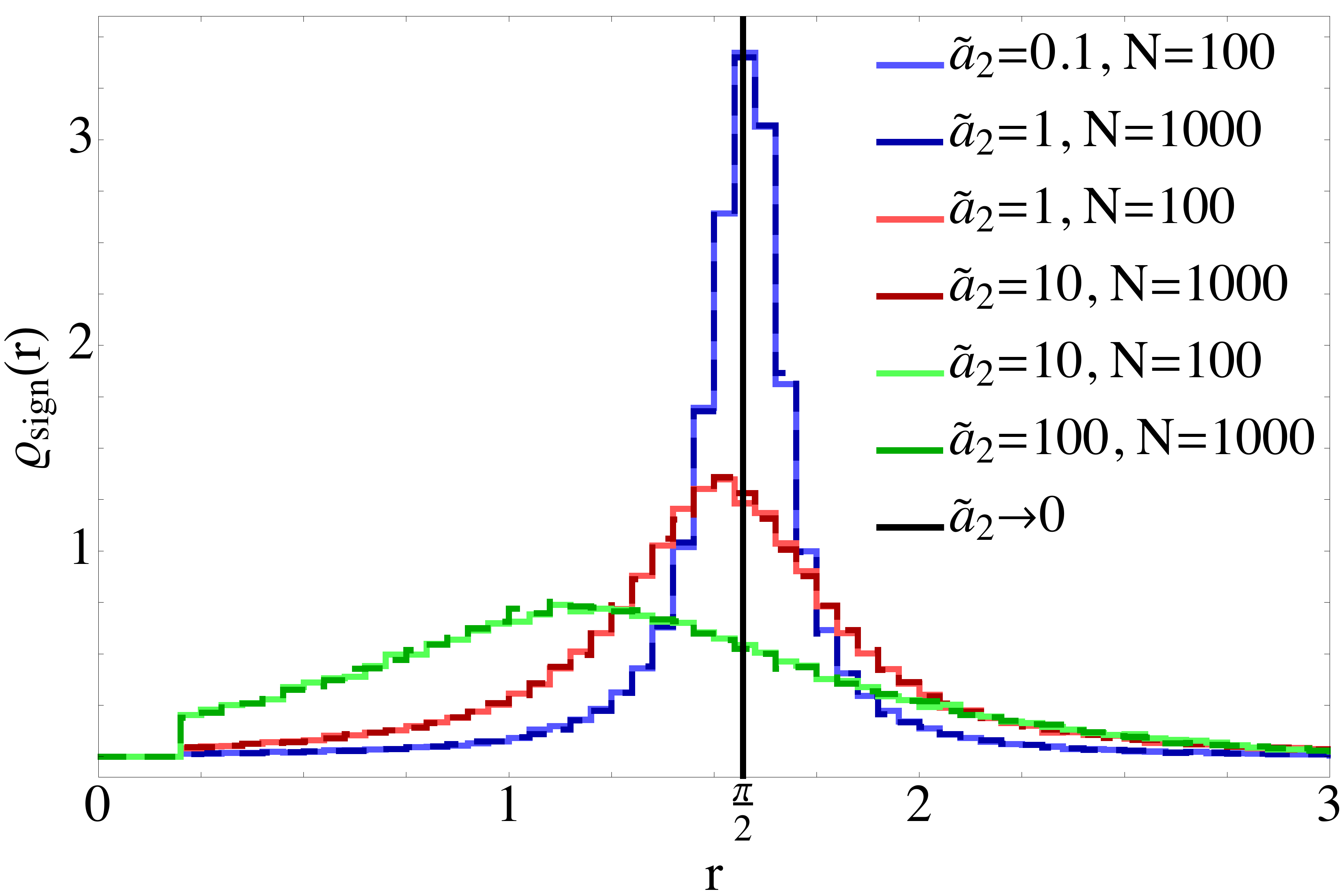}
	\caption{\label{fg:sign-ratio} Distribution $\rho_{\rm sign}(r)$ of the ratio $r=\Tr A/\Tr \sign(A)$ for $k=0$ and $N_{\rm f}=0$ generated by Monte Carlo simulations. The matrix size $N$ is chosen to be one of the two values $N=100$ (bright solid histograms) and $N=1000$ (dark dashed histograms) and the parameter $\alpha_2=1-\widetilde{a}_2/N$ takes three values with the ratios $\widetilde{a}_2/N=10^{-1}$ (green),  $\widetilde{a}_2/N=10^{-2}$ (red),  and $\widetilde{a}_2/N=10^{-3}$ (blue). The ensemble size varies because we omitted those configurations with $r=\infty$, i.e. $\Tr \sign(A)=0$. The number of these configurations decreases according to $\sqrt{\widetilde{a}_2/N}$. The limit $\widetilde{a}_2/N\rightarrow 0$ (the physical case) becomes a Dirac delta distribution at $r=\pi/2$ (black vertical line). Since the histograms with fixed quotient $\widetilde{a}_2/N$ agree almost perfectly, they are barely distinguishable, we are save to assume that the plots show the limiting large $N$ behavior.}
\end{figure}
%%%%%%%%%%%%%%%%%%%%%%%%%%%%%%%
\subsection{Chern-Simons term and $\eta$-invariant}\label{sec:eta}

The parameter $k$ is not the only source of the Chern-Simons-level as we will show for the Dyson index $\beta=2$.
For $M\to 0$ the phase of the fermion determinant also contributes by the
$\eta$ invariant $\eta(A)=\sum_{j=1}^{N} \sign(\lambda_j)$, see~\cite{AlvarezGaume:1984nf,Leutwyler:1990an,Redlich:1983dv,Deser:1997gp}, as follows
\be
\prod_{f=1}^{2N_{\rm f}}\det (iA+ m_f\1_N) \overset{|M|\to0}{\approx} |\det A|^{2N_{\rm f}}e^{i\pi N_{\rm f}\sum_{j=1}^N\sign(\lambda_j)}.
\ee
This can be combined with the imaginary part of the first term in the exponent of Eq.~\eref{eq:Zbeta2}, which
can be written as
\be
-2ik \alpha_2 {\rm Tr } A
\overset{N\gg1, |1-\alpha_2|\ll1}{\approx} -\pi i  k  \sum_{j=1}^{N} \sign(\lambda_j).
\label{rmt-cs}
\ee
This approximation can be seen by considering the integral \eref{eq:Zbeta2} for $k =0$ which
gives a quenched approximation for $\Tr A$. The integral can be rewritten as
\be
Z^{\beta = 2} = c \int \rmd A \int \rmd x~ \rme^{-\frac {\alpha_2(1-\alpha_2)}{2} x^2
-\frac N2 \Tr (A - \alpha x/ N)^2}.
  \ee
  For the physically interesting limit of $\alpha = 1 - O(1/N)$, we have that
  $ x \sim O(\sqrt N)$, so that $\Tr A$ fluctuates with a magnitude of
  $O(1/\sqrt N) $, which is much larger than the average level spacing of
  $O(1/N)$. These fluctuations are collective, meaning that all eigenvalues of
  $A$ move up and down with $x$ in the same way. Therefore, starting with
  a configuration with an equal number of eigenvalues on the left
  and right of zero, a fluctuation  where $k$ eigenvalues
  move to the right  changes  $\Tr A$  by
  \be
  \delta \Tr A =  N k \Delta \lambda
  \ee
  and $\Delta \lambda= \pi/N$ the level spacing so that
\be
  N \Delta \lambda k =  \pi k,
  \ee
  The sum of the sign of the eigenvalues changes by
  \be
   \delta   \sum_i \sign \lambda_i = 2k.
  \ee
  This results in the ratio
  \be
  \frac{ \sum_i \lambda_i}{  \sum_i \sign \lambda_i}
  =\frac{ \sum_i \delta \lambda_i}{  \sum_i \delta \sign \lambda_i} = \frac \pi 2,
  \ee
  which is the desired relation \eref{rmt-cs}.
    This relation \eref{rmt-cs} has also been checked numerically,
  see Figure~\ref{fg:sign-ratio}, where the distribution of the ratio $r=\sum_{j=1}^{N} \lambda_j/\sum_{j=1}^{N} \sign (\lambda_j)$ has been generated with Monte Carlo simulations.

 Summarizing, the phase of the fermion determinant renormalizes the
bare Chern-Simons level as
\be
-\pi i k_{\rm rn} \eta(A)=-\pi i(k-N_{\rm f} ) \eta(A).
\ee
Therefore, the action occurs as
in  three-dimensional QCD, see Eq.~(1.4) in \cite{Komargodski:2017keh}\footnote{The relation between the number of flavors $\tilde{N}_{\rm f}$ in~\cite{Komargodski:2017keh}) and our choice $N_{\rm f}$ is $\tilde{N}_{\rm f}=2N_{\rm f}$.}.
We will see in the ensuing discussion that the pattern of chiral symmetry  breaking does not depend on $\alpha_\beta$,
as long as the integrals are convergent, which is certainly the case for any value of $\alpha_\beta <N/(N+4N_{\rm f}/\beta)<1$. The slightly smaller bound than $1$
avoids that the integral over $Tr A$ does not diverge (this can be seen
after splitting $A$ into its trace and a traceless part).

\subsection{\boldmath Derivation of the sigma model at large $N$}
\label{sec:sigma}

The derivations of the partitions functions
\eqref{eq:zN2}, \eqref{eq:zN1} and \eqref{eq:zN4} are 
similar, and we, therefore, outline only the $\beta=2$ class here. The procedure 
follows standard steps~\cite{Shuryak:1992pi,Halasz:1995qb}.  
First we linearize the squared trace term at the expense of a new 
Gaussian integral over an auxiliary variable $x$ and, afterwards,
 shift $A\to A-x\1_N$ to eliminate the linear term in $A$. 
This makes it clear that the partition function of our model is nothing but 
a reweighted integral of the ordinary Gaussian matrix model%
\footnote{Matrix models having a similar structure were studied in   
\cite{Bertuola2004,AbulMagd2005,AkeVivo2008,Kanazawa:2016nlh}.} 
\ba
	Z^{\beta=2} &=\frac{N}{\sqrt{2\pi\alpha_2}}
	\int_{-\infty}^{\infty}\!\!\! \rmd x\; 
	\exp\left[-\frac{N^2} 2 \frac{1-\alpha_2}{\alpha_2}  x^2+2iNkx\right]\calZ_{N,N_{\rm f}}(M-ix)\,,
	\label{eq:Zxhms}
	\\
	\calZ_{N,N_{\rm f}}(M-ix) &\equiv \int \rmd A\; 
	\exp\left[-\frac{N}{2}\Tr A^2\right]
	\prod_{f=1}^{2N_{\rm f}}\det \big[iA+(m_f-ix) \1_{N}\big]\,.  
	\label{eq:calZfasd}
\ea
Upon rewriting the determinant in terms of Grassmann variables,
\begin{equation}
\det \big[i\wt{A}+(m_f-ix) \1_{N}\big]=\frac{\int \rmd \psi_f \rmd \bar{\psi}_f\exp[- \sum_{a,b=1}^{N}\bar{\psi}_f^a (i\wt{A}+(m_f-ix) \1_{N})_{ab}\psi_f^b]}{\int \rmd \psi_f \rmd \bar{\psi}_f\exp[- \sum_{a=1}^{N}\bar{\psi}_f^a \psi_f^a]},
\end{equation}
one can easily integrate out $\wt{A}$,
\be
\begin{split}
	Z^{\beta=2} =&~
	2^{N/2}\left(\frac{\pi}{N}\right)^{N^2/2}\frac{N}{\sqrt{2\pi\alpha_2}}\int_{-\infty}^{\infty} \rmd x \;
	\exp\left[-\frac{N^2} 2 \frac{1-\alpha_2}{\alpha_2} x^2+2iNkx\right]\\
	&\times	\frac{\int \rmd \psi \rmd \bar{\psi} \exp[- \sum_{a=1}^{N}\sum_{f,g=1}^{2N_{\rm f}}\bar{\psi}_f^a (M-ix\1_{2N_{\rm f}})^{}_{fg} \psi_g^a
		+\sum_{f,g=1}^{2N_{\rm f}}(\sum_{a=1}^{N}\bar\psi_f^a\psi_g^a)(\sum_{b=1}^{N}\bar\psi_g^b\psi_f^b)/(2N)]}{\int \rmd \psi_f \rmd \bar{\psi}_f\exp[- \sum_{a=1}^{N}\sum_{f=1}^{2N_{\rm f}}\bar{\psi}_f^a \psi_f^a]} .
\end{split}
\ee
The quartic term in the fermions can be brought into bilinear form by means of the 
Hubbard-Stratonovich transformation. For this purpose we introduce an auxiliary Hermitian $2N_{\rm f}\times 2N_{\rm f}$ matrices $H$. This allows to integrate out 
the Grassmann variables leading to the result
\be
\begin{split}
	Z^{\beta=2} =&~ 2^{(N-2N_{\rm f})/2}\left(\frac{\pi}{N}\right)^{(N^2-4N_{\rm f}^2)/2}\frac{N}{\sqrt{2\pi\alpha_2}}\\
	&\times
	\int_{-\infty}^{\infty} \!\!\! \rmd x 
	\exp\left[-\frac{N^2} 2 \frac{1-\alpha_2}{\alpha_2} x^2+2iNkx\right] \int \rmd H \; 
	\exp\bigg[ - \frac{N}{2}\Tr H^2 
	\bigg]\; {\det}^N (H-ix\1_{2N_{\rm f}}+M).
	\label{eq:Z834}
\end{split}
\ee
After shifting $H\to H+ix\1_{2N_{\rm f}}$ via analytic deformation of the contours, we 
perform the $x$-integral for which we need the stricter bound $\alpha_2<N/(N+2N_{\rm f})$. Then, we obtain
\ba
	Z^{\beta=2}	=2^{(N-2N_{\rm f})/2}\left(\frac{\pi}{N}\right)^{(N^2-4N_{\rm f}^2)/2}\sqrt{\frac{\widetilde{\alpha}_2}{\alpha_2}}\int \rmd H\;
	\exp\bigg[
		- \frac{N}{2}\widetilde{\alpha}_2 (\Tr H-2k)^2 - \frac{N}{2}\Tr H^2
	\bigg]\;{\det}^N (H+M)
	\label{eq:ZNsig}
\ea
with
\be
\frac{1}{\widetilde{\alpha}_2}=N\frac{1-\alpha_2}{\alpha_2}-2N_{\rm f}>0.
\ee
This is the finite $N$ result that is still exact without any approximations. When employing the choice~\eqref{alfbet} the parameter $\widetilde{\alpha}_2$ simplifies, i.e., $\widetilde{\alpha}_2=1$. In the following we keep $\widetilde{\alpha}_2$ fixed.

Let us consider the large-$N$ limit with $\widehat{M}=NM$ fixed. The integral is dominated by saddle point manifolds and fluctuations around them. 
When diagonalizing $H=U^\dagger \Lambda U$ with a real diagonal matrix 
$\Lambda=\diag(\lambda_1,\ldots,\lambda_{N_{\rm f}})$,
 the saddle-point equation in the chiral limit $M=0$ reads 
\begin{equation}
\begin{split}
	& 0\overset{!}{=}\frac{\der}{\der\lambda_n}S_\Lambda \quad{\rm with}\quad 
	S_\Lambda \equiv 
		\frac{\widetilde{\alpha}_2}{2}\bigg(\sum_{i=1}^{2N_{\rm f}}\lambda_i - 2k\bigg)^2 
		+ \frac 12\sum_{i=1}^{2N_{\rm f}}
		(\lambda_i^2 -  \log \lambda_i^2 )
	\\
	& \quad \Longleftrightarrow ~~\frac{1}{\lambda_n} - \lambda_n 
	= \widetilde{\alpha}_2\bigg(\sum_{i=1}^{2N_{\rm f}}\lambda_i - 2k\bigg) 
	\qquad \quad  \text{for}~~n=1,\cdots,2N_{\rm f}\,.
	\end{split}
	\label{eq:sadeq}
\end{equation}
In general, there are multiple real solutions to this equation. However,
we look for the minimum of the real part of $S_\Lambda$ which is achieved for $\Lambda = \Lambda_k$, cf. Eq.~\eqref{eq:Lpdef},
when $k\in\ZZ$ with $|k|\leq N_{\rm f}$. Indeed, the lower bound to the real part is
\ba
	\Re~S_\Lambda & = \frac{\widetilde{\alpha}_2}{2}
		\bigg(\sum_{i=1}^{2N_{\rm f}}\lambda_i - 2k \bigg)^2 
		+ \sum_{i=1}^{2N_{\rm f}}
		\bigg(\frac{1}{2}\lambda_i^2 - \log| \lambda_i|\bigg)
	\overset{\widetilde{\alpha}_2>0}{\geq} \sum_{i=1}^{2N_{\rm f}}
	\bigg(\frac{1}{2}\lambda_i^2 - 
	\log |\lambda_i|\bigg) \geq N_{\rm f} 
\ea
which is saturated only by $\Lambda_k$ and permutation of its diagonal elements. The second inequality follows from the fact that $\lambda_i^2/2 - 
	\log |\lambda_i|$ is a concave function with its two minimums at $\lambda_i=\pm1$. The fluctuations about $\Lambda_k$
give an overall constant, that comprises the sign $(-1)^{Nk}$ from ${\det}^N\Lambda_k$, and can be 
incorporated exactly. As a result we obtain to first order in $M$,
\ba
	Z^{\beta=2} & \sim \int\limits_{\U(2N_{\rm f})} \ms~ \rmd \mu(U)\; 
	{\det}^N\bigg( \1_N+ \frac{U^\dagger \Lambda_k^{-1} U\widehat{M}}{N}\bigg)
	\sim \int\limits_{\U(2N_{\rm f})} \ms~ \rmd \mu(U)\; 
	\exp\big[ \Tr( U \Lambda_k U^\dagger \widehat{M} ) \big].
	\label{eq:Zeffp}
        \ea
where we exploited that $\Lambda_k^{-1}=\Lambda_k$ because $k$ is an integer, cf.~\eqref{eq:Lpdef}. This result hold regardless
of the value of $\alpha_2$  as long as $\widetilde{\alpha}_2$ is fixed. The latter implies that $1-\alpha_2$ is of the order $1/N$. The proper normalization of this partition function is computed in Appendix~\ref{sec:Aunquenched}.

The result~\eqref{eq:Zeffp} realizes the symmetry breaking pattern
\eqref{eq:ssblist} for the $\beta=2$ class.  
The integration for $U$ is effectively over the coset 
$\U(2N_{\rm f})/[\U(N_{\rm f}+k)\times\U(N_{\rm f}-k)]$. This modified symmetry breaking 
pattern is evidently enforced by the squared trace term 
in~\eqref{eq:ZNsig}. The idea of using squared trace terms to constrain the  
symmetry realization is similar to 
the squared trace deformation in Polyakov-loop models 
\cite{Pisarski:2006hz,Myers:2007vc,Unsal:2008ch}. 
The symmetry breaking patterns for $\beta=1$ and $4$ in \eqref{eq:ssblist} 
are realized by the same mechanism. 

Spectral sum rules for the eigenvalues of the Dirac operator 
can be derived by matching the quark mass expansion of the effective 
finite-volume partition function~\eqref{eq:Zeffp} with that of the 
partition function in QCD$_3$. 
They have already been obtained in eq.~(5.17) of \cite{Akemann:2000df} for 
general $k$ from a mathematical perspective. Adapting 
\cite{Akemann:2000df} to our convention, we obtain
\ba
	\left\langle \sum_n \frac{1}{\zeta_n} \right\rangle_k = 
	\frac{ik}{N_{\rm f}}\,, 
	\quad
	\left\langle \sum_n \frac{1}{\zeta_n^2} \right\rangle_k = 
	\frac{2(N_{\rm f}^2-k^2)}{N_{\rm f}(4N_{\rm f}^2-1)}\,, 
	\quad
	\left\langle \bigg( \sum_n \frac{1}{\zeta_n}\bigg)^2 \right\rangle_k = 
	\frac{1-4k^2}{4N_{\rm f}^2-1}
\ea
for a few low order sum-rules. Here $\{i\zeta_n\}$ are the 
Dirac eigenvalues rescaled by the average level spacing with $\zeta_n\in\RR$, while 
$\langle\cdots\rangle_k$ denotes the average with respect to the 
QCD$_3$ action with a Chern-Simons term at level $k$.
These sum-rules are a direct consequence of the pattern of spontaneous symmetry
breaking and are independent of the specific details of the random matrix model.

\subsection{\boldmath Phase transitions at non-integer $k$}
\label{sec:phase}

As the symmetry breaking pattern changes when $k$ is shifted with unit 
increment, there must be a phase transition 
at non-integer values of $k$ for $|k|<N_{\rm f}$.%
\footnote{Note that $k$ in the matrix model can be varied continuously even though 
$k$ in QCD$_3$ is quantized to integers.}
To determine the locus of phase transitions, we have to solve 
the $2N_{\rm f}$ coupled saddle point equations~\eqref{eq:sadeq}.

In the first step we consider
\be 
\lambda_0 \equiv - \frac{\widetilde{\alpha}_2}{2}\bigg(
\sum_{n=1}^{2N_{\rm f}}\lambda_{n} - 2k \bigg)
\ee
as the $(2N_{\rm f}+1)$'st variable. Then, the equations~\eqref{eq:sadeq} for $\lambda_1,\ldots,\lambda_{2N_{\rm f}}$ decouple and can be solved in terms of the auxiliary variable $\lambda_0$, yielding
\ba 
	\label{sol2}
	\lambda_{n}=\lambda_0+L_n\sqrt{\lambda_0^2+1}
\ea 
with $L_n=\pm1$ a sign which is not fixed yet. 
The sum of these signs is denoted by 
$2k_L=\sum_{j=1}^{2N_{\rm f}}L_j$, which plays the role of $2k$ 
for a non-integer $k$. To obtain the solution for $\lambda_0$,
we sum over all $n$ in~\eqref{sol2} and find

\ba 
	\label{epsdf}
	\sum_{n=1}^{2N_{\rm f}}\lambda_n
        =2k - \frac{2\lambda_0}{\tilde{\alpha}_2} = 2N_{\rm f} \lambda_0 + 
	2k_L \sqrt{\lambda_0^2+1}\,.
\ea 
This equation has a unique real solution, because
the right hand side plus $2\lambda_0/\tilde{\alpha}_2$  is strictly monotonically increasing.
The unique solution is
\ba 
	\label{sol3}
	\lambda_0 = \frac{k(N_{\rm f}+1/\tilde{\alpha}_2) - k^{}_L 
	\sqrt{(N_{\rm f}+1/\tilde{\alpha}_2)^2+k^2-k_L^2}}{(N_{\rm f}+1/\tilde{\alpha}_2)^2-k_L^2}\,.
\ea

Summarizing, equations~\eqref{sol2} and \eqref{sol3} yield 
all $2^{2N_{\rm f}}$ saddle points. The solutions only 
depend on the still free integer $k^{}_L=-N_{\rm f},\dots,N_{\rm f}$. 
The solutions for a fixed $k_L^{}$ are 
$\left(\begin{array}{c} 2N_{\rm f} \\ N_{\rm f}+k^{}_L \end{array}\right)$ degenerate. 
The real part of the action $S_\Lambda$ for a fixed $k^{}_L$ is
\begin{equation}
\begin{split}
	\Re\big[S_\Lambda^{(k^{}_L)}\big] 
	=&\; \frac{2}{\widetilde{\alpha}_2}\lambda_0^2 + \sum_{i=1}^{2N_{\rm f}}
	\bigg\{
		\frac{1}{2}\Big(\lambda_0+L_i\sqrt{\lambda_0^2+1}\,\Big)^2 - 
		\log \Big|\lambda_0+L_i\sqrt{\lambda_0^2+1} \,\Big| 
	\bigg\}
	\\
	=&\; 2\bigg(N_{\rm f}+\frac{1}{\widetilde{\alpha}_2}\bigg)\lambda_0^2+2k^{}_L \lambda_0\sqrt{\lambda_0^2+1} - 
	2k_L^{}\log \Big|\lambda_0+\sqrt{\lambda_0^2+1} \,\Big| + N_{\rm f}
%	\\
%	\overset{\eqref{epsdf}}{=}& 2 k \lambda_0 - 
%	2k_L^{}\log \Big|\lambda_0+\sqrt{\lambda_0^2+1} \,\Big|
%	+ N_{\rm f}
%	\\
%	=& 2k \frac{k(N_{\rm f}+1/\tilde{\alpha}_2)-k^{}_L\sqrt{(N_{\rm f}+1/\tilde{\alpha}_2)^2+k^2-k_L^2}}{(N_{\rm f}+1/\tilde{\alpha}_2)^2 - k_L^2} - 
%	2k^{}_L\log \left|\frac{k+\sqrt{(N_{\rm f}+1/\tilde{\alpha}_2)^2+k^2-k_L^2}}{N_{\rm f}+1/\tilde{\alpha}_2+k^{}_L}\right| + N_{\rm f}\,.
\label{eq:salbe}
\end{split}
\end{equation}
This quantity has to be minimized in the integer $k^{}_L=-N_{\rm f},\dots,N_{\rm f}$. 
%Let us define $\gamma=k^{}_L/(N_{\rm f}+1/\tilde{\alpha}_2)$ and $\delta=k/(N_{\rm f}+1/\tilde{\alpha}_2)$. 
%Then we have to look for the minimum of
%\ba 
%	\label{eq:salbe}
%	\frac{\Re\big[S_\Lambda^{(k^{}_L)}\big]-N_{\rm f}}{2(N_{\rm f}+1/\tilde{\alpha}_2)}
%	& = \delta\frac{\delta-\gamma\sqrt{1+\delta^2-\gamma^2}}{1-\gamma^2} - 
%	\gamma\log \left|
%		\frac{\delta+\sqrt{1+\delta^2-\gamma^2}}{1+\gamma}
%	\right|
%\ea 
%as a function of $\gamma\in(-1,1)$.

When $k$ is an integer with $|k|\leq N_{\rm f}$, \eqref{eq:salbe} has 
a unique minimum at $k^{}_L=k$ when $\lambda_0 =0$. 
Note that this minimum is completely independent of $N_{\rm f}$
and $\widetilde{\alpha}_2$. Thus the discussion is valid for any number of
flavors. The $N_{\rm f}$ and $\widetilde{\alpha}_2$ 
dependence enters the game when we studying the phase transition point with a real-valued $k$.
Then we have to compare the actions $S_\Lambda^{(\lfloor k\rfloor)}$ and 
$S_\Lambda^{(\lceil k\rceil)}$ with the floor function $\lfloor.\rfloor$ 
and the ceiling function $\lceil.\rceil$ yielding the largest integer 
smaller than or equal to $k$ and the smallest integer larger than 
or equal to $k$, respectively. 
The phase transition then happens when
\ba 
	\label{phasetranseq}
\begin{split}
	\Re\big[S_\Lambda^{(\lfloor k\rfloor)}\big]=\Re\big[S_\Lambda^{(\lceil k\rceil)}\big].
\end{split}
\ea
This is a transcendental equation in $k$ which always has one solution 
in each interval $[j,j+1]$ with $j=-N_{\rm f},-N_{\rm f}+1,\ldots,N_{\rm f}-1$. 
The change of $S_\Lambda^{(\lfloor k\rfloor)}$ to 
$S_\Lambda^{(\lceil k\rceil)}$ 
obviously exhibits a kink in the parameter $k^{}_L$. 
Therefore these phase transitions are of first order.
\begin{table}[tb]
	\setlength{\tabcolsep}{5pt}
	\begin{tabular}{c|cccccc}
		%\hline \hline 
		$N_{\rm f}$ & $[0,1]$ & $[1,2]$ & $[2,3]$ & $[3,4]$ & $[4,5]$&$[5,6]$  
		\\\hline 
		0 & 0.52770 & --- & --- & --- & --- & -- 
		\\
		1 & 0.50551 & 1.51674 & --- & --- & ---  & -- 
		\\
		2 & 0.50237 & 1.50713 & 2.51194 & --- & ---   & -- 
		\\
		3 & 0.50132 & 1.50396 & 2.50661 & 3.50927 & ---  & -- 
		\\
		4 & 0.50084 & 1.50252 & 2.50420 & 3.50589 & 4.50759  & -- 
		\\
		5 & 0.50058 & 1.50175 & 2.50291 & 3.50408 & 4.50525  & 5.50642
		%\\
		%\hline \hline 
	\end{tabular}
	\caption{\label{tb:ptpoints} Location of first-order phase transitions in each interval of $k$ for the Dyson index $\beta=2$ and the choice $\widetilde{\alpha}_2=1$, in particular we have chosen Eq.~\eqref{alfbet} for the ``Chern-Simons'' coupling $\alpha_2$.
          The critical point in the interval $[N_{\rm f},N_{\rm f}+1]$
        is obtained by solving \eref{phasetranseq} for 
         $N_{\rm f}+\epsilon$ with $\epsilon \to 0$. }
\end{table}

Phase transition points for general $N_{\rm f}$ are 
located symmetrically on the positive and negative sides of the $k$ axis, 
and hence it is sufficient to look for solutions to \eqref{phasetranseq} 
for $k>0$. Table~\ref{tb:ptpoints} is a summary for $1\leq N_{\rm f}\leq 5$ with the Dyson index $\beta=2$ and the ``Chern-Simons'' coupling $\alpha_2$ chosen as in~\eqref{alfbet}.  
Only in the limit $N_{\rm f}\to\infty$ do the half integers 
$k=n+\frac{1}{2}\;(n\in\ZZ)$ become the phase transition points. 
For a finite number of flavors we get corrections, which can be computed 
via a large-$N_{\rm f}$ expansion of \eqref{phasetranseq} and
the solution $k=(n+1/2)\sum_{j=0}^{\infty}c_j/(N_{\rm f}+1/\tilde{\alpha}_2)^j$.  
Assuming $n=\calO(1)$, the corrected transition points are
\ba 
	\label{papproxf}
	k = \left[ 1 + \frac{1}{24}\frac{1}{(N_{\rm f}+1/\tilde{\alpha}_2)^2} 
	+ \frac{17}{1920}\frac{1}{(N_{\rm f}+1/\tilde{\alpha}_2)^4}
	+ \calO\big((N_{\rm f}+1/\tilde{\alpha}_2)^{-6}\big)\right] 
	\left(n+\frac{1}{2}\right) 
\ea
for $n\in\ZZ$. The residue  
$\calO\big((N_{\rm f}+1/\tilde{\alpha}_2)^{-6}\big)$ may depend on $n$, as well, though it seems to be a very weak dependence.

\subsection{\boldmath Partition function at finite and large $N$}
\label{sec:part}

In this section, we evaluate the partition function at finite $N$ and use this
result to derive its large-$N$ limit. 
This discussion serves two purposes. First and foremost,
the  finite $N$ results enable us to study
the approach to the thermodynamic limit.
Second, it provides an  independent consistency check of the large $N$ result~\eqref{eq:Zeffp}.

The partition function~\eqref{eq:Zxhms} is a one-parameter 
integral over a GUE partition function with $2N_{\rm f}$ flavors.
The latter will be rewritten using the identity \cite{Akemann:2002vy,Strahov:2002zu},
\ba
	\frac{\int \rmd A\;\exp[-\frac{N}{2}\Tr A^2]
	\prod_{\ell=1}^{L} \big [\det (\lambda_\ell \1_{N}-A) \det (\mu_\ell \1_{N}-A) \big]}{\int \rmd A\;\exp[-\frac{N}{2}\Tr A^2]}= \frac{C_{N,L}}{\Delta_L(\lambda)\Delta_L(\mu)} \det_{1\leq a,b\leq L}
	\kkakko{ \KK^{(N)}_{N+L}(\lambda_a, \mu_b) },
	\label{eq:qcd3part}
\ea
where the integration is over hermitian $N\times N$ matrices, 
and we employ the notation $\Delta_L(.)$ for the Vandermonde determinant with the exceptional case $\Delta_1\equiv 1$. 
The kernel is given by
\ba
	\KK^{(N)}_{j}(\lambda_a,\,\mu_b) \equiv \frac{
		P^{(N)}_{j} (\lambda_a) P^{(N)}_{j-1} (\mu_b) - 
		P^{(N)}_{j} (\mu_b) P^{(N)}_{j-1} (\lambda_a)
	}{\lambda_a-\mu_b}\,, 
	\label{eq:Kdddfe}
        \ea
with the monic polynomials
\ba
	P^{(N)}_{\ell}(s) \equiv \frac{1}{(2N)^{\ell/2}}
	H_{\ell} \bigg(\sqrt{\frac{N}{2}}\,s \bigg)\,, 
	\qquad k=0,1,2,\ldots 
\ea 
where $H_{\ell}(x)=\rme^{x^2}(-\partial)^\ell\rme^{-x^2}$ 
are the Hermite polynomials. The polynomials $P^{(N)}_{\ell}(s)$ satisfy the orthogonality relation 
\ba
	\int_{-\infty}^{\infty} \!\! \rmd s \; \rme^{-Ns^2/2} P^{(N)}_{\ell}(s)P^{(N)}_{m}(s) 
	= \delta_{\ell m} h^{(N)}_{m},
	\qquad  
	h^{(N)}_{m} \equiv \frac{\sqrt{2\pi}\,m!}{N^{m+\frac{1}{2}}} \,,
\ea        
and the  normalization constant is given by
\ba
	C_{N,L} \equiv \kkakko{h_{N+L-1}^{(N)}}^{-L}\prod_{i=N}^{N+L-1} h_{i}^{(N)}.
\ea

In our case we have $2N_{\rm f}$ flavors, each with its own mass $m_f$,
so that we have no natural splitting into two sets of masses.
We choose $\lambda_j=x+i m_j$ and
$\mu_j=x+i\wt{m}^{}_j\equiv x+i m^{}_{N_{\rm f}+j}$ for $j=1,\cdots,N_{\rm f}$. 
Applying~\eqref{eq:qcd3part} to the partition function~\eqref{eq:Zxhms}, we obtain,
 up to irrelevant normalization,
 \ba
	\ms ~Z^{\beta=2} 
	& \sim 
	(-1)^{N_{\rm f}N}\int_{-\infty}^{\infty}\!\! \rmd x\;
        \frac{\exp[-N\mkakko{N_{\rm f}+\frac12}x^2 + 2iNkx]}{\Delta_{N_{\rm f}}(x+im)\Delta_{N_{\rm f}}(x+i\wt{m})}
	\det_{1\leq a,b\leq N_{\rm f}}\kkakko{ \KK^{(N)}_{N+N_{\rm f}}(x+im_a, x+i\wt{m}_b) }
	\label{eq:Z2kern}
	\\
	& \sim  \frac{(-1)^{N_{\rm f}(N_{\rm f}-1)/2+N_{\rm f}N}}{\Delta_{N_{\rm f}}(m)\Delta_{N_{\rm f}}(\wt{m})}
	\int_{-\infty}^{\infty}\!\! \rmd x\; \exp\left[-N\mkakko{N_{\rm f}+\frac12}x^2 + 2iNkx\right]
	\notag
	\\
	& \quad \times 
	\det_{1\leq a,b\leq N_{\rm f}} \!\! \kkakko{
		\frac{
			\scalebox{0.85}{$\displaystyle 
			H_{N+N_{\rm f}} \bigg( \sqrt{\frac{N}{2}}\,(x+im_a) \bigg) 
			H_{N+N_{\rm f}-1} \bigg( \sqrt{\frac{N}{2}}\,(x+i\wt{m}_b) \bigg) - 
			H_{N+N_{\rm f}} \bigg( \sqrt{\frac{N}{2}}\,(x+i\wt{m}_b) \bigg) 
			H_{N+N_{\rm f}-1} \bigg( \sqrt{\frac{N}{2}}\,(x+im_a) \bigg)
			$}
		}{i(m_a-\wt{m}_b)}
	}
	\!, \ms \nn
	\label{eq:ZfinNmain}
\ea
where we used the translation symmetry and homogeneity of the Vandermonde determinant
$\Delta_{N_{\rm f}}(x+im)=\Delta_{N_{\rm f}}(im)=i^{N_{\rm f}(N_{\rm f}-1)/2}\Delta_{N_{\rm f}}(m)$
and likewise for $\wt{m}$. The sign $(-1)^{N_{\rm f}}$ results from pulling a factor $i$ out
of each determinant in~\eqref{eq:Zxhms}.

In the simplest case $k=0$,  due to the the Gaussian factor,
the variable $x$ scales as $\calO(1/\sqrt{N})$    
while the masses $m$ and $\wt{m}$ are of order $\calO(1/N)$.  When exploiting the asymptotic form,
\be
H_N(t)\sim \rme^{t^2/2}\cos\Big[\big(2N+\frac{1}{2}\big)\frac{t}{\sqrt{2N}}
-\frac{N\pi}{2}\Big]
\ee
at $N\gg 1$ and $t=\calO(1)$, we obtain
\ba
	\frac{\scalebox{0.75}{$\displaystyle 
			H_{N+N_{\rm f}} \bigg( \sqrt{\frac{N}{2}}\,(x+im_a) \bigg) 
			H_{N+N_{\rm f}-1} \bigg( \sqrt{\frac{N}{2}}\,(x+i\wt{m}_b) \bigg) - 
			H_{N+N_{\rm f}} \bigg( \sqrt{\frac{N}{2}}\,(x+i\wt{m}_b) \bigg) 
			H_{N+N_{\rm f}-1} \bigg( \sqrt{\frac{N}{2}}\,(x+im_a) \bigg)
			$}
	}{i(m_a-\wt{m}_b)} 
	\sim  \rme^{Nx^2/2}\frac{\sinh[N(m_a-\wt{m}_b)]}{m_a-\wt{m}_b}. \!\!\!  
\ea 
Thereupon, the integral over $x$ factorizes, and we are left with the simpler expression,
\ba\label{asympk0}
	Z^{\beta=2} \sim \frac{(-1)^{N_{\rm f}(N_{\rm f}-1)/2+N_{\rm f}N}}
	{\Delta_{N_{\rm f}}(m)\Delta_{N_{\rm f}}(\wt{m})}
	\det_{1\leq a,b\leq N_{\rm f}} \!\! \ckakko{
		\frac{\sinh[N(m_a-\wt{m}_b)]}{m_a-\wt{m}_b}
	}.
\ea
For a parity-invariant mass $\wt{m}=-m$ it coincides  with 
the partition function obtained previously \cite{Szabo:2000qq,Szabo:2005gi}.

One can also obtain the result~\eqref{asympk0} from the low-energy limit of
the partition function given in  ~\eqref{eq:zN2} for $k=0$. For this purpose we exploit the parameterization~\cite{Splittorff:2003cu,KiebKana}
\begin{equation}
U^\dagger\Lambda_0U=\diag(U_1^\dagger,U_2^\dagger)\left(\begin{array}{cc} \cos\Phi & \sin\Phi \\ \sin\Phi & \cos\Phi \end{array}\right)\diag(U_1,U_2),\ {\rm with}\ \Phi=\diag(\varphi_1,\ldots,\varphi_{N_{\rm f}})\in[0,\pi]^{N_{\rm f}}\ {\rm and}\ U_1,U_2\in{\rm U}(N_{\rm f}).
\end{equation}
The measure is then
\begin{equation}
\rmd\mu(U)\sim \Delta_{N_{\rm f}}^2(\cos\Phi)\prod_{j=1}^{N_{\rm f}}\sin\varphi_j\rmd\varphi_j \rmd\mu(U_1)\rmd\mu(U_2),
\end{equation}
and the Lagrangian takes the form
\begin{equation}
\Tr U^\dagger\Lambda_0U \widehat{M}=N(\Tr U_1^\dagger \cos\Phi U_1 m-\Tr U_2^\dagger \cos\Phi U_2 \wt{m}).
\end{equation}
The integrals over $U_1$ and $ U_2$ are each Harish-Chandra--Itzykson--Zuber integrals~\cite{HC,IZ},
\begin{equation}
\int \rmd\mu(U_1)\exp[N\Tr U_1^\dagger \cos\Phi U_1 m]\sim\frac{\det_{1\leq a,b\leq N_{\rm f}}[\exp[Nm_a\cos\varphi_b]]}{\Delta_{N_{\rm f}}(m)\Delta_{N_{\rm f}}(\cos\Phi)}
\end{equation}
and similar for $\wt{m}$. The remaining integrals in
\begin{equation}
  Z^{\beta=2} \sim \int\prod_{j=1}^{N_{\rm f}}\sin\varphi_j\rmd\varphi_j
  \frac{\det_{1\leq a,b\leq N_{\rm f}}[\exp[Nm_a\cos\varphi_b]]\det_{1\leq a,b\leq N_{\rm f}}[\exp[-N\wt{m}_a\cos\varphi_b]]}{\Delta_{N_{\rm f}}(m)\Delta_{N_{\rm f}}(\wt{m})}
\end{equation}
over $\Phi$ can be performed with the Andr\'eief identity~\cite{Andreief} yielding~\eqref{asympk0}.

\begin{figure}[t!]
	\centering
	\includegraphics[width=.45\textwidth]{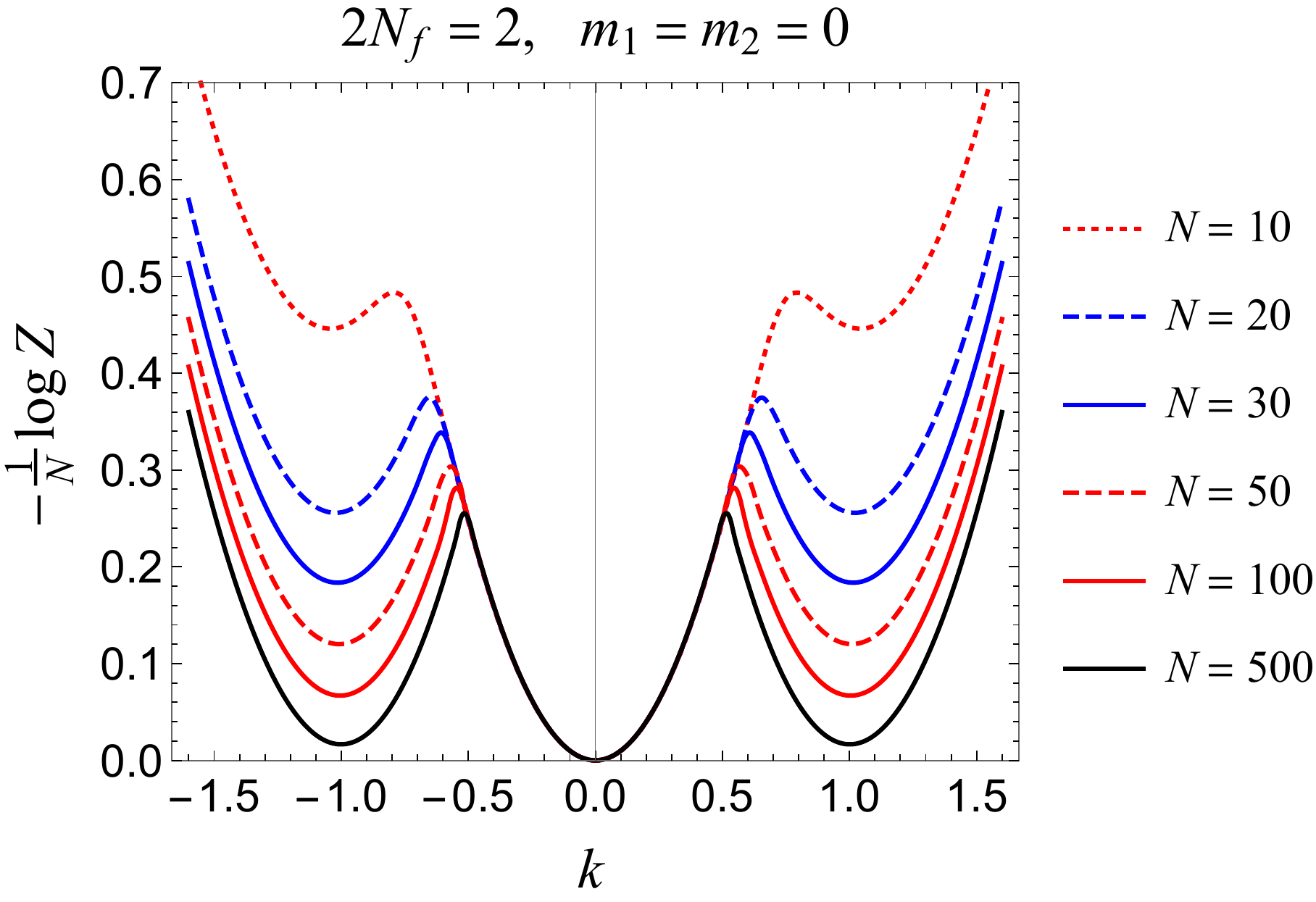}\hfill
	\includegraphics[width=.45\textwidth]{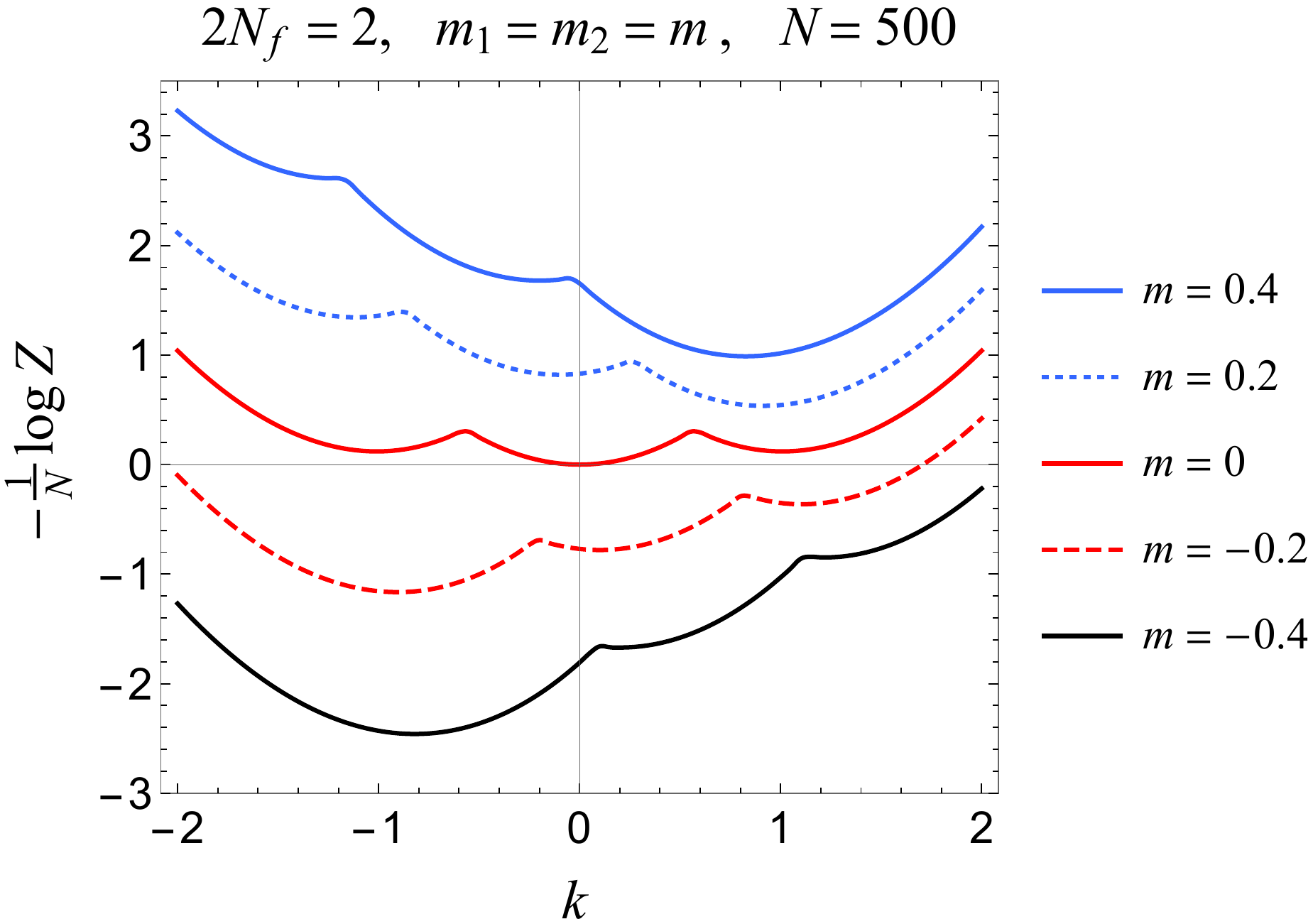}
\caption{\label{fg:pot_finite_N}
          The free energy for two flavors with equal mass in the chiral limit for varying $N$
          for zero mass (left) and different masses for $N = 100$ (right). 
	  The value at $k=0$ is subtracted for each $N$.
	The spacing between curves in the right plot has been adjusted by hand 
	for better visibility.
        }
\end{figure}

Let us examine the simplest case of two flavors ($2N_{\rm f}=2$) with a $\U(2)$-invariant mass $m_1=\wt{m}_1=m$. Then,
Eq.~\eqref{eq:ZfinNmain} simplifies to
\ba
	Z_{2N_{\rm f}=2}^{\beta=2}(m;k) & \sim
	(-1)^{N_{\rm f}N +1} \rme^{Nm(4k+3m)/2}\int_{-\infty}^{\infty}\!\! \rmd x\; \rme^{-3x^2}
	\cos \big(\sqrt{2N}(2k+3m)x \big)
	\Big[ 
		H_{N+2}(x)H_N(x) - H_{N+1}(x)^2 
	\Big] 
	\label{eq:Z2sfd}
\ea
after substituting $x\to\sqrt{2/N}x-i m$.
This integral can be carried out with the help of a  formula 
in \cite[Sec.~7.374, eq.~9]{GRmathbook2007}, with the result 
\ba
	Z_{2N_{\rm f}=2}^{\beta=2}(m;k) & \sim \rme^{- 2Nk^2/3}
	\sum_{\ell=1}^{N+1}  
	\frac{3^\ell}{(\ell-1)!(N+2-\ell)!(N+1-\ell)!}
	H_{2N+2-2\ell}\bigg( \frac{i\sqrt{3N}(2k+3m)}{6} \bigg)\,.
	\label{eq:fastZ2s}
\ea
This form is suited for fast numerical evaluation.
Since $Z(m;k)=Z(-m;-k)$, one may assume $m\geq0$ without loss of generality. 
The $k$ dependence of the partition function can now be investigated 
via \eqref{eq:fastZ2s}.

%%%%%%%%%%%%%%%%%%%%%%%%%%%%%%%
\begin{figure}[b!]
        \hfill\includegraphics[width=.42\textwidth]{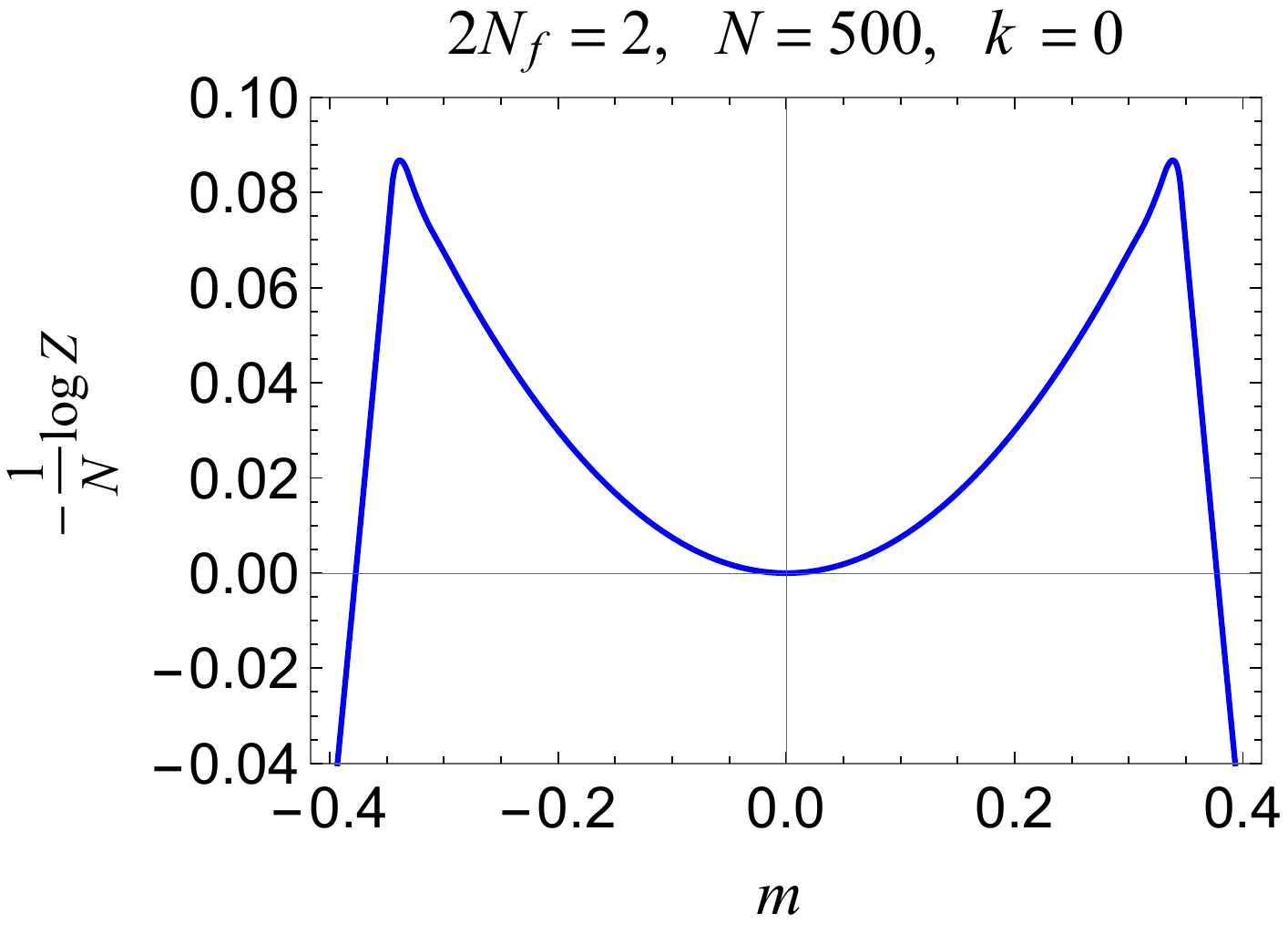}\hfill
   \includegraphics[width=.45\textwidth]{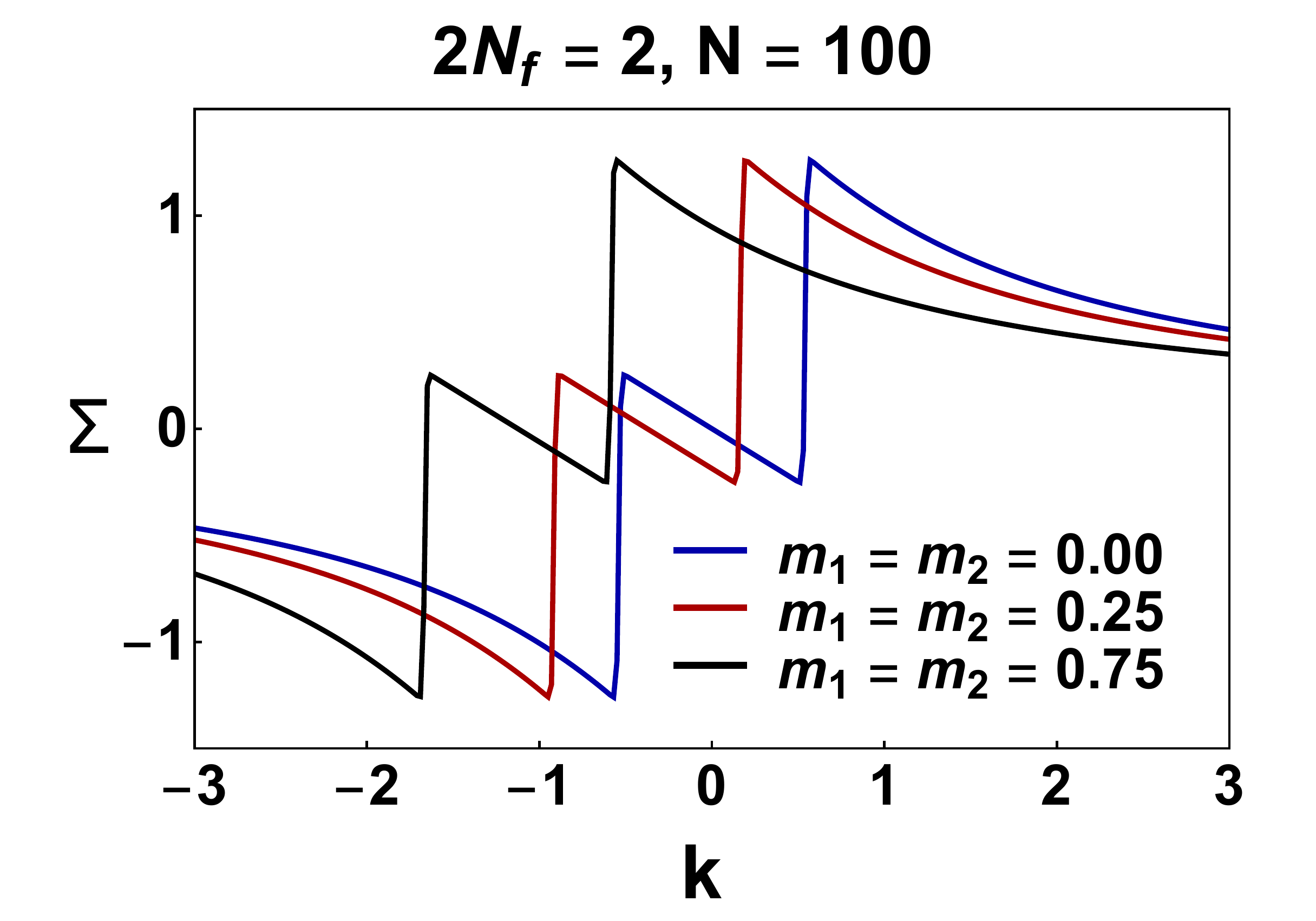}
  \caption{\label{fg:cond}The mass dependence of the
    free energy  for $k=0$ for two flavors with mass $(m_1,m_2)=(m,m)$ (left)
    and the $k$ dependence of the chiral
    condensate, $\Sigma$, for three different values of the mass also for
    two flavors with equal mass.
  } 
%        Note that the masses are not in the microscopic limit because $Nm\gg1$. Therefore, the phase transition points shift.
\end{figure}

Figure~\ref{fg:pot_finite_N} (left) displays the evolution of the free energy
with $N$ in the microscopic limit. Evidently there are quite strong deviations from the large $N$ limit even for quite moderate matrix sizes like $N=50$ which usually yields close to perfect agreement for the microscopic level density.
As $N$ grows, there appear kinks that get sharper. They represent  
first order transitions  in the thermodynamical limit. The region around the origin is 
the flavor symmetry broken phase $\U(2)\to \U(1)\times\U(1)$ 
with massless Nambu-Goldstone modes. 
The other two regions are the symmetry-restored phases. 
The right plot of Figure~\ref{fg:pot_finite_N}
 illustrates the shift of 
of the phase transition points  as a function  of the masses that are of order $\mathcal{O}(1)$ instead of $\mathcal{O}(1/N)$, and hence they are outside the microscopic domain.
The two kinks move 
towards negative $k$ (positive $k$) for $m>0$ ($m<0$), respectively. 

Figure~\ref{fg:cond} (left plot) shows the mass dependence of the free energy 
at $k=0$, again at an $N$ independent mass, i.e., $Nm\gg1$. The two pronounced kinks at 
$m\approx \pm 0.34$ indicate  a strong first-order transition that corresponds to the passage of a kink over the origin in the right plot of Fig.~\ref{fg:pot_finite_N}. 
The middle region corresponds to a  symmetry-broken phase with massless
Nambu-Goldstone bosons% 
\footnote{Note that the pions remain massless in the presence of flavor-symmetric 
fermion masses. This is an important difference from four-dimensional 
QCD where the quark mass inevitably breaks the flavor symmetry of quarks.} 
whereas the outer regions are symmetric gapped phases. 
Such phase transitions at nonzero masses 
were argued to exist in \cite{Komargodski:2017keh} and our matrix 
model serves as a toy model for this phenomenon.  

Finally, in Fig.~\ref{fg:cond} (right)  we show the $k$ dependence of the quark-antiquark condensate defined as
\be
\Sigma = \frac 1{2N} \frac{\der}{\der m} \log  Z(m;k)
\ee
for three different values of the mass, $m_1=m_2=0.0$, $m_1=m_2=0.25$ an $m_1=m_2=0.75$. Hence,  also in Fig.~\ref{fg:cond}  we do not show the microscopic limit of the chiral condensate. The masses are of order $\mathcal{O}(1)$ and not $\mathcal{O}(1/N)$. In the microscopic limit the two kinks are at exactly the same position as in Fig.~\ref{fg:pot_finite_N} (left).  The condensate has still a discontinuity at the values of $k$ for which the free
energy has a kink for masses which are of order one. Yet, for increasing mass the kinks move to infinity, approximately as $\sim \frac 32 m$,
and the quenched result is recovered for $m \to \infty$. For masses with opposite
sign, $m_1 = -m_2$, the $k$-dependence of the free energy remains symmetric
about zero, but the kinks move away from zero, again like $\sim \frac 32 m$ for large $m$,
so that for $m\to \infty$ the quenched limit is recovered.

%%%%%%%%%%%%%%%%%%%%%%%%%%%%%%%

\section{\label{sc:specf}Spectral Correlation Functions}

This section is mostly devoted to the level density, the quark-antiquark condensate and their microscopic large-$N$ limit, though in Subsection~\ref{sec:general}  we also study all $k$-point correlation functions at finite $N$. 
In this subsection we obtain an exact expression for the spectral
density at finite $N$. To illustrate  what happens
when $N$ is increasing at fixed Chern-Simons coupling $k$, we consider
the quenched level density ($N_{\rm f}=0$) at finite $N$ in Subsection~\ref{sec:quenched}. 
This result is amenable to
a saddle point approximation which allows us  to obtain the microscopic limit
of the spectral density (see section \ref{sec:micro}) defined as
\be
\widehat R(\widehat \lambda; k) \equiv
\lim_{N\to \infty} \frac 1N R\bigg(\frac{\widehat \lambda}{N}; k\bigg)\,.
\ee

For more flavors the microscopic limit can be derived much more easily 
from an expression where the spectral density is given by a ratio
of an $N_{\rm f}+2$ flavor partition function and an $N_{\rm f}$ flavor partition
function as is discussed in section~\ref{sec:part}.  Starting from
this result we obtain in section \ref{sec:unquenched} the microscopic
spectral density for $N_{\rm f}$ flavors. In particular, we will work out the one-flavor
case in detail.

For $N_{\rm f}=1$ the partition function is given by the sum of three saddle points
(see section~\ref{sec:phase})
\be
Z_{N_{\rm f}=1}(m_1,m_2;k) = Z_{N_{\rm f}=1,k_L=-1}(m_1,m_2;k)+ Z_{N_{\rm f}=1,k_L=0}(m_1,m_2;k)+ Z_{N_{\rm f}=1,k_L=1}(m_1,m_2;k).
\ee
Because of the reweighted structure of the partition function,
the spectral density is given by
\be
R(\lambda;k) =\frac 1{Z_{N_{\rm f}=1}(m_1,m_2;k)}
 \sum_{k_L=-N_{\rm f}}^{N_{\rm f}} [Z_{N_{\rm f}=1,k_L}(m_1,m_2;k)R_{N_{\rm f}=1,k_L}(\lambda;k)]
  \ee
  with the level density corresponding to the saddle point $k_L$ given by
  \be
  R_{N_{\rm f}=1,k_L}(\lambda;k) = \frac{\big\langle\sum_k\delta(\lambda-\lambda_k) \big\rangle_{Z_{N_{\rm f}=1,k_L}} }{Z_{N_{\rm f}=1,k_L}}.
      \ee
      Since the partition functions behave as
      \be
      Z_{N_{\rm f}=1,k_L} \sim e^{-N f_{k_L}},
      \ee
      where $f_{k_L} $ is the free energy of the $k_L$'th saddle point, we have
      that in the large-$N$ limit, the spectral density is dominated by
      one saddle point (unless $k$ is exactly at the phase transition
      point).
      To derive the results of subsection~\ref{sec:unquenched},
      we make use of  results obtained in Appendix~\ref{sec:app}.

\subsection{\boldmath General $N_{\rm f}$ at finite $N$}\label{sec:general}

The matrix model approach gives us a way to investigate 
spectral fluctuations of the Dirac operator in QCD$_3$. The simplest way 
to compute the $n$-point correlation functions $R(\lambda_1,\dots,\lambda_n;M;k)$ 
of the matrix $A$ in the ensemble \eqref{eq:Zbeta2} 
would be to make use of the reweighted 
structure \eqref{eq:Zxhms}. If $\wt{R}(\lambda_1,\dots,
\lambda_n;M;x)$ is the $n$-point spectral correlation function 
of $A$ for the GUE ensemble $\calZ_{N,N_{\rm f}}(M-ix)$ with $2N_{\rm f}$
flavors, 
we simply have
\ba
	\label{eq:Rreweig}
	R(\lambda_1,\dots,\lambda_n;M;k) \sim  
	\frac{
		\displaystyle
		\int_{-\infty}^{\infty}\!\!\! \rmd x\; 
		\rme^{-N(N_{\rm f}+\frac12)x^2+2iNkx}\calZ_{N,N_{\rm f}}(M-ix) 
		\wt{R}(\lambda_1,\dots,\lambda_n;M;x)
	}{
		\displaystyle \int_{-\infty}^{\infty}\!\!\! \rmd x\; 
		\rme^{-N(N_{\rm f}+\frac12)x^2+2iNkx}\calZ_{N,N_{\rm f}}(M-ix)
	}\,.
\ea
The spectral correlation functions corresponding
to the partition function
$\calZ_{N,N_{\rm f}}(M-ix)$ have been computed in~\cite{NagaoSlevin93,Verbaarschot:1994ip} 
for the massless case, and in~\cite{Damgaard:1997pw,Akemann:1998ta,
Szabo:2000qq,Szabo:2005gi} for nonzero masses with pairwise
 opposite signs. By substituting the results from~\cite{Damgaard:1997pw,Akemann:1998ta,Szabo:2000qq,Szabo:2005gi} 
into $\wt{R}$ we obtain the spectral functions for our matrix model. 

In this paper we only  explicitly work out the one-point function 
(spectral density) for arbitrary $2N_{\rm f}$ masses. 
Recalling the shift $A\to A-x\1_N$, we get from \eqref{eq:calZfasd}
\begin{equation}
\begin{split}
	\wt{R}(\lambda; M;x)  =&~ 
	\frac{1}{\calZ_{N,N_{\rm f}}(M-ix)}\int \rmd A\; 
	\Tr \delta(\lambda+x-A)
	\rme^{-\frac{N}{2}\Tr A^2}
	\prod_{f=1}^{2N_{\rm f}}\det \big[iA+(m_f-ix) \1_{N}\big]
	\\
	\sim &~ \frac{(-1)^{N_{\rm f}N}}{\calZ_{N,N_{\rm f}}(M-ix)}
	\int_{\RR^N} \!\! \rmd a^{}_1\cdots \rmd a^{}_N~ 
	\delta(\lambda+x-a_N^{})
	\rme^{-\frac{N}{2}\sum_{n=1}^{N}a_n^2}\Delta_N^2(a)
	\prod_{f=1}^{2N_{\rm f}}\prod_{n=1}^{N}(a_n-(x+im_f))
	\\
	=& ~ \frac{(-1)^{N_{\rm f}N}
		\rme^{-\frac{N}{2}(\lambda+x)^2}
		\prod_{f=1}^{2N_{\rm f}}(\lambda-im_f)
	}{\calZ_{N,N_{\rm f}}(M-ix)}
	\int_{\RR^{N-1}} \!\! \rmd a^{}_1\cdots \rmd a^{}_{N-1} 
	\rme^{-\frac{N}{2}\sum_{n=1}^{N-1}a_n^2}\Delta_{N-1}^2(a)
	\\
	& \times \prod_{n=1}^{N-1} 
	\bigg[(a_n-x-\lambda)^2\prod_{f=1}^{2N_{\rm f}}(a_n-(x+im_f))
	\bigg]\,.
	\end{split}
\end{equation}
The remaining integral is nothing but the partition function of GUE with 
$2N_{\rm f}+2$ flavors. Thus, it can be expressed as an integral over  a
hermitian $(N-1)\times(N-1)$ 
matrix $B$,
\ba
	\label{eq:R451}
	\wt{R}(\lambda; M;x) & \sim  (-1)^{N_{\rm f}-N+1}
	\frac{
		\rme^{-\frac{N}{2}(\lambda+x)^2}
		\prod_{f=1}^{2N_{\rm f}}(\lambda-im_f)
	}{\calZ_{N,N_{\rm f}}(M-ix)} \int \rmd B\;
	\rme^{-\frac{N}{2}\Tr B^2}\prod_{f=1}^{2N_{\rm f}+2}
	\det [iB+(m_f-ix)\1_{N-1}]
\ea
where we have defined $m^{}_{2N_{\rm f}+1}=m^{}_{2N_{\rm f}+2}=-i\lambda$. 
This matrix integral can be computed with the help of \eqref{eq:qcd3part}. 
In doing so, the degeneracy of $m^{}_{2N_{\rm f}+1}$ and $m^{}_{2N_{\rm f}+2}$ 
must be lifted slightly to avoid an apparent singularity in \eqref{eq:qcd3part}. Labeling  
\be
	\begin{split}
		(\kappa^{}_1,\cdots,\kappa^{}_{N_{\rm f}},\kappa^{}_{N_{\rm f}+1}) 
		& \equiv (m^{}_1, \cdots, m^{}_{N_{\rm f}}, -i\lambda)\,,
		\\
		({\wt\kappa}^{}_1,\cdots,{\wt\kappa}^{}_{N_{\rm f}},{\wt\kappa}^{}_{N_{\rm f}+1}) 
		& \equiv (\wt{m}^{}_{1},\cdots, \wt{m}^{}_{N_{\rm f}}, -i\lambda + \epsilon) 
		= (m^{}_{N_{\rm f}+1},\cdots,m^{}_{2N_{\rm f}}, -i\lambda + \epsilon)\,, 
		\qquad \epsilon=0^+
	\end{split}
\ee
and combining~\eqref{eq:qcd3part},~\eqref{eq:Z2kern},~\eqref{eq:Rreweig}
 and~\eqref{eq:R451}, the spectral density is finally obtained as
\ba
	R(\lambda; M;k) & \sim 
	\rme^{-\frac{N}{2}\lambda^2}\prod_{f=1}^{2N_{\rm f}}(\lambda-im_f) 
	\notag
	\\
	& \quad \times 
	\lim_{\epsilon\to0}\frac{
			\displaystyle \frac{(-1)^{(N_{\rm f}+1)N_{\rm f}/2}}{\Delta_{N_{\rm f}+1}(\kappa)\Delta_{N_{\rm f}+1}(\wt{\kappa})}
			\int_{-\infty}^{\infty} \!\! \rmd x\;
			\rme^{- N[(N_{\rm f}+1)x^2 - (2ik-\lambda ) x]}
			\det_{1\leq a,b\leq N_{\rm f}+1}\kkakko{ \KK^{(N)}_{N+N_{\rm f}}(x+i\kappa_a, x+i\wt{\kappa}_b) }
		}{
			\displaystyle 
			\frac{(-1)^{N_{\rm f}(N_{\rm f}-1)/2}}{\Delta_{N_{\rm f}}(m)\Delta_{N_{\rm f}}(\wt{m})}
			\int_{-\infty}^{\infty}\!\! \rmd x\; \rme^{-N\big[\mkakko{N_{\rm f}+\frac12}x^2 - 2ikx\big]}
			\det_{1\leq a,b\leq N_{\rm f}}\kkakko{ \KK^{(N)}_{N+N_{\rm f}}(x+im_a, x+i\wt{m}_b) }
		}
	\notag
	\\
	& \sim (-1)^{N_{\rm f}}\rme^{-\frac{N}{2}\lambda^2}
	\notag
	\\
	& \quad \times
	\scalebox{0.9}{$\displaystyle 
	\frac{
		\displaystyle \int_{-\infty}^{\infty}\!\! \rmd x\;
		\rme^{- N[(N_{\rm f}+1)x^2 - ( 2ik-\lambda ) x]}
		\det \kkakko{ 
			\begin{array}{cc} \left\{ 
				\KK^{(N)}_{N+N_{\rm f}}(x+im_a, x+i\wt{m}_b) 
				\right\}_{\substack{1\leq a \leq N_{\rm f} \\ 1\leq b \leq N_{\rm f}}} 
				& 
				\left\{ 
				\KK^{(N)}_{N+N_{\rm f}}(x+im_a, x+\lambda) 
				\right\}_{1\leq a \leq N_{\rm f}}
				\\
				\left\{ 
				\KK^{(N)}_{N+N_{\rm f}}(x+\lambda, x+i\wt{m}_b) 
				\right\}_{1\leq b \leq N_{\rm f}} 
				& [\der\KK]^{(N)}_{N+N_{\rm f}}(x+\lambda)
			\end{array}
		}
	}{
		\displaystyle \int_{-\infty}^{\infty}\!\! \rmd x\; \rme^{-N \big[\mkakko{N_{\rm f}+\frac12}x^2 - 2ikx\big]}
		\det_{1\leq a,b\leq N_{\rm f}}\kkakko{ \KK^{(N)}_{N+N_{\rm f}}(x+im_a, x+i\wt{m}_b) }
	}
	$},
	\label{eq:R1res}
\ea
where we employ the notation
\ba
	[\der\KK]^{(N)}_{n}(\mu)  \equiv 
	\lim_{\lambda\to\mu}\KK^{(N)}_{n}(\lambda,\mu)= \frac{1}{2(2N)^{n-1}}\bigg[
		H_{n}^2 \bigg(\sqrt{\frac{N}{2}}\,\mu\bigg) - 
		H_{n+1}\bigg(\sqrt{\frac{N}{2}}\,\mu\bigg) H_{n-1}
		\bigg(\sqrt{\frac{N}{2}}\,\mu\bigg)
	\bigg] \,.
\ea
The overall normalization of $R$ is fixed by 
$\displaystyle \int_{-\infty}^{\infty} \ms \rmd \lambda\;
R(\lambda;M;k)=N$.
Equation~\eqref{eq:R1res} implies that for real masses
the complex conjugate of the level density acts as a
reflection of the spectrum, i.e.,
\ba
	\big[ R(\lambda;M;k) \big]^* =  
	R(-\lambda;M;k).
\ea
Hence, $\Re\big[R(\lambda;M;k)\big]$ 
is an even function of $\lambda$ while 
$\Im\big[R(\lambda;M;k)\big]$ is odd.

Another representation of the level density, which is convenient for
the derivation of the microscopic limit, is
\begin{equation}\label{leveldens-ratio.b}
R(\lambda; M;k)= \frac{(-\pi)^{N-1}}{(N-1)!}\rme^{-\frac{N}{2}\lambda^2} \prod_{f=1}^{2N_{\rm f}}(i\lambda+m_f) \frac{\int \rmd B\exp[\frac{\alpha_2}{2}(\Tr B+\lambda-2i k)^2-\frac{N}{2}\Tr B^2]\prod_{f=1}^{2N_{\rm f}+2}\det[i B+m_f\1_{N-1}]}{\int \rmd A\exp[\frac{\alpha_2}{2}(\Tr A-2i k)^2-\frac{N}{2}\Tr A^2]\prod_{f=1}^{2N_{\rm f}}\det[i A+m_f\1_{N}]}.
\end{equation}
This result is obtained by shifting back to $A\to A+x\1_N$ and $B\to B+x\1_{N-1}$ and then integrating over $x$ in the denominator as well as in the numerator.
The normalization follows from integration over $\lambda$ and combining this integral with the $B$-integral to the $A$-integral in the denominator. We recall the value of $\alpha_2=N/(N+2N_{\rm f}+1)$, cf.~\eqref{alfbet}. Indeed this result could be directly derived from the partition function~\eqref{eq:Zbeta2},
by setting one of the eigenvalues of $A$ equal to $\lambda$.

When additionally shifting  $B\to B+\lambda'\1_{N-1}$ with $\lambda'=\lambda/(2N_{\rm f}+2)$ in Eq.~\eqref{leveldens-ratio.b}, we reduce
the level density to a quotient of two almost identical partition functions,
\begin{equation}\label{leveldens-ratio}
\begin{split}
R(\lambda; M;k)=&~(-1)^{N-1}\frac{\pi^{N-1}}{(N-1)!}\exp\left[-\frac{(2N_{\rm f}+1)N}{4(N_{\rm f}+1)}\lambda^2-i\frac{N}{N_{\rm f}+1}k\lambda\right] \prod_{f=1}^{2N_{\rm f}}(i\lambda+m_f) \\
&\times\frac{\int \rmd B\exp[\frac{\alpha_2}{2}(\Tr B-2i k)^2-\frac{N}{2}\Tr B^2]\prod_{f=1}^{2N_{\rm f}+2}\det[i B+(m_f+i\lambda')\1_{N-1}]}{\int \rmd A\exp[\frac{\alpha_2}{2}(\Tr A-2i k)^2-\frac{N}{2}\Tr A^2]\prod_{f=1}^{2N_{\rm f}}\det[i A+m_f\1_{N}]}.
\end{split}
\end{equation}

%%%%%%%%%%%%%%%%%%%%%%%%%%%%%%%%
\begin{figure}[t!]
	\centering
	\includegraphics[width=.4\textwidth]{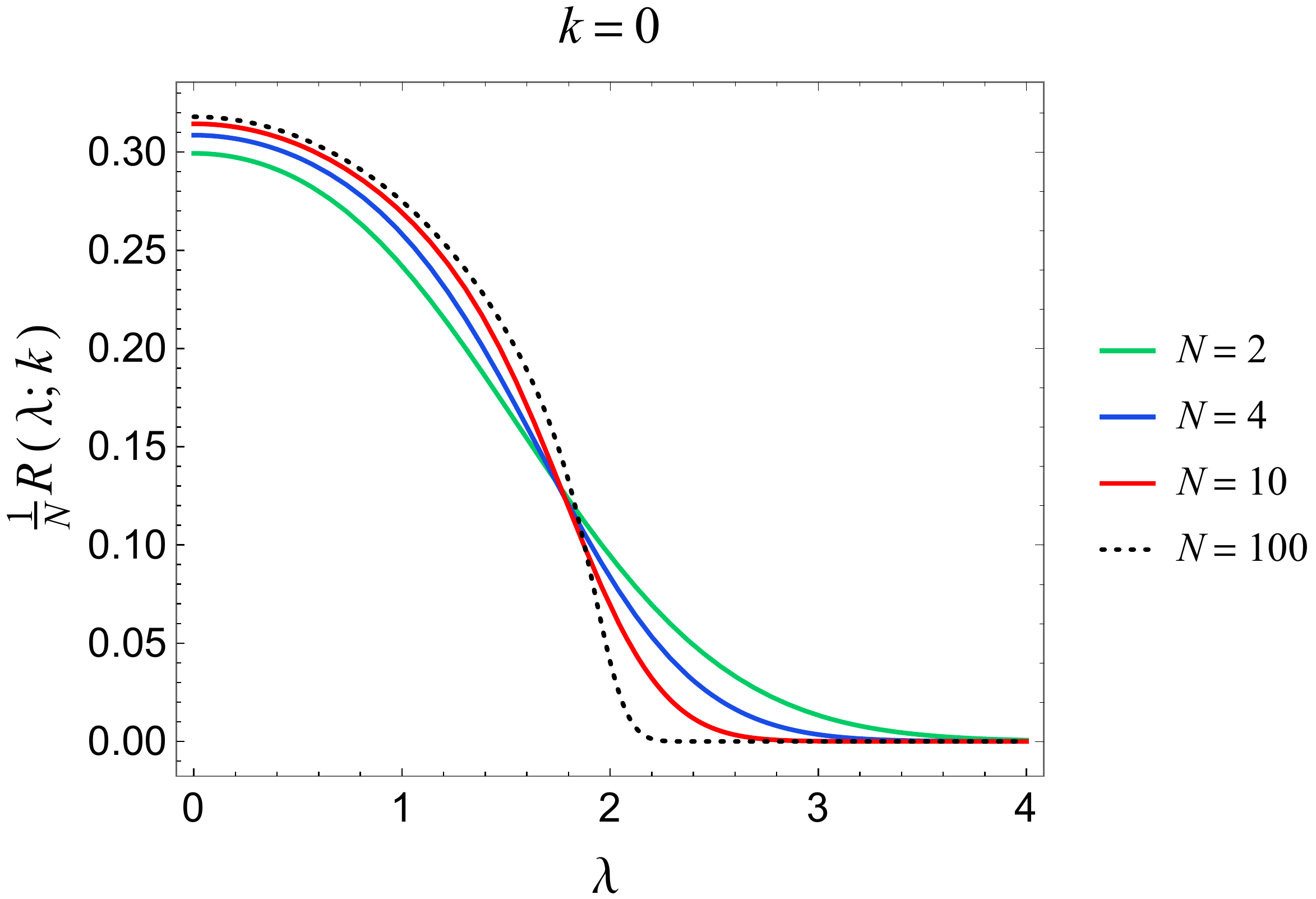}
	\vspace{-.5\baselineskip}
	\caption{\label{fg:R_quench_p0}Quenched spectral density at $k=0$. 
	It is an even function of $\lambda$.}
\end{figure}
%%%%%%%%%%%%%%%%%%%%%%%%%%%%%%%%

\subsection{\boldmath Quenched limit at finite $N$\label{sec:quenched}}

In the quenched limit $N_{\rm f}=0$, equation~\eqref{eq:R1res} 
reduces to
\ba
	R(\lambda;k) & \sim  \rme^{-\frac{N}{2}\lambda^2}
	\int_{-\infty}^{\infty}\!\! \rmd x\;\rme^{- N(x-2ik)(x-\lambda)}
	\bigg[
		H_{N}^2 \bigg(\sqrt{\frac{N}{2}}\,x \bigg) - 
		H_{N+1}\bigg(\sqrt{\frac{N}{2}}\,x \bigg) 
		H_{N-1}\bigg(\sqrt{\frac{N}{2}}\,x \bigg)
	\bigg]
	\label{eq:wertrds}
\ea
after shifting $x\to x-\lambda$.
Using the orthogonality relations for the Hermite polynomials, the normalization 
can be easily evaluated. The $x$ integral may also be performed 
with the help of the formula \cite[Sec.~7.374, eq.~9]{GRmathbook2007}, leading to the result
\ba
	R(\lambda;k) & = \frac{N!}{2^{N+1}}\sqrt{\frac{N}{\pi}} 
	\rme^{N\big(\frac{\lambda}{2}-ik\big)^2
	-\frac{N}{2}\lambda^2+2Nk^2} 
	\sum_{\ell=1}^{N}\frac{2^{2\ell-N}}{(\ell-1)!(N+1-\ell)!(N-\ell)!}H_{2N-2\ell}
	\bigg(  \frac{\sqrt{N}}{2} ( \lambda + 2ik )  \bigg)
	\,.
	\label{eq:Rqex}
\ea
Since $H_{2N-2k}$ is an even function, 
$R(\lambda;k)=R(-\lambda;-k)$, and  one can assume $k\geq 0$ 
without loss of generality.

As can be seen from Figure~\ref{fg:R_quench_p0} for the quenched spectral
density at $k=0$, in contrast to the standard GUE, the
oscillatory structure of the spectral 
density due to peaks of individual eigenvalues is not present even for small $N$. 
This feature was also seen in other one-parameter-reweighted ensembles~\cite{AkeVivo2008,Kanazawa:2016nlh}. 
We expect that this feature will
 carry over to three-dimensional QCD as well. This figure also shows that
the large $N$-limit, given by the semi-circle $\rho(x)=\sqrt{1-(x/2)^2}/\pi$, is  already well-approximated for $N=100$. 

%%%%%%%%%%%%%%%%%%%%%%%%%%%%%%%%
\begin{figure}[tb]
	\centering
	\includegraphics[width=.4\textwidth]{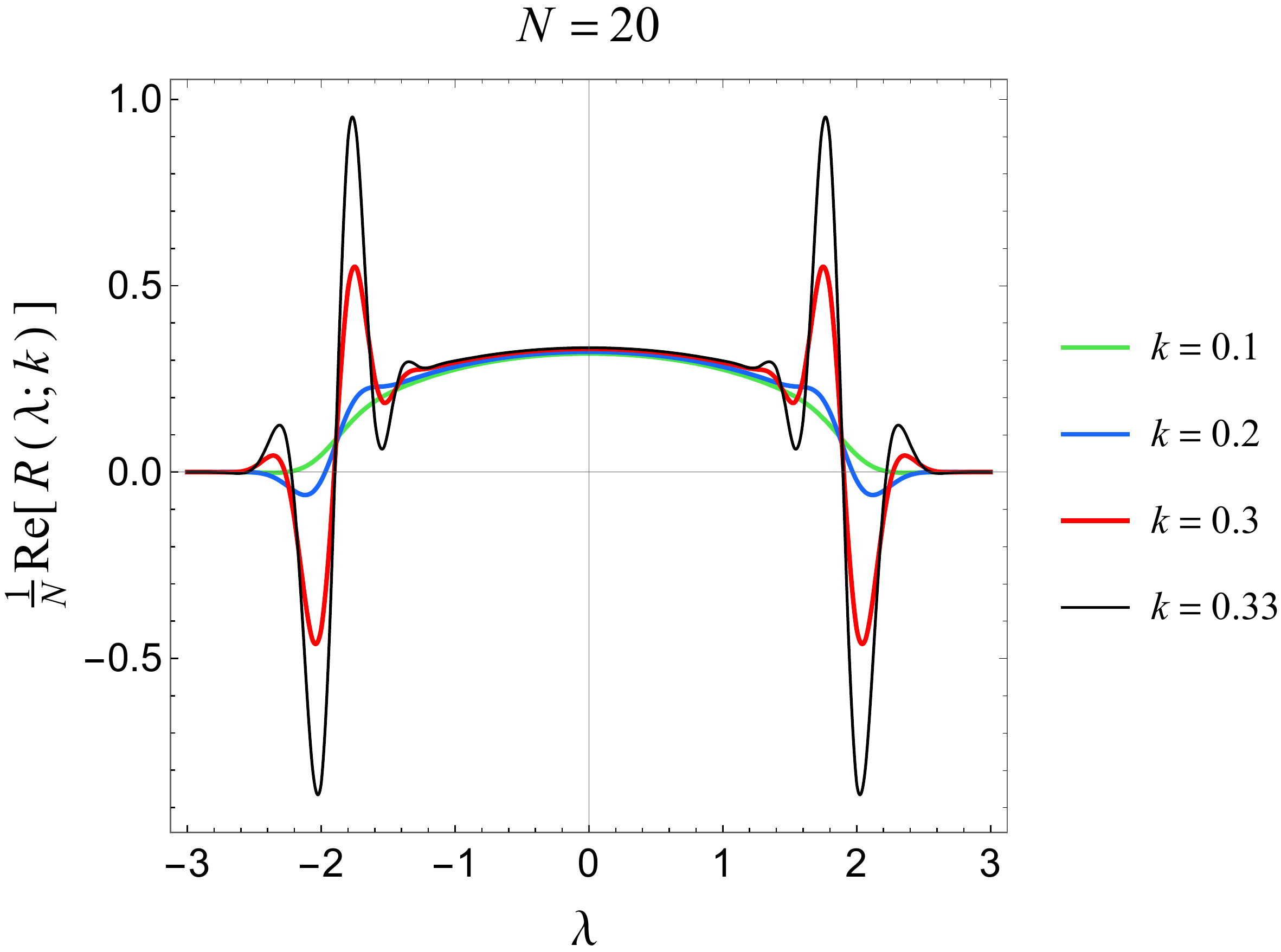}\quad
	\includegraphics[width=.4\textwidth]{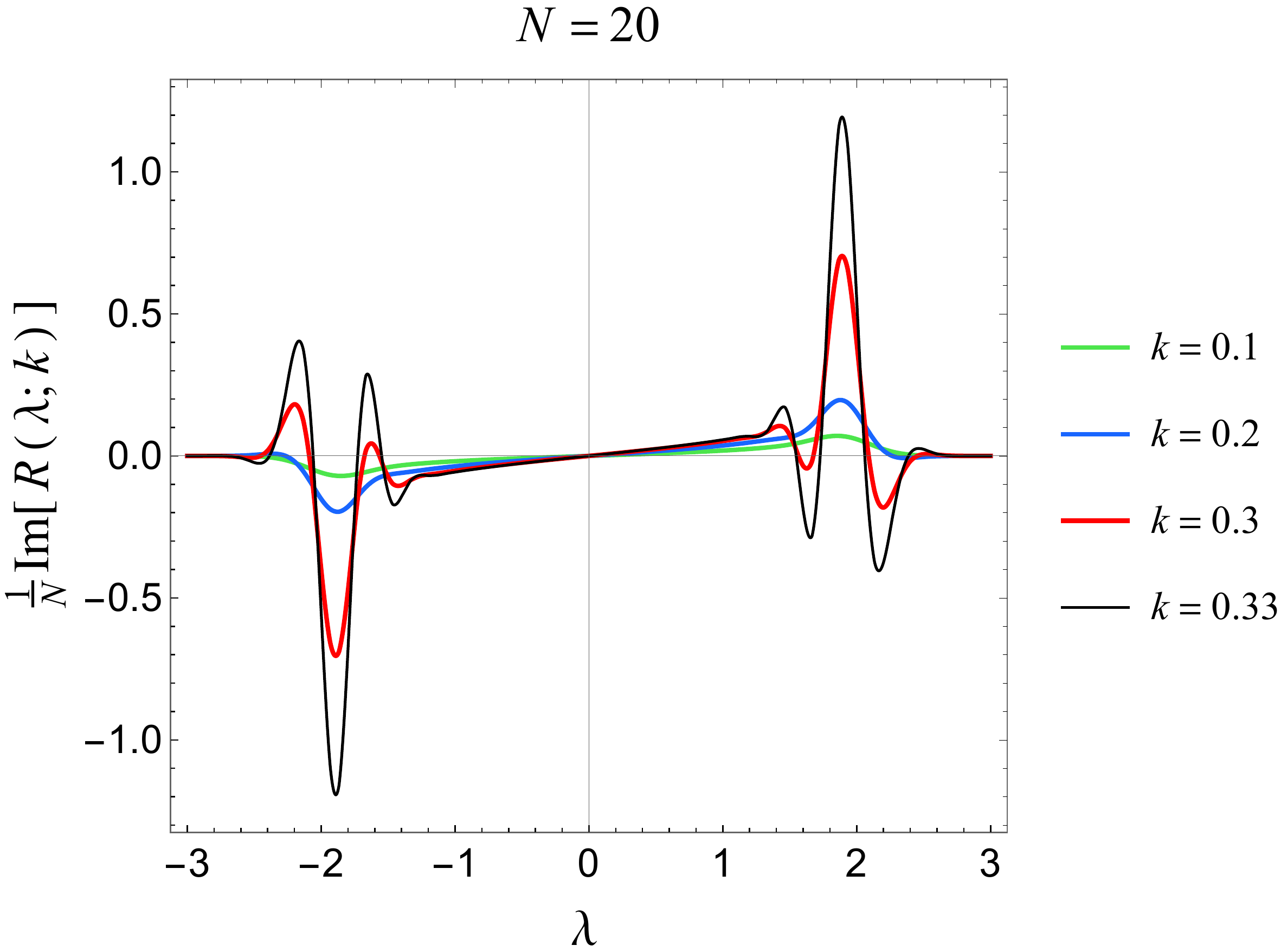}
	\vspace{-.5\baselineskip}
	\caption{\label{fg:R_quench_compl}The real (left) and imaginary (right) 
	part of the quenched spectral density for $N=20$.}
	\vspace{15pt}
	\includegraphics[width=.4\textwidth]{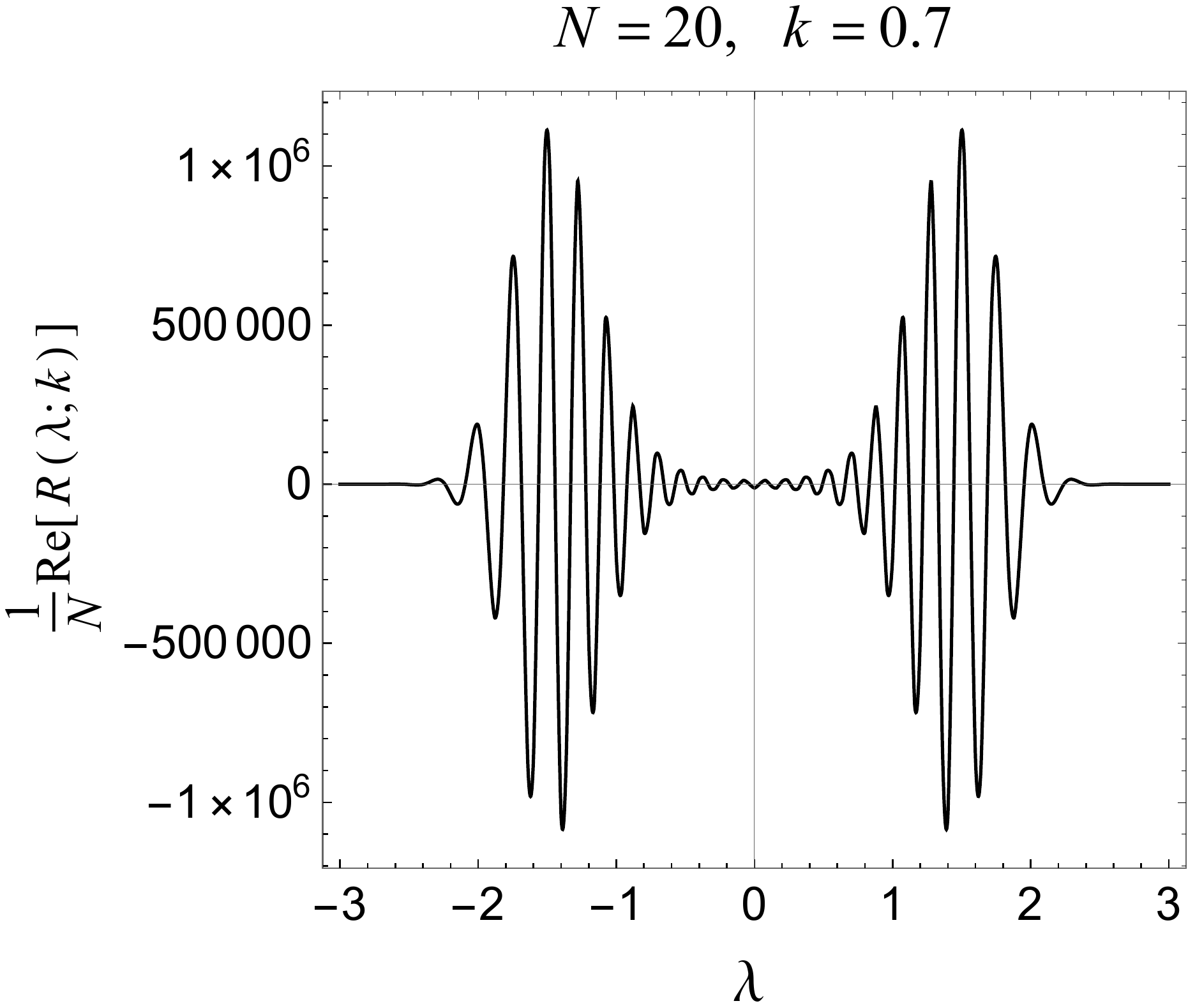}\qquad
	\includegraphics[width=.4\textwidth]{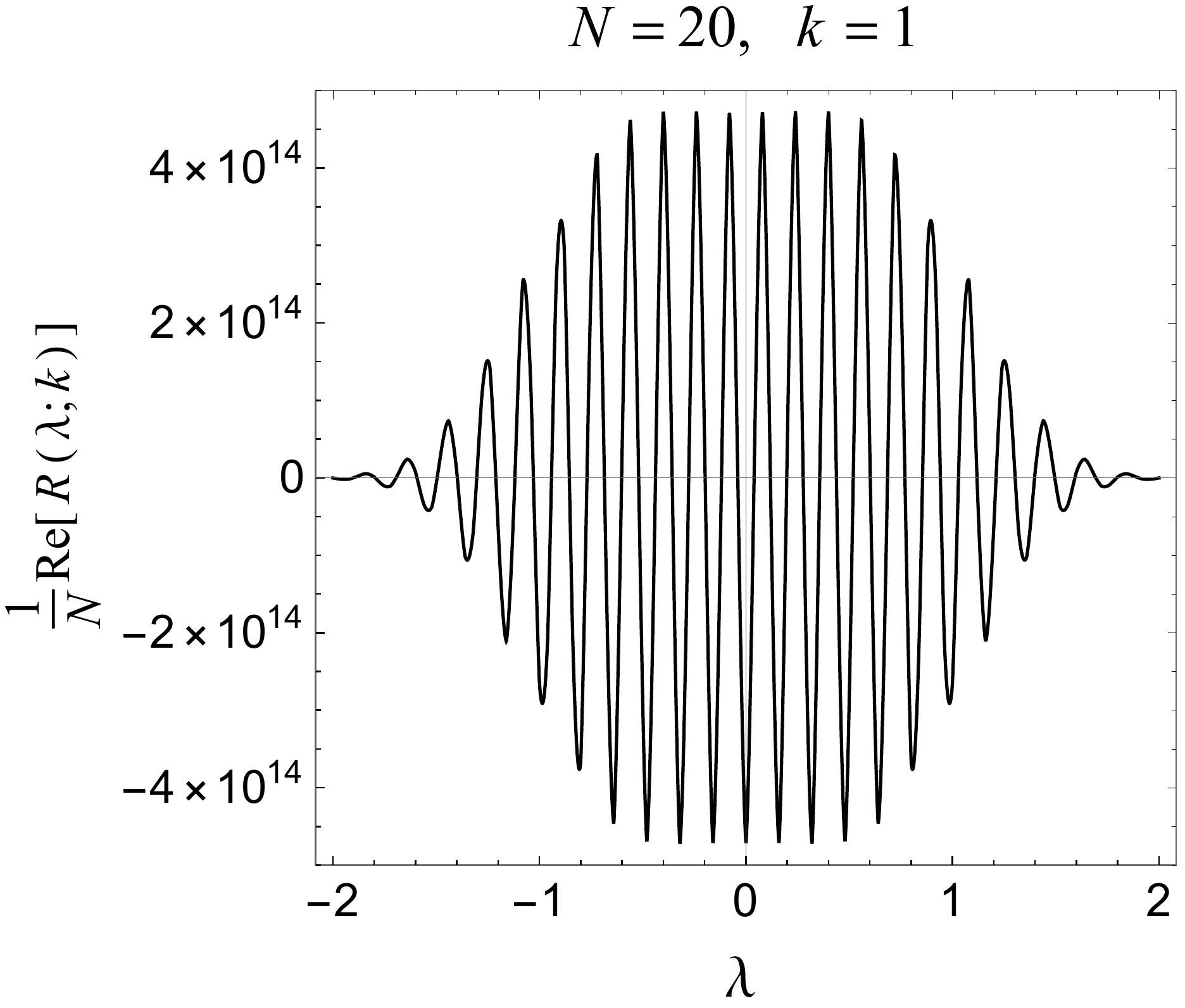}
	\vspace{-.5\baselineskip}
	\caption{\label{fg:R_quench_p07}The real part of the quenched 
	spectral density for $N=20$ at $k=0.7$ and $1$.}
\end{figure}
%%%%%%%%%%%%%%%%%%%%%%%%%%%%%%%%

When increasing $k$ for a  fixed matrix dimension $N$, say $N=20$, the spectral ``density'' becomes 
complex-valued. We illustrate this in Figure~\ref{fg:R_quench_compl} where the real and imaginary parts of the level density $R(\lambda; k)$ are shown. At nonzero
$k>0$,  the semi-circle  undergoes a dramatic 
deformation of its shape. First, small oscillations at the two edges appear.
They even change the sign of the spectral density for small regions regardless
of how small $k$ is. The amplitude of these oscillations
grows with $k$. While keeping $k$ below a threshold $k_c$, see the ensuing subsections, the oscillations die out around the origin and we can expect a well-defined microscopic limit. Yet, when increasing $k$ beyond $k_c$, the oscillations intensify and move into the bulk of the spectrum, see Figure~\ref{fg:R_quench_p07}, such that even there the spectral density does not remain strictly positive.
The amplitudes of the oscillations grow rapidly with $k$, even though the 
normalization condition $\int \rmd \lambda\;R(\lambda;k)=N$ is strictly satisfied. 
A similar oscillation of the spectral density was also observed in matrix models for QCD at nonzero
chemical potential~\cite{Akemann:2004dr,Osborn:2005ss,Akemann:2010tv,Kanazawa:2012zzr} 
and for QCD with nonzero theta angle~\cite{Kanazawa:2011tt,Verbaarschot:2014upa}. 

The question is how this oscillatory behavior carries over to the large-$N$ limit while keeping $k$ fixed. Three things may happen. Either the oscillations do not reach the origin; then we expect the universal results from GUE with a possible reweighting since the level spacing is changing. Second, the oscillations reach the origin but are not strong enough to make the microscopic limit ill-defined, in particular the amplitude  does not grow with the matrix dimension $N$. And third,
the oscillations become so dominant that the microscopic limit is not well-defined at the origin. The latter will usually happen at about $k_{c}\approx N_{\rm f}+1/2$ as we will see below.

\subsection{\boldmath Quenched microscopic large-$N$ limit}\label{sec:micro}

\begin{figure}[t!]
	\centering
	\includegraphics[width=.35\textwidth]{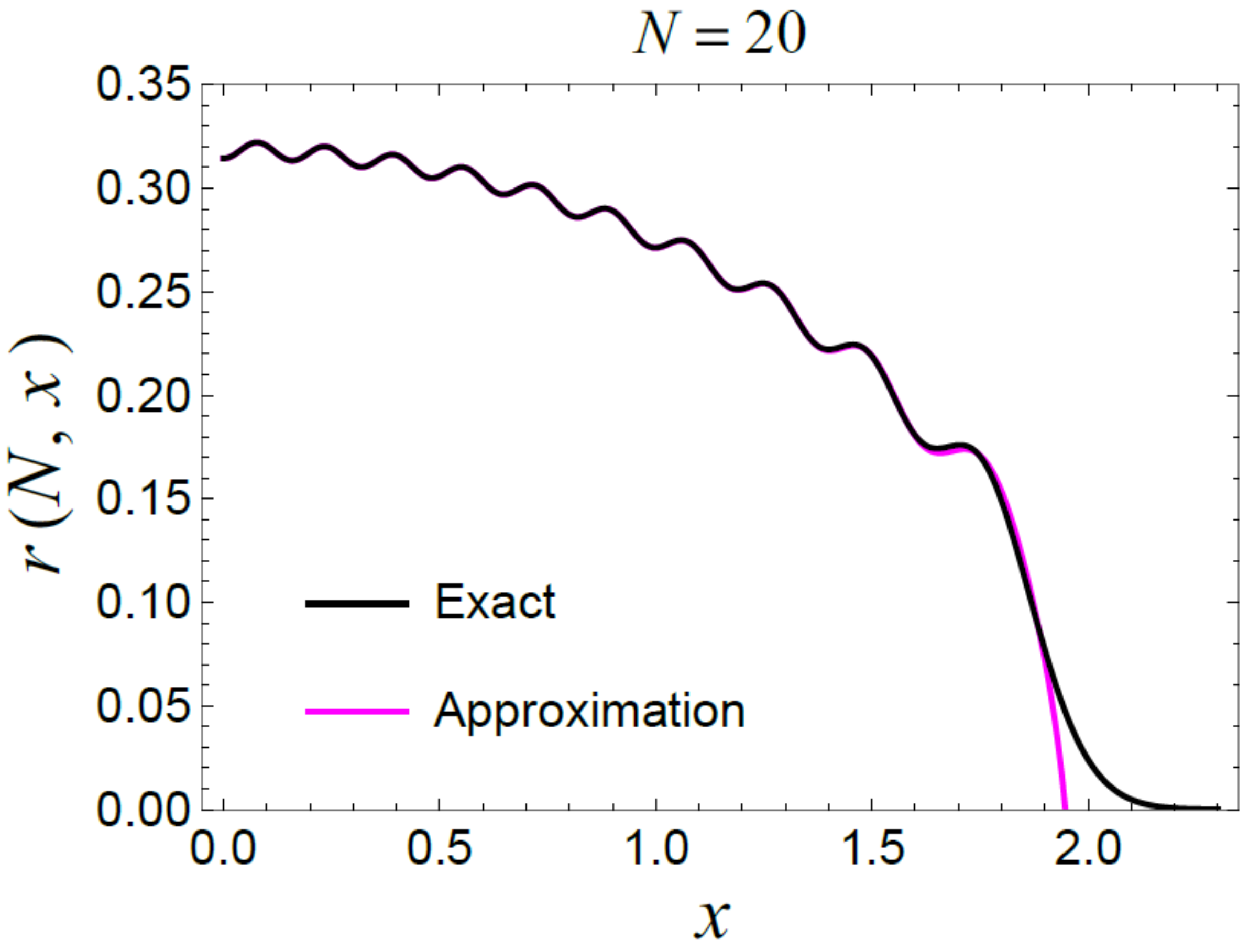}
	\caption{\label{fg:r_semicircle}$r(N,x)$ and its large-$N$ 
	approximation \eqref{eq:rhoappr} for $N=20$.}
\end{figure}

The next task is to evaluate the microscopic limit of the quenched 
density \eqref{eq:wertrds} where we evaluate its large-$N$ limit at
fixed $\hat \lambda \equiv \lambda N$.
Incorporating the normalization factor, 
we write~\eref{eq:wertrds} as
\ba
	\frac{1}{N}R(\lambda;k) & = 
	\frac {{\mathcal N}(N,0)}{{\mathcal N}(N,k)} 
	\sqrt{\frac N{2\pi}} \rme^{-\frac N 2 \lambda^2} \!\! 
	\int_{-\infty}^\infty  \!\!\!  \rmd x\,
	\rme^{-N(x-2ik)(x-\lambda)+\frac N2 x^2} r(N,x) \,,
\ea
where 
\be
	{\mathcal N}(N,k) = 2^{N+1} (N-1)!\;\pi\,\rme^{-2Nk^2}
\ee
and
\be
	r(N,x) & \displaystyle 
	= \sqrt{\frac{2\pi}N}\frac 1{{\mathcal N}(N,0)}
	\rme^{-\frac N2 x^2}
	\bigg[
		H_{N}^2 \bigg(\sqrt{\frac{N}{2}}\,x \bigg) - 
		H_{N+1}\bigg(\sqrt{\frac{N}{2}}\,x \bigg) 
		H_{N-1}\bigg(\sqrt{\frac{N}{2}}\,x \bigg) 
	\bigg]\,,
	\label{eq:rhoHH}
\ee
which  is normalized as $\int_{-\infty}^{\infty} \rmd x\; r(N,x)=1$. 
While $r(N,x)$ is similar to that of the Wigner-Dyson ensemble, 
the $N$ dependence is slightly different. 
However, for large $N$ it also approaches a semi-circle. 
To obtain more quantitative results we use the uniform asymptotic
expansion of the Hermite polynomials \cite{Dominici2007}
\begin{equation}
\begin{split}
	H_N(x)\overset{N\gg1}{\approx}& \sqrt{2} \, \frac {(2N)^{N/2}\rme^{-N/2}}
	{\left( 1 - \frac {x^2}{2N} \right )^{1/4}}
	\rme^{ x^2/2 }\cos \left[
	N \ckakko{
		\frac 12 \sin\mkakko{2\arcsin \left(\frac{x}{\sqrt{2N}}\right)} + 
		\arcsin \left(\frac{x}{\sqrt{2N}}\right) 
		 - \frac \pi 2
	} + \frac 12 \arcsin \left(\frac{x}{\sqrt{2N}}\right)
	\right]\\
	=~&\sqrt{2} \, \frac {(2N)^{N/2}\rme^{-N/2}}
	{\left( 1 - \frac {x^2}{2N} \right )^{1/4}}
	\rme^{ x^2/2 }\cos \left[	
		\sqrt{N-\frac{x^2}{2}}\frac{x}{\sqrt{2}} + 
		\left(N+\frac{1}{2}\right)\arcsin \left(\frac{x}{\sqrt{2N}}\right) 
		 - \frac{N\pi}{ 2}\right],
\end{split}
\end{equation}
valid for $|x|<\sqrt{2N}$. For large $N$, we thus obtain
\ba
	r(N,x) & \overset{N\gg1}{\approx} \frac 1{\pi} \sqrt{1-\frac{x^2}{4}} 
	- \frac{1}{\pi N}
	\frac{\cos\Big[N\Big(x\sqrt{1-\frac{x^2}{4}}
	- 2\,\text{arccos}\,\frac{x}{2}\Big)\Big]}{4-x^2}\,.
	\label{eq:rhoappr}
\ea
In Figure~\ref{fg:r_semicircle} we compare 
\eqref{eq:rhoHH} and \eqref{eq:rhoappr} for $N=20$. 
The agreement is  excellent except near the edge of the semi-circle. 
%%%%%%%%%%%%%%%%%%%%%
%%%%%%%%%%%%%%%%%%%%%
Returning to the quenched density, it is comprised of two pieces 
\ba
	\frac{1}{N}R(\lambda;k) & \overset{N\gg1}{\approx} \frac{1}{N}R^{(a)}(\lambda;k) 
	+ \frac{1}{N}R^{(b)}(\lambda;k) \,,
\ea
associated with the two terms in \eqref{eq:rhoappr}, where $R^{(a)}(\lambda;k)$ corresponds to  the semi-circle part and $R^{(b)}(\lambda;k)$ to the oscillatory part. The prefactor $1/N$ results from the scale on which we want to zoom in about the origin.

To evaluate the 
first contribution at large $N$ we note
\ba
	-N(x-2ik)(x-\lambda)+\frac N2 x^2 = 
	N\bigg(\!\!-2k^2+\frac{\lambda^2}{2}\bigg) - \frac{N}{2}(x-2ik-\lambda)^2 \,.
\ea
The saddle point can be approximated as $x=2ik$ since $\lambda=\widehat{\lambda}/N$, so 
\ba
	\frac{1}{N}R^{(a)}(\lambda;k) &  =
	\sqrt{\frac{N}{2\pi}}\rme^{2Nk^2-\frac{\widehat{\lambda}^2}{2N}}
	\int_{-\infty}^\infty  \!\!\!  \rmd x\,
	\rme^{-N(x-2ik)(x-\widehat{\lambda}/N)+\frac N2 x^2}
	\frac 1{\pi} \sqrt{1-\frac{x^2}{4}}
	\overset{N\gg1}{\approx} \frac{\sqrt{1+k^2}}{\pi}\,.
\ea
The average over the oscillatory part can be evaluated as
\ba
	\frac{1}{N}R^{(b)}(\lambda;k) & = - \frac{1}{\pi \sqrt{2\pi N}} 
	\rme^{2Nk^2-\frac{\widehat{\lambda}^2}{2N}} \!\! 
	\int_{-2}^{2} \!  \rmd x\,
	\rme^{-N(x-2ik)(x-\widehat{\lambda}/N)+\frac N2 x^2} 
	\frac{\cos\Big[N\Big(x\sqrt{1-\frac{x^2}{4}}
	- 2\,\text{arccos}\,\frac{x}{2}\Big)\Big]}{4-x^2} 
	\\
	& \overset{N\gg1}{\approx} - \frac{1}{2\pi \sqrt{2\pi N}} 
	\rme^{2Nk^2} \!\!\int_{-2}^{2} \!  \rmd x\, 
	\frac{\exp[N f_+(x)]+\exp[N f_-(x)]}{4-x^2}\rme^{(x-2ik)\widehat{\lambda}}.
\ea
We have dropped the term $\exp[-N\lambda^2/2]$ since 
$\lambda\sim 1/N$ in the microscopic large-$N$ limit. Furthermore we introduced the function 
\ba
	f_\pm(x) = - (x-2ik)x+\frac{x^2}{2}
	\pm i \bigg(x\sqrt{1-\frac{x^2}{4}}
	- 2\,\text{arccos}\,\frac{x}{2}\bigg)\,. 
\ea
For large $N$ the integral can be evaluated with the saddle point method. 
Solving
\begin{equation}
f_\pm'(x)=-x+2i k\pm 2i\sqrt{1-\frac{x^2}{4}}=0
\end{equation}
 yields the solutions
\ba
	\left\{\begin{array}{l}
		f'_+(x_c)=0,\quad ~\text{for \;$k<0$},
		\\
		f'_-(x_c)=0,\quad ~\text{for \;$k>0$},
	\end{array}\right.
	\qquad \text{with} \quad 
	x_c = i\frac{k^2-1}{k}\,.
\ea
To check this, we want to point out that
\begin{equation}
\sqrt{-\left(\frac{y}{2}-\frac{2}{y}\right)^2}=-i\,\sign\left({\rm Im}\left[\frac{y}{2}-\frac{2}{y}\right]\right)\left(\frac{y}{2}-\frac{2}{y}\right)\ {\rm for}\ y\notin\mathbb{R}
\end{equation}
which forbids a saddle point of $f_+(x)$ for $k>0$ and of $f_-(x)$ for $k<0$.

The saddle point expansion leads to
\ba
	\frac{1}{N}R^{(b)}(\lambda;k) & \overset{N\gg1}{\approx}
	- \frac{1}{2\pi \sqrt{2\pi N}} 
	\frac{k^2}{(k^2+1)^2}\exp\left[N(2k^2+f_{-\sign(k)}(x_c))-i\frac{k^2+1}{k}\widehat{\lambda}\right]\int_{-\infty}^\infty  \!\!\!  \rmd x\, 
	 \exp\left[-\frac{N}{1+k^2}(x-x_c)^2\right]\nonumber
	\\
	& = - \frac{1}{\sqrt{8}\pi N} 
	\frac{k^2}{(k^2+1)^{3/2}}\exp\left[N(2k^2+f_{-\sign(k)}(x_c))-i\frac{k^2+1}{k}\widehat{\lambda}\right].
\ea
The explicit form of the function $f_{-\sign(k)}(x_c)$ is given by
\be
	\begin{split}
		f_{-\sign(k)}(x_c) & = 1-k^2 +2{\rm arcsinh}\left(\frac{k^2-1}{2|k|}\right)+\sign(k)\pi i.
	\end{split}
\ee
To derive this intermediate result we used the identity
\begin{equation}
\arccos(ix)=\frac{\pi}{2}-i{\rm arcsinh}(x)\ {\rm for}\ x\in\mathbb{R}.
\end{equation}

\begin{figure}[t!]
	\centerline{
		\includegraphics[width=.3\textwidth]{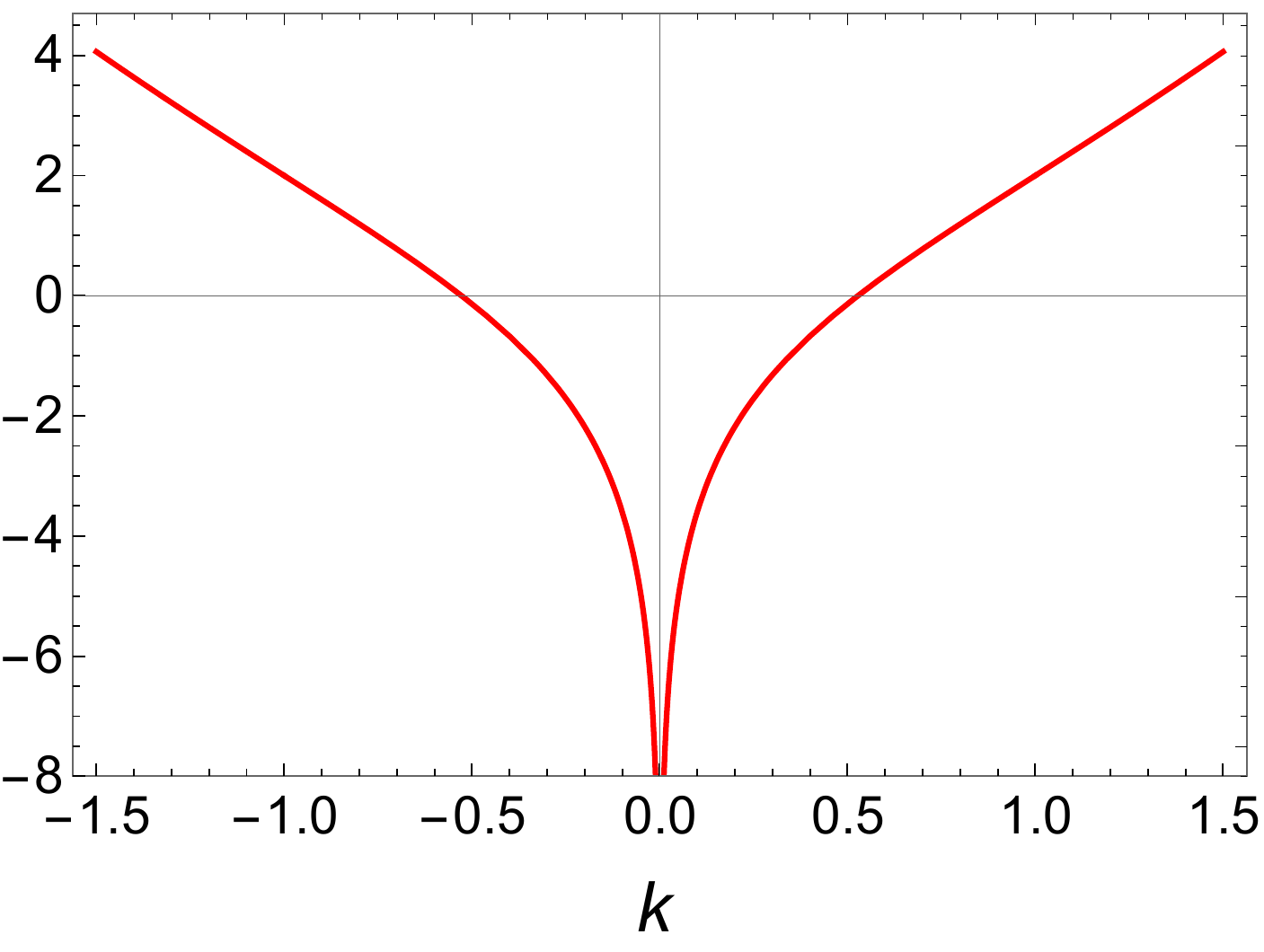}
	}
	\caption{\label{fg:g_plot} Shown is the function 
	$2k^2+{\rm Re}[f_{-\sign(k)}(x_c)]$.}
\end{figure}

Let us collect all results, so that the microscopic level density reads
\ba
	\frac{1}{N}R(\lambda;k) & \overset{N\gg1}{\approx} 
	\frac{\sqrt{1+k^2}}{\pi} - \frac{(-1)^N}{\sqrt{8}\pi N} 
	\frac{k^2}{(k^2+1)^{3/2}}\exp\left[N\left(1+k^2 +2{\rm arcsinh}\biggl(\frac{k^2-1}{2|k|}\biggl)\right)-i\frac{k^2+1}{k}\widehat{\lambda}\right]\,.
	\label{eq:Rasympt}
\ea
This indicates that the amplitude of the oscillation is controlled by $1+k^2 +2{\rm arcsinh}[(k^2-1)/2|k|]$, which is shown in Figure~\ref{fg:g_plot}. 
The function changes sign at $k=\pm k_c$ with 
\ba
	k_c =\int_0^\infty\Theta\left[2{\rm arcsinh}\left(\frac{1-k^2}{2|k|}\right)-k^2-1\right]\rmd k=0.527697\cdots. \label{eq:kcval}
\ea 
Thus, for $N\gg 1$, the amplitude of the oscillations grows
exponentially for $|k|>k_c$, but it
dies out for $|k|<k_c$. In the limit $k\to 0$ we smoothly recover 
the well-known microscopic spectral density of GUE, $\rho_{\rm mic}(\widehat{\lambda})=1/\pi$, see~\cite{Verbaarschot:1994ip}. 
It is intriguing to note that the value of $k_c$  in Eq.~\eqref{eq:kcval} coincides 
with the phase transition point obtained from \eqref{phasetranseq} 
in the limit $N_{\rm f}\to +0$. This observation is not surprising when considering an
alternative derivation given in the Appendix~\ref{sec:a3}.
%%%%%%%%%%%%%%%%%%%%%
%%%%%%%%%%%%%%%%%%%%%

\begin{figure}[t!]
	\centerline{
		\includegraphics[height=.32\textwidth]{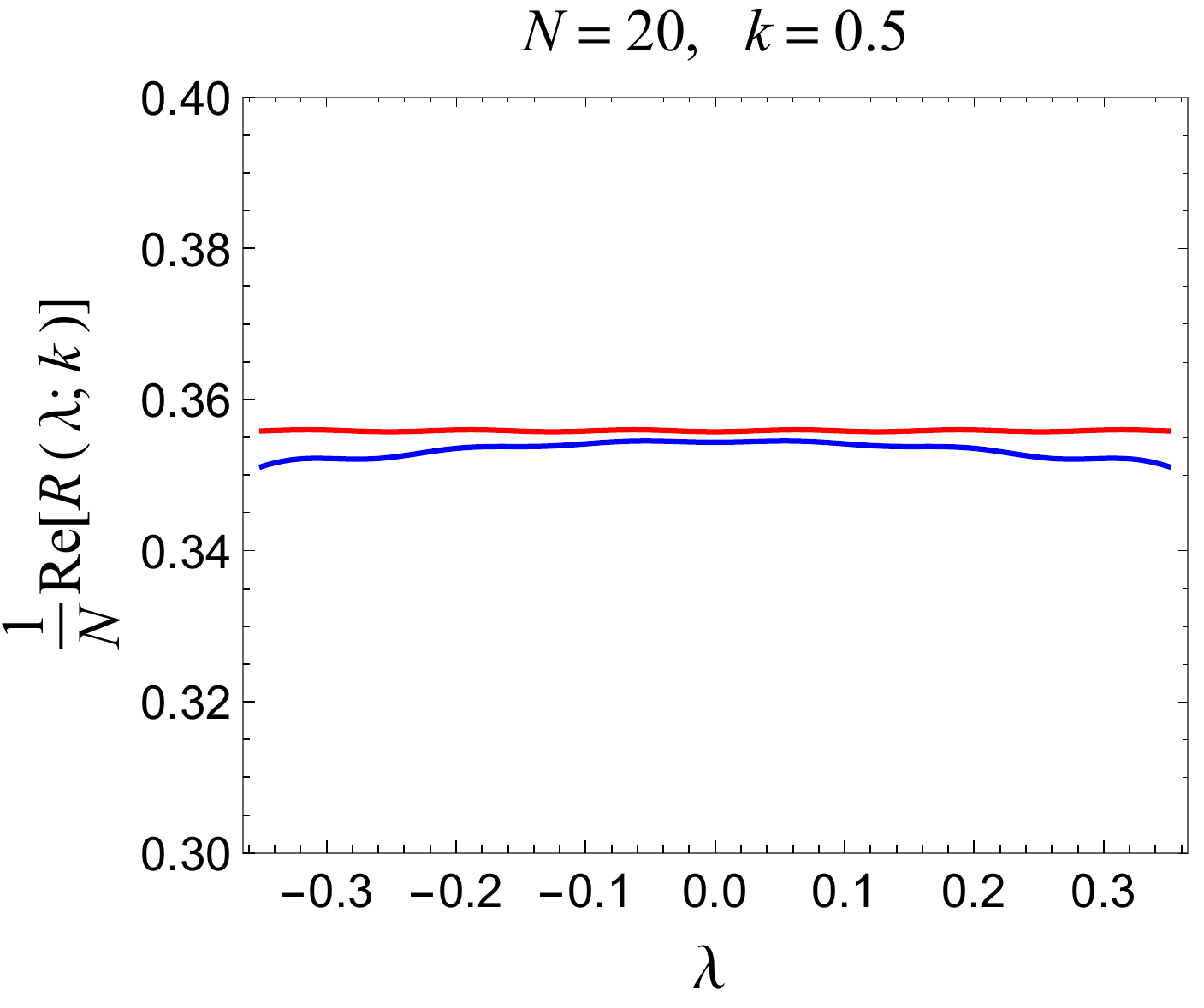}
		\quad
		\includegraphics[height=.32\textwidth]{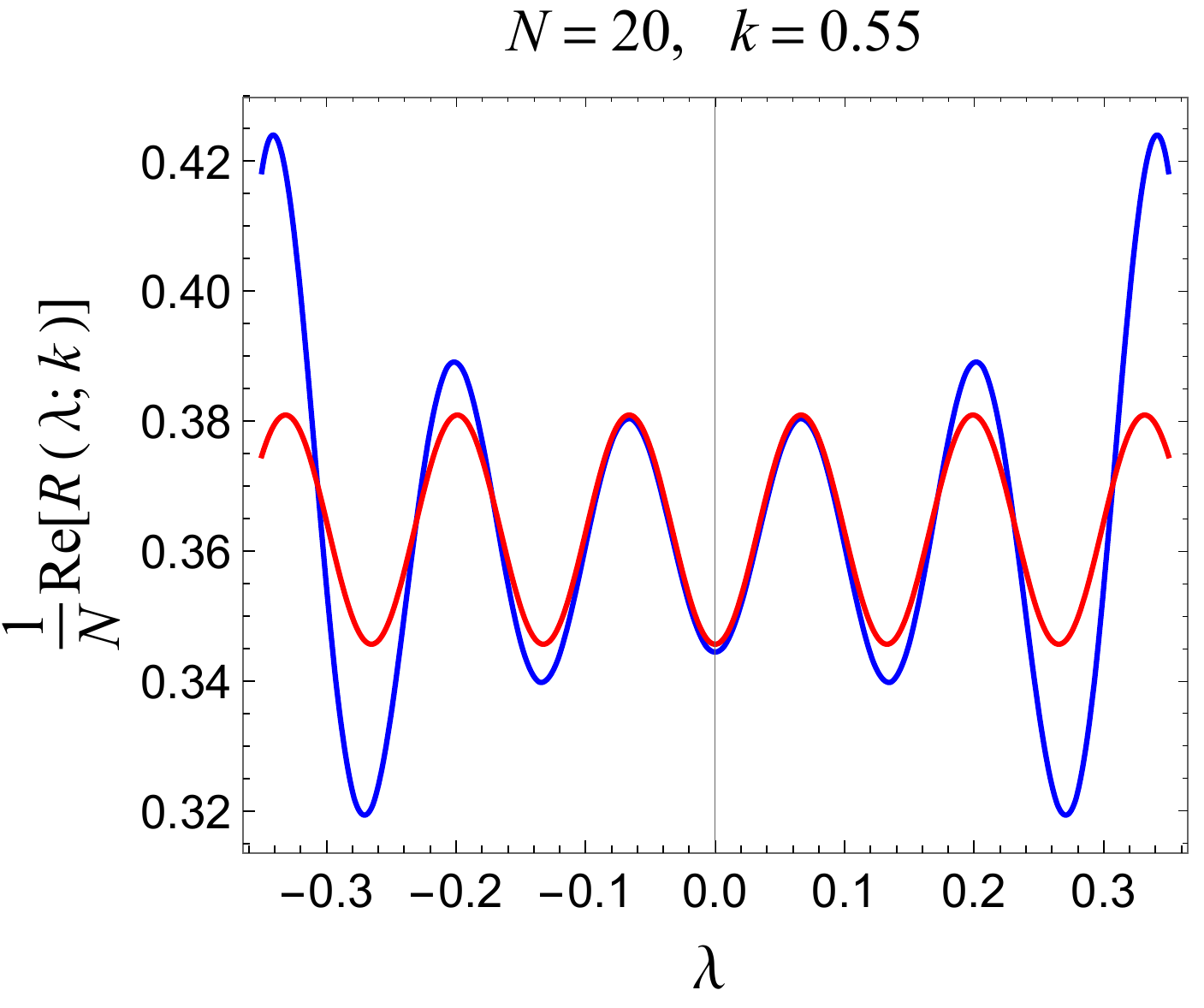}
	}
	\vspace{2mm}
	\centerline{
		\includegraphics[height=.32\textwidth]{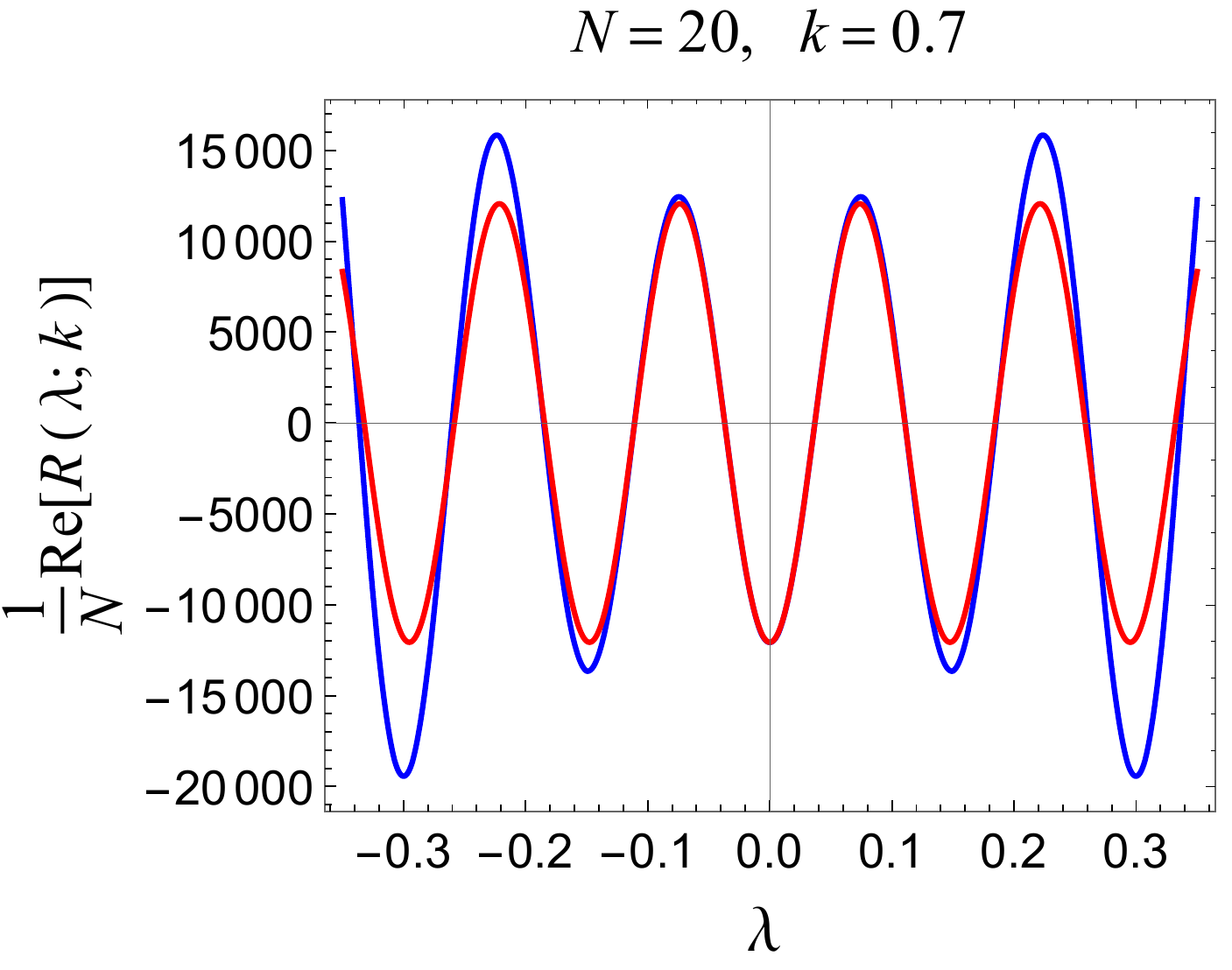}
		\quad
		\includegraphics[height=.32\textwidth]{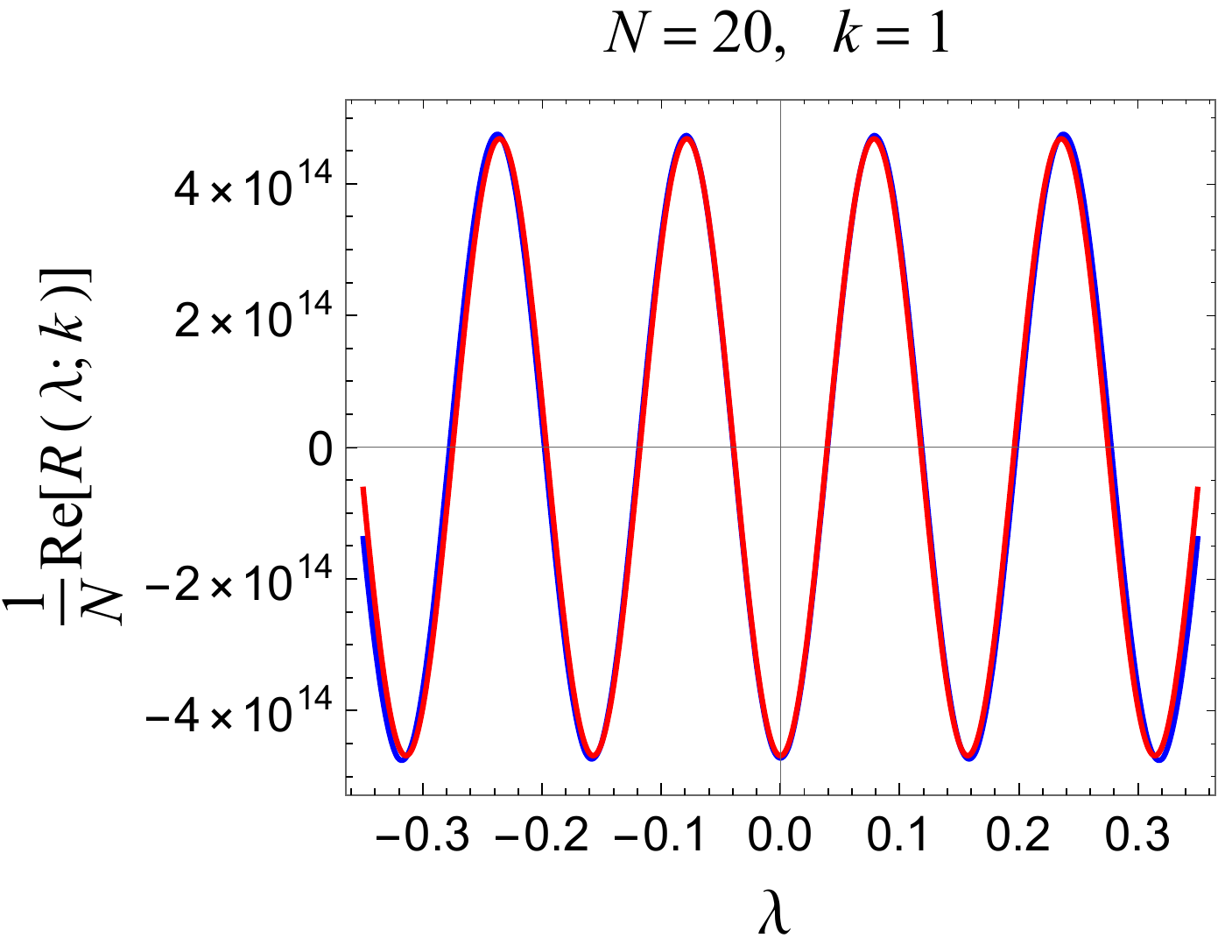}
	}
	\caption{\label{fgg_verygood}Comparison of the real parts of the exact quenched density~\eqref{eq:Rqex} (blue line) 
	and its asymptotic approximation~\eqref{eq:Rasympt} (red line).}
\end{figure}

In Figure~\ref{fgg_verygood} we numerically compare 
\eqref{eq:Rasympt} with the exact density~\eqref{eq:Rqex} 
for various $k$.  In all cases they show good agreement 
in the region $\lambda\sim 1/N$ despite the relatively small matrix dimension $N=20$. When increasing $N$ for $|k|>k_c$ the oscillations become dominant; the amplitude grows exponentially, and a microscopic limit does not exist.

%%%%%%%%%%%%%%%%%%%%%
%%%%%%%%%%%%%%%%%%%%%

The quark-antiquark condensate in the quenched case can be readily calculated since the microscopic spectral density is $\widehat{R}(\widehat{\lambda};k)=\sqrt{1+k^2}/\pi$. Hence, the quark-antiquark condensate is equal to
\begin{equation}
\Sigma V=\int_{-\infty}^\infty \frac{\widehat{R}(\widehat{\lambda};k)}{i\widehat{\lambda}+\widehat{m}}d\widehat{\lambda}=\sign(\widehat{m})\sqrt{1+k^2}.
\end{equation}
This result only holds for $|k|<k_c$.

\subsection{Unquenched microscopic level density}\label{sec:unquenched}

To derive the microscopic level density with dynamical quarks, we start from
Eq.~\eqref{leveldens-ratio} where we have to compute two kinds of partition functions.
For this purpose, we first  need to approximate the prefactor
in Eq.~\eqref{leveldens-ratio} which in the microscopic limit simplifies to 
\begin{equation}\label{a:3.1}
\begin{split}
&(-1)^{N-1}\frac{\pi^{N-1}}{(N-1)!}\exp\left[-\frac{(2N_{\rm f}+1)N}{4(N_{\rm f}+1)}\lambda^2-i\frac{N}{N_{\rm f}+1}k\lambda\right] \prod_{f=1}^{2N_{\rm f}}(i\lambda+m_f)\\
\overset{N\gg1}{\approx}&(-1)^{N-1}\frac{\pi^{N-3/2}e^N}{\sqrt{2}N^{N-1/2+2N_{\rm f}}}\rme^{-i\frac{k\widehat{\lambda}}{N_{\rm f}+1}} \prod_{f=1}^{2N_{\rm f}}(i\widehat{\lambda}+\widehat{m}_f)\,.
\end{split}
\end{equation}
We recall that in this limit $\lambda N\equiv \widehat \lambda$  and  $m_f N \equiv\widehat{m}_f$ with $\widehat{\lambda}$ and $\widehat{m}_f$ fixed in the limit $N\to\infty$. The two partition functions in the numerator and denominator of~\eqref{leveldens-ratio} are computed in detail in the Appendix~\ref{sec:app} with the aid of
random matrix  methods.

Let us briefly revisit the quenched density. To obtain this quantity
we combine Eqs.~\eqref{a:0.3},~\eqref{a:2.11} and~\eqref{a:3.1}, the latter two for $N_{\rm f}=0$, in~\eqref{leveldens-ratio}. Since the partition function $Z_{N_{\rm f}=0}=Z_{\rm q}$ is always in the trivial phase $k_L=0$ for $|k| < k_c\approx 0.527697$ and $\widehat{\wt{M}}=-i\widehat{\lambda}/2\1_2$, we obtain
\begin{equation}\label{a:3.2}
\begin{split}
\frac{1}{N}R(\lambda;M;k)\overset{N\gg1}{\approx}&\frac{\exp\left[2Nk^2-N(\lambda_+^2+\lambda_-^2)/2+N\right]}{\pi}\sqrt{1+k^2}\exp\left[i\frac{\lambda_++\lambda_--2k}{2}\widehat{\lambda}\right].
\end{split}
\end{equation}
Since $\lambda_\pm=k\pm\sqrt{1+k^2}$ in the present case (see eq. \eref{a:1.8} for the
definition for arbitrary $N_f$), we can simplify this expression to
\begin{equation}\label{a:3.3}
\begin{split}
\frac{1}{N}R(\lambda;M;k)\overset{N\gg1}{\approx}&\frac{\sqrt{1+k^2}}{\pi}.
\end{split}
\end{equation}
This is the leading order term in Eq.~\eqref{eq:Rasympt}.
The oscillatory
part which becomes dominant for $|k|>k_c$
can be obtained from Eq.~\eqref{a:2.11}
in the limit $k_L\to1$ instead of $k_L=0$. The limit is important since some terms seem to diverge; nevertheless they cancel with other terms that vanish so that one needs to employ l'Hospital's rule several times.

In the case of dynamical quarks, we collect the terms in Eqs.~\eqref{a:1.14},~\eqref{a:2.11} and~\eqref{a:3.1} and find
\begin{equation}\label{a:3.4_}
\begin{split}
\frac{1}{N}R(\lambda;M;k)
\overset{N\gg1}{\approx}&\frac{(-1)^{k_L}}{2\pi } \frac{(N_{\rm f}+k_L)!(N_{\rm f}-k_L)!}{(2N_{\rm f}+1)!(2N_{\rm f})!}\left(\frac{\sqrt{(N_{\rm f}+1)^2+k^2-k_L^2}-k}{\sqrt{(N_{\rm f}+1)^2+k^2-k_L^2}+k}\;\frac{N_{\rm f}+1+k_L}{N_{\rm f}+1-k_L}\right)^{2k_L}\\
&\times \left(2\frac{(N_{\rm f}+1)\sqrt{(N_{\rm f}+1)^2+k^2-k_L^2}-kk_L}{(N_{\rm f}+1)^2-k_L^2}\right)^{2N_{\rm f}+1}\rme^{-i(\lambda_++\lambda_-)\widehat{\lambda}}\prod_{f=1}^{2N_{\rm f}}(\widehat{\lambda}-i\widehat{m}_f)\\
&\times \frac{\int_{\U(2N_{\rm f}+2)}  \rmd\mu(U)\rme^{-\Tr U^\dagger \diag(\lambda_-\1_{N_{\rm f}+1+k_L},\lambda_+\1_{N_{\rm f}+1-k_L}) U\diag(\widehat{m}_1,\ldots,\widehat{m}_{2N_{\rm f}},-i\widehat{\lambda},-i\widehat{\lambda})}}{\int_{\U(2N_{\rm f})} \rmd\mu(U)\rme^{-\Tr U^\dagger \diag(\lambda_-\1_{N_{\rm f}+k_L},\lambda_+\1_{N_{\rm f}-k_L}) U\diag(\widehat{m}_1,\ldots,\widehat{m}_{2N_{\rm f}})}}.
\end{split}
\end{equation}
Now, we combine the saddle point solutions $\lambda_\pm$ into
\begin{equation}\label{a:3.5}
\omega=\frac{\lambda_+-\lambda_-}{2}=\frac{(N_{\rm f}+1)\sqrt{(N_{\rm f}+1)^2+k^2-k_L^2}-kk_L}{(N_{\rm f}+1)^2-k_L^2}
\end{equation}
and their sum, and we perform the two Harish-Chandra--Itzykson--Zuber integrals~\cite{HC,IZ} in~\eqref{a:3.4_} which yields the simplification
\begin{equation}\label{a:3.4}
\begin{split}
\frac{1}{N}R(\lambda;M;k)
\overset{N\gg1}{\approx}&\frac{(-1)^{k_L+1}}{2\pi } \left(\frac{\sqrt{(N_{\rm f}+1)^2+k^2-k_L^2}-k}{\sqrt{(N_{\rm f}+1)^2+k^2-k_L^2}+k}\;\frac{N_{\rm f}+1+k_L}{N_{\rm f}+1-k_L}\right)^{2k_L} \omega\\
&\times \frac{\det\limits_{\substack{1\leq a\leq N_{\rm f}+1+k_L \\ 1\leq b\leq N_{\rm f}+1-k_L \\ 1\leq c\leq 2N_{\rm f}}}\left[\begin{array}{c|c|c} (\omega \widehat{m}_c)^{a-1}e^{\omega \widehat{m}_c} & (-i\omega \widehat{\lambda})^{a-1}e^{-i\omega \widehat{\lambda}} & (a-1-i\omega \widehat{\lambda}) (-i\omega \widehat{\lambda})^{a-2}e^{-i\omega \widehat{\lambda}} \\  (\omega \widehat{m}_c)^{b-1}e^{-\omega \widehat{m}_c} & (-i\omega \widehat{\lambda})^{b-1}e^{i\omega \widehat{\lambda}} & (b-1+i\omega \widehat{\lambda})(-i\omega \widehat{\lambda})^{b-2}e^{i\omega \widehat{\lambda}}\end{array}\right]}{\prod_{f=1}^{2N_{\rm f}}(\omega \widehat{\lambda}-i\omega \widehat{m}_f)\ \det\limits_{\substack{1\leq a\leq N_{\rm f}+k_L \\ 1\leq b\leq N_{\rm f}-k_L \\ 1\leq c\leq 2N_{\rm f}}}\left[\begin{array}{c} (\omega \widehat{m}_c)^{a-1}e^{\omega \widehat{m}_c}  \\  (\omega \widehat{m}_c)^{b-1}e^{-\omega \widehat{m}_c} \end{array}\right]}.
\end{split}
\end{equation}
Thence, the density for $k_L=0$ is essentially the one with no Chern-Simons term. Interestingly, at the phase transition points the sign of the level density jumps.
Especially in the case of $k$ being an integer, the result reduces to
\begin{equation}\label{a:3.5}
\begin{split}
\frac{1}{N}R(\lambda;M;k)
\overset{N\gg1}{\approx}&\frac{(-1)^{k+1}}{2\pi }  \frac{\det\limits_{\substack{1\leq a\leq N_{\rm f}+1+k \\ 1\leq b\leq N_{\rm f}+1-k \\ 1\leq c\leq 2N_{\rm f}}}\left[\begin{array}{c|c|c} \widehat{m}_c^{a-1}e^{\widehat{m}_c} & (-i\widehat{\lambda})^{a-1}e^{-i\widehat{\lambda}} & (a-1-i\widehat{\lambda}) (-i\widehat{\lambda})^{a-2}e^{-i\widehat{\lambda}} \\  \widehat{m}_c^{b-1}e^{-\widehat{m}_c} & (-i\widehat{\lambda})^{b-1}e^{i\widehat{\lambda}} & (b-1+i\widehat{\lambda})(-i\widehat{\lambda})^{b-2}e^{i\widehat{\lambda}}\end{array}\right]}{\prod_{f=1}^{2N_{\rm f}}(\widehat{\lambda}-i\widehat{m}_f)\ \det\limits_{\substack{1\leq a\leq N_{\rm f}+k \\ 1\leq b\leq N_{\rm f}-k \\ 1\leq c\leq 2N_{\rm f}}}\left[\begin{array}{c} \widehat{m}_c^{a-1}e^{\widehat{m}_c}  \\  \widehat{m}_c^{b-1}e^{-\widehat{m}_c} \end{array}\right]}.
\end{split}
\end{equation}

Here, we want to underline as before that we assume that $|k|$ is smaller than a critical value $k_c>N_{\rm f}+1/2$. Above this value, the microscopic level density is governed by the oscillations as in the unquenched system, and a microscopic limit does not exist. The corresponding oscillatory part can be obtained by
choosing $k_L=N_{\rm f}$ in Eq.~\eqref{a:1.14}   and $k_L=N_{\rm f}+1$
in Eq.~\eqref{a:2.11}. The amplitude will again grow exponentially with $N$. The critical value $k_c$ for a fixed number of flavors $2N_{\rm f}$ can be obtained
from~\eqref{phasetranseq} by comparing $\lfloor k\rfloor=N_{\rm f}$ and $\lceil k\rceil=N_{\rm f}+1$.

\begin{figure}[t!]
	\centering
	\includegraphics[width=1\textwidth]{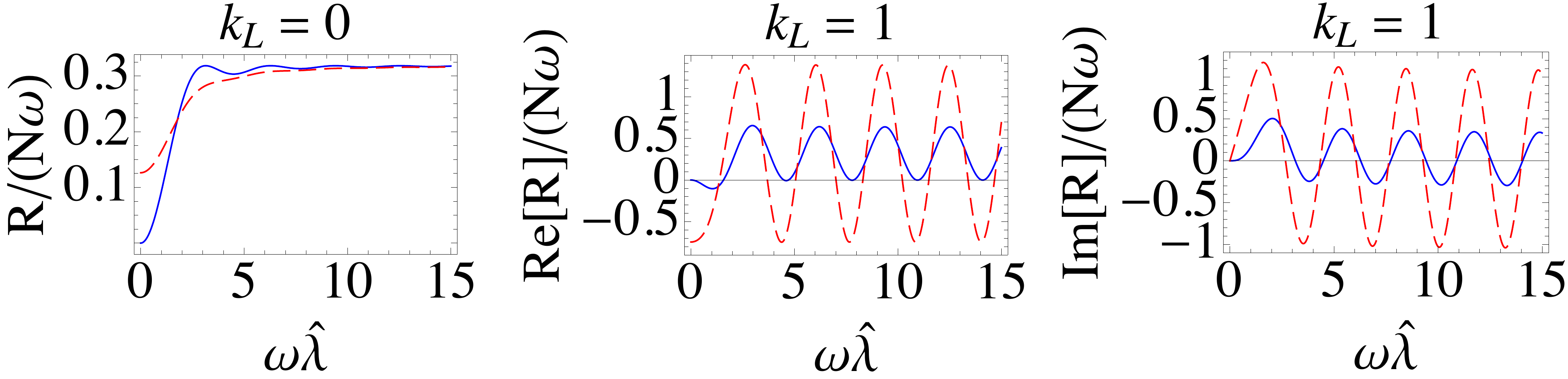}
	\vspace{-.5\baselineskip}
	\caption{\label{fg:2flavor} Microscopic level density for two flavors, $2N_{\rm f}=2$, at the masses $Nm_1=-Nm_2=0$ (blue solid curves) and at $Nm_1=-Nm_2=1.5$ (red dashed curves). The left plot shows the level density in the phase $k_L=0$, see~\eqref{a:3.7}, which is the standard situation without or very weak $|k|<0.50551\ldots$ Chern-Simons term. In particular the level density is real and positive. The middle and right plot represent the real and imaginary part of the microscopic level density in the phase $k_L=1$, cf., Eq.~\eqref{a:3.9}. Since only the normalization constant in front of the level density changes with $k$ inside this phase (note the unfolding with $\omega$), we have chosen $k=1$. We recall that the imaginary part is an odd function around the origin while the real part is an even function.}
\end{figure}

Finally we want to present the result for two flavors ($2N_{\rm f}=2$). For the phase $k_L=0$, the level density is equal to
\begin{equation}\label{a:3.6}
\begin{split}
\frac{1}{N}R_{2N_{\rm f}=2}(\lambda;m_1,m_2;k)
\overset{N\gg1}{\approx}&\frac{\omega}{\pi} \left(1+\frac{\omega (\widehat{m}_1-\widehat{m}_2)}{\sinh[\omega(\widehat{m}_1-\widehat{m}_2)]}\frac{\sinh[\omega(\widehat{m}_1+i \widehat{\lambda})]\sinh[\omega(\widehat{m}_2+i \widehat{\lambda})]}{\omega^2(\widehat{\lambda}-i\widehat{m}_1)(\widehat{\lambda}-i\widehat{m}_2)}\right)\\
=~&\omega\rho_{\rm mic}^{2N_{\rm f}=2}(\omega\widehat{\lambda};\omega\widehat{m}_1,\omega\widehat{m}_2;k=0)
\end{split}
\end{equation}
with $\omega=\sqrt{1+k^2/4}$, cf.~Eq.~\eqref{a:3.5}, and $\rho_{\rm mic}^{2N_{\rm f}=2}(k=0)$ being the microscopic level density without a Chern-Simons term. Interestingly, the whole spectrum is only rescaled by the factor $\omega$, in particular the mean level spacing is not anymore $\pi$ but $\pi/\omega$. We recall that $|k|$ has to be smaller than
the critical value of
$0.50551\ldots$, see Table~\ref{tb:ptpoints}. When $m_1=-m_2=m$ the density is evidently real
\begin{equation}\label{a:3.7}
\begin{split}
\frac{1}{N}R_{2N_{\rm f}=2}(\lambda;m_1=-m_2=m;k)
\overset{N\gg1}{\approx}&\frac{\omega}{\pi }  \left(1-\frac{2\omega \widehat{m}}{\sinh[2\omega\widehat{m}]}\frac{|\sinh[\omega(\widehat{m}+i \widehat{\lambda})]|^2}{\omega^2(\widehat{\lambda}^2+\widehat{m}^2)}\right)=\omega\rho_{\rm mic}^{2N_{\rm f}=2}(\omega\widehat{\lambda};\omega\widehat{m};0).
\end{split}
\end{equation}
It is shown in the left plot of Figure~\ref{fg:2flavor} for two different quark masses. When taking the masses to infinity $\widehat{m}\to\infty$ the quarks  decouple and we recover the quenched case $\rho_{\rm mic}(\widehat{\lambda})=1/\pi$ as expected.

The corresponding quark-antiquark condensates for $\widehat{m}=\widehat{m}_1=\widehat{m}_2$ and $\widehat{m}=\widehat{m}_1=-\widehat{m}_2$ are in this phase equal to
\begin{equation}
\begin{split}
\Sigma V=\frac{1}{2}\frac{\partial}{\partial \widehat{m}}\log Z_{2N_{\rm f}=2}^{\beta=2}(\widehat{m},\widehat{m};k,k_L=0)=\frac{1}{2}\frac{\partial}{\partial \widehat{m}}\log \left[e^{-(\lambda_++\lambda_-)\widehat{m}}\right]=- \frac{\lambda_++\lambda_-}{2}
\end{split}
\end{equation}
and
\begin{equation}
\begin{split}
\Sigma V=\frac{1}{2}\frac{\partial}{\partial \widehat{m}}\log Z_{2N_{\rm f}=2}^{\beta=2}(\widehat{m},-\widehat{m};k,k_L=0)=\frac{1}{2}\frac{\partial}{\partial \widehat{m}}\log \left[\frac{\sinh[2\omega\widehat{m}]}{\widehat{m}}\right]=\frac{\omega}{\tanh[2\omega\widehat{m}]}-\frac{1}{2\widehat{m}},
\end{split}
\end{equation}
respectively.

%%%%%%%%%%%%%%%%%%%%%%%%%%%%%%%%
%%%%%%%%%%%%%%%%%%%%%%%%%%%%%%%%

In the nontrivial phase $k_L=1$ (the case $k_L=-1$ is very similar), the level density has the form
\be\label{a:3.8}
\frac{1}{N}R_{2N_{\rm f}=2}(\lambda;m_1,m_2;k)
&\overset{N\gg1}{\approx}&\frac{9\omega}{\pi}
\left(\frac{\sqrt{3+k^2}-k}{\sqrt{3+k^2}+k}\right)^2
\biggl(
1+\frac i2 \left [\frac 1{\omega(\widehat{\lambda}-i \widehat{m}_1)}+\frac 1{\omega(\widehat{\lambda}-i \widehat{m}_2)} \right ] \nn\\
&&
    +\frac{e^{-\omega(\widehat{m}_1+\widehat{m}_2)-2i\omega\widehat{\lambda}}}{\omega(\widehat{m}_1-\widehat{m}_2)}
      \left [\frac{\widehat{\lambda}-i \widehat{m}_1}{\widehat{\lambda}-i \widehat{m}_2}e^{\omega(\widehat{m}_1-\widehat{m}_2)}-\frac{\widehat{\lambda}-i \widehat{m}_2}{\widehat{\lambda}-i \widehat{m}_1}e^{\omega(\widehat{m}_2-\widehat{m}_1)} \right  ] 
\biggl)
\ee
with $\omega=(2\sqrt{3+k^2}-k)/3$ [cf.~Eq.~\eqref{a:3.5}]. This spectral
density has always a non-trivial imaginary part even when we set $m_1=-m_2=m$
in which case it simplifies to
\begin{equation}\label{a:3.9}
\begin{split}
\frac{1}{N}R_{2N_{\rm f}=2}(\lambda;m_1=-m_2=m;k)
\overset{N\gg1}{\approx}&\frac{9\omega}{\pi} \left(\frac{\sqrt{3+k^2}-k}{\sqrt{3+k^2}+k}\right)^2\biggl(1+i\frac{\omega\widehat{\lambda}}{\omega^2(\widehat{\lambda}^2+\widehat{m}^2)}\\
&+\frac{\omega^2(\widehat{\lambda}^2-\widehat{m}^2)}{\omega^2(\widehat{\lambda}^2+\widehat{m}^2)}\frac{\sinh[2\omega\widehat{m}]}{2\omega\widehat{m}}e^{-2i\omega\widehat{\lambda}} -i\frac{\omega\widehat{\lambda}}{\omega^2(\widehat{\lambda}^2+\widehat{m}^2)}\cosh[2\omega\widehat{m}]e^{-2i\omega\widehat{\lambda}}\biggl)\,.
\end{split}
\end{equation}
From these results we can read off several things. First of all, the level density exhibits  complex oscillations in the phase $k_L=1$, cf., middle and right plot in Figure~\ref{fg:2flavor}, which also holds for any $|k_L|>0$ phase when considering even more flavors. Additionally, the amplitude of these oscillations grows exponentially in the quark masses. Therefore, we cannot expect that the
quenched limit
exists in this phase; especially the reduction of the number of flavors does not work anymore. Finally, the change from one phase to another, say $k_L\to k_L+1$,
drastically changes the microscopic spectral density. There is not a smooth transition of the microscopic level densities in the various phases and it completely breaks down when $|k|$ crosses a critical $k_c$, see discussion above.

%%%%%%%%%%%%%%%%%%%%%%%%%%%%%%%%
\begin{figure}[t!]
	\centering
	\includegraphics[width=0.45\textwidth]{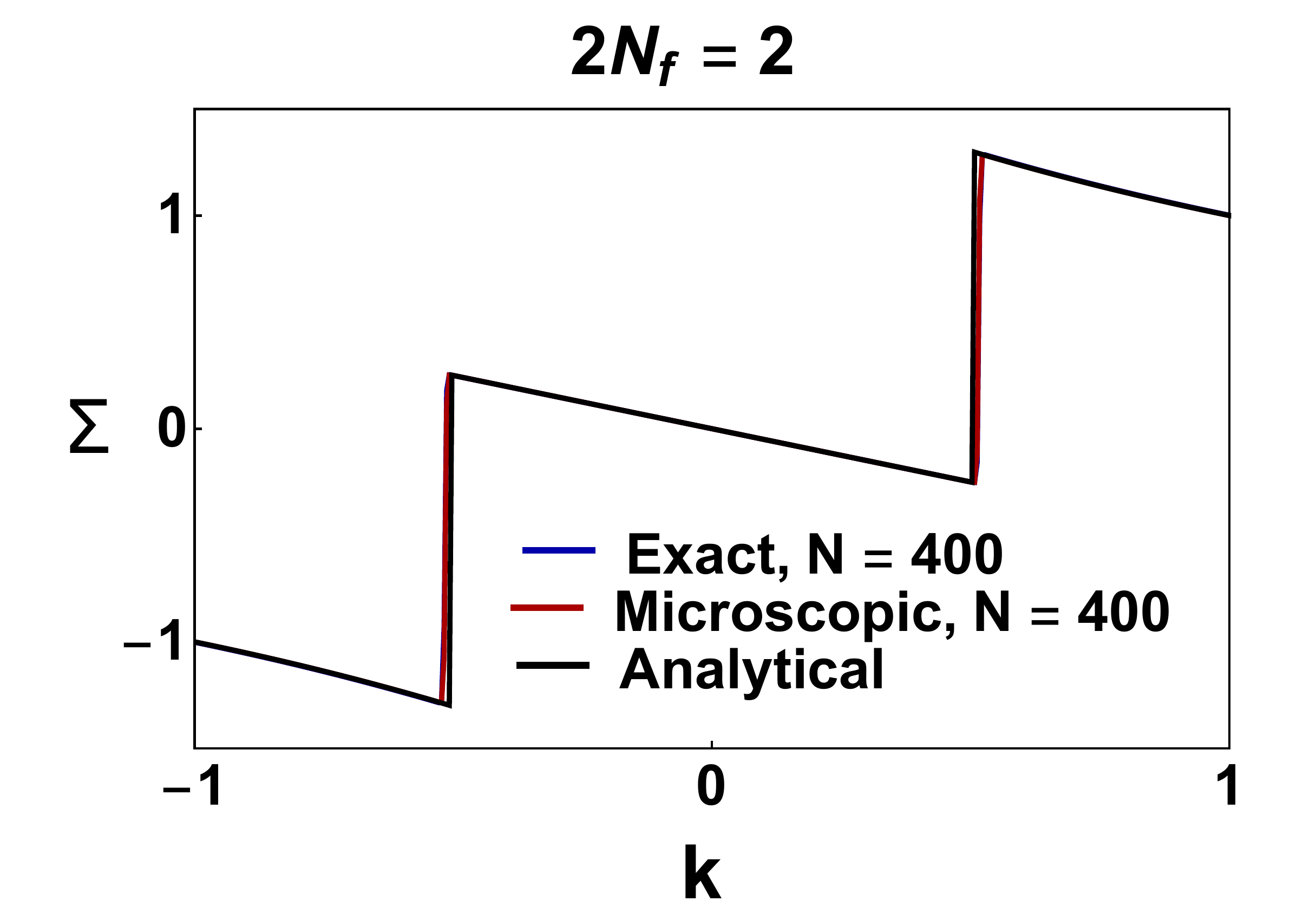}
	\vspace{-.5\baselineskip}
	\caption{\label{fg:micro-cond}The $k$ dependence of the  condensate
          in the limit of zero quark masses. The exact result (blue curve)
          and microscopic
          result (red curve)  have been obtained from the partition functions \eref{eq:fastZ2s}
          and \eref{a:1.14}, for $N=400$ and $m_1=m_2 = 10^{-4}$, respectively.
Note that the two curves  completely overlap.
          The black curve
          represents the analytical result \eref{micro-cond}.
}
\end{figure}
%%%%%%%%%%%%%%%%%%%%%%%%%%%%%%%%

The quark-antiquark condensate as a function of $k$ readily follows for
from the partition function which for $2N_{\rm f}$ has three components
\be
Z_{2N_{\rm f}=2}^{\beta=2}(\widehat{m}_1,\widehat{m}_2;k,k_L=1)+
Z_{2N_{\rm f}=2}^{\beta=2}(\widehat{m}_1,\widehat{m}_2;k,k_L=0)+
Z_{2N_{\rm f}=2}^{\beta=2}(\widehat{m}_1,\widehat{m}_2;k,k_L=-1).
\ee
The explicit expressions are given in Eq. \eref{a:1.14}. In the thermodynamic limit
only one of the three components contributes to the condensate depending
on the value of $k$. For $m_1 =m_2$ the integrand in Eq. \eref{a:1.15} does not
depend on $U$ resulting in a pure exponential mass dependence.
In the microscopic limit, 
we thus find a mass independent chiral condensate given by
\be
\Sigma(k) &=& -\theta(k-k_c) \lambda_-(k,k_L=1) -\theta(-k-k_c) \lambda_+(k,k_L=-1)
\nn\\ &&- \frac 12 \theta(k_c-k)\theta(k+k_c)(\lambda_-(k,k_L=0)+\lambda_+(k,k_L=0))
\label{micro-cond}
    \ee
    with $k_c = 0.50551$ (see Table  \ref{tb:ptpoints}).
Up to $1/N $ corrections, 
this result (black curve in Fig. \ref{fg:micro-cond})
is in agreement with the $k$ dependence of the
condensate for close to
massless quarks obtained from the exact partition function
\eref{eq:fastZ2s} (blue curve in Fig. \ref{fg:micro-cond}) which coincides with
the result from the microscopic partition function \eref{a:1.14}  (red curve in Fig. \ref{fg:micro-cond}). The small discrepancy between the last two curves and the analytical
result is due to $1/N$ corrections -- taking the quark masses closer to zero does not
change the curves.

\section{\label{sc:concl}Conclusion and outlook}

We have constructed a random matrix theory for QCD in three dimensions (QCD$_3$) with
a Chern-Simons term of level $k$ that reproduces the pattern of
spontaneous symmetry  breaking according to U($2N_{\rm f}$) $\to $
U($N_{\rm f}+k$)$\times$U($N_{\rm f}-k$) as proposed recently by Komargodski and
Seiberg \cite{Komargodski:2017keh}. This random matrix model is an extension of  the random matrix for
QCD$_3$ without a Chern-Simons term ($k=0$).
The Chern-Simons term of the random matrix model is in some aspects different in character
from the Chern-Simons term of QCD in 3 dimensions but agrees in other aspects. In particular, the
level $k$ is not quantized, which is not surprising since the random matrix
model does not have a local gauge invariance. However, the effect of the
Chern-Simons term on the eigenvalues is similar --- it adds a phase
proportional to $k$ to the phase of the fermion determinant.
It is remarkable that,
in all cases we know of, random matrix theories with global
symmetries of QCD-like theories reproduce their pattern of spontaneous
symmetry breaking and break the symmetry in such a way that the corresponding
condensate has the maximum global symmetry, see Ref.~\cite{Peskin:1980gc}.
The present work shows that a complex action can violate 
this feature, even in the case of random matrix theory.

What we have learned from earlier work on random matrix theory
with a complex action  is that the imaginary
part of the action can move the phase boundaries of the phase quenched
theory. For QCD at nonzero chemical potential the phase of the fermion
determinant moves the critical chemical potential of half the pion
mass to 1/3 of the baryon mass. For QCD at nonzero theta angle, the chiral
condensate does not change sign when one of the quark masses does not
change sign. Keeping this in mind, it is not unexpected that the phase due to
the  Chern-Simons term can change the phase of the theory: the imaginary part
of the action nullifies the leading phase so that the subleading phase
becomes dominant. At $k \ne 0$, the phase with the standard pattern of
chiral symmetry breaking is canceled, so that phases with asymmetric
breaking of spontaneous symmetry breaking becomes dominant.

\acknowledgments

J.V. acknowledges partial support from  U.S. DOE Grant  No.~DE-FAG-88FR40388, and M.K. acknowledges the support by the German research council (DFG) via the CRC 1283: ``Taming uncertainty and profiting from randomness and low regularity in analysis, stochastics and their applications''. Ideas related to this paper
are also discussed in ``Symmetry Breaking, Phase Transitions and Duality in Large-$N$ QCD$_3$''' by Adi Armoni, Thomas Dumitrescu, Guido Festuccia and Zohar Komargodski.
Zohar Komargodski is thanked for discussions and  sharing a draft the paper.

\appendix

\section{Derivation of some partition functions}\label{sec:app}

In this appendix we work out  the explicit computation of
the two partition functions in Eq.~\eqref{leveldens-ratio}. The one in the denominator has to be dealt separately for the quenched (Subsection~\ref{sec:Aquenched}) and unquenched (Subsection~\ref{sec:Aunquenched}) ensemble while this distinction is not
relevant for the partition function in the numerator, which is evaluated in Subsection~\ref{sec:Bint}.

\subsection{\boldmath Quenched $A$ integral}\label{sec:Aquenched}

We first consider the quenched partition function
\begin{equation}\label{a:0.1}
Z_{\rm q}=\int \rmd A\exp\left[\frac{\alpha_2}{2}(\Tr A-2i k)^2-\frac{N}{2}\Tr A^2\right],
\end{equation}
and linearize the squared term by introducing an auxiliary $x$-integral,
\begin{equation}\label{a:0.2}
Z_{\rm q}=\int_{-\infty}^\infty\frac{\rmd x}{\sqrt{2\alpha_2 \pi}} \int \rmd A\exp\left[-\frac{x^2}{2\alpha_2}-x(\Tr A-2i k)-\frac{N}{2}\Tr A^2\right].
\end{equation}
Next, we shift $A\to A-x/N\1_N$ and integrate over $A$,
\begin{equation}\label{a:0.3}
Z_{\rm q}=2^{N/2}\left(\frac{\pi}{N}\right)^{N^2/2}\int_{-\infty}^\infty\frac{\rmd x}{\sqrt{2\alpha_2 \pi}} \exp\left[-\frac{1-\alpha_2}{2\alpha_2}x^2+2i kx\right].
\end{equation}
Finally, we perform the integration over $x$ and arrive at
\begin{equation}\label{a:0.4}
Z_{\rm q}=\frac{2^{N/2}\left(\pi/N\right)^{N^2/2}}{\sqrt{1-\alpha_2}}\exp\left[-\frac{2\alpha_2}{1-\alpha_2}k^2\right]\overset{N\gg1}{\approx}2^{N/2}\sqrt{N}\left(\pi/N\right)^{N^2/2}\exp\left[-2Nk^2\right].
\end{equation}

\subsection{\boldmath Unquenched $A$ integral}\label{sec:Aunquenched}

The next quantity we consider is the unquenched partition function
\begin{equation}\label{a:1.1}
Z_{N_{\rm f}}=\int \rmd A\exp\left[\frac{\alpha_2}{2}(\Tr A-2i k)^2-\frac{N}{2}\Tr A^2\right]\prod_{f=1}^{2N_{\rm f}}\det[i A+m_f\1_{N}].
\end{equation}
As in the previous section, we first linearize the squared term with the help of an $x$-integral and then introduce a Gaussian integral over a  complex Grassmann valued $N\times 2N_{\rm f}$ matrix $V$,
\begin{equation}\label{a:1.2}
Z_{N_{\rm f}}=\int_{-\infty}^\infty\frac{\rmd x}{\sqrt{2\alpha_2 \pi}} \int \rmd A\exp\left[-\frac{x^2}{2\alpha_2}-x(\Tr A-2i k)-\frac{N}{2}\Tr A^2\right]\frac{\int \rmd V \exp[-i\Tr AVV^\dagger +\Tr V^\dagger V M]}{\int \rmd V \exp[\Tr V^\dagger V]}.
\end{equation}
Here, we used the anticommuting property of Grassmann variables.
The integral over $A$ can again be performed after the shift $A\to A-x/N\1_N$ yielding
\begin{equation}\label{a:1.3}
Z_{N_{\rm f}}=2^{N/2}\left(\frac{\pi}{N}\right)^{N^2/2}\int_{-\infty}^\infty\frac{\rmd x}{\sqrt{2\alpha_2 \pi}} \exp\left[-\frac{1-\alpha_2}{2\alpha_2}x^2+2ikx\right]\frac{\int \rmd V \exp[\frac{1}{2N}\Tr (V^\dagger V)^2 +\Tr V^\dagger V (M-i\frac{x}{N}\1_{2N_{\rm f}})]}{\int \rmd V \exp[\Tr V^\dagger V]}.
\end{equation}
In the next stage, we integrate over $x$ and find
\begin{equation}\label{a:1.4}
Z_{N_{\rm f}}=\frac{2^{N/2}\left(\pi/N\right)^{N^2/2}}{\sqrt{1-\alpha_2}}\frac{\int \rmd V \exp[\frac{1}{2N}\Tr (V^\dagger V)^2 +\Tr V^\dagger VM -\frac{N}{2(2N_{\rm f}+1)}(2k-\frac{1}{N}\Tr V^\dagger V)^2]}{\int \rmd V \exp[\Tr V^\dagger V]}.
\end{equation}

Now we are ready to apply the bosonization formula~\cite{Som,LSZ,KSG}  and replace $V^\dagger V$ by $N \wt{U}$ with $\wt{U}\in\U(2N_{\rm f})$. The scaling factor $N$ is chosen for convenience of the saddle point analysis. Thus, we have
\begin{equation}\label{a:1.5}
\begin{split}
Z_{N_{\rm f}}&=\frac{2^{N/2}\left(\pi/N\right)^{N^2/2}}{\sqrt{1-\alpha_2}}\frac{\int \rmd\mu(\wt{U}) \det^{-N}\wt{U} \exp[\frac{N}{2}\Tr \wt{U}^2 +\Tr \wt{U}\widehat{M} -\frac{N}{2(2N_{\rm f}+1)}(2k-\Tr \wt{U})^2]}{\int \rmd\mu(\wt{U}) \det^{-N}\wt{U}\exp[N\Tr \wt{U}]},
\end{split}
\end{equation}
where $\widehat{M}=NM$ is fixed in the microscopic limit. Next we diagonalize the matrix $\wt{U}=U^\dagger z U$ with $z$ a diagonal matrix of complex phases,
\begin{equation}\label{a:1.6}
\begin{split}
  Z_{N_{\rm f}}&=\frac{2^{N/2}\left(\pi/N\right)^{N^2/2}}{\sqrt{1-\alpha_2}}
  \frac{\int \frac{\rmd z}{\det^{N+1}z} |\Delta_{2N_{\rm f}}(z)|^2
\int \rmd \mu(U)
    \exp[\frac{N}{2}\Tr z^2 -\frac{N}{2(2N_{\rm f}+1)}(2k-\Tr z)^2 +\Tr U^\dagger z U\widehat{M}]}{\int \frac{\rmd z}{\det^{N+1}z} |\Delta_{2N_{\rm f}}(z)|^2\exp[N\Tr z]}
    \\
&=\frac{2^{N/2}\left(\pi/N\right)^{N^2/2}}{(2\pi i)^{2N_{\rm f}}N^{2N_{\rm f}N} (2N_{\rm f})!\sqrt{1-\alpha_2}}\prod_{j=0}^{2N_{\rm f}-1}\frac{(N+j)!}{j!}\\
&\quad \times \int \frac{\rmd z}{\det^{N+1}z} |\Delta_{2N_{\rm f}}(z)|^2 \int \rmd\mu(U)\exp\left[\frac{N}{2}\Tr z^2 -\frac{N}{2(2N_{\rm f}+1)}(2k-\Tr z)^2 +\Tr U^\dagger z U\widehat{M}\right].
\end{split}
\end{equation}
Let $z_k$ be one saddle point solution of the saddle point equation
\begin{equation}\label{a:1.7}
z-z^{-1}+\frac{1}{2N_{\rm f}+1}(2k-\Tr z)\1_{2N_{\rm f}}=0
\end{equation}
and maximizing the integrand. The relation to $\Lambda_k$, discussed in section~\ref{sec:phase}, is $z_k=\Lambda_k^{-1}$ which indeed yields the saddle point equation~\eqref{eq:sadeq} after plugging this relation into~\eqref{a:1.7}.
Therefore, we already know that there are $(2N_{\rm f})!/[(N_{\rm f}+k_L)!(N_{\rm f}-k_L)!]$ points satisfying these two conditions, where $k_L$ is either the integer above $k$ or below $k$ depending on the phase the system is in. In particular we can choose $z_k=\diag(\lambda_+^{-1}\1_{N_{\rm f}+k},\lambda_-^{-1}\1_{N_{\rm f}-k})$ with
\begin{equation}\label{a:1.8}
\lambda_\pm=\frac{k\pm\sqrt{(N_{\rm f}+1)^2+k^2-k_L^2}}{N_{\rm f}+1\pm k_L}.
\end{equation}
We underline that these solutions are always real because $k^2-k_L^2\geq-|k+k_L|/2\geq-\lceil |k|\rceil\geq -N_{\rm f}$. Moreover we have always $\lambda_+>0$ and $\lambda_-<0$.

The contribution from the fluctuations about the saddle point can be obtained
from the expansion $z=z_k+i \diag(\delta z_+,-\delta z_-)/\sqrt{N}$ with $\delta z_+\in\mathbb{R}^{N_{\rm f}+k_L}$ and  $\delta z_-\in\mathbb{R}^{N_{\rm f}-k_L}$, where the phase pre-factors reflect the direction of the original contour. Then, the measure transforms as follows
\begin{equation}\label{a:1.9}
 |\Delta_{2N_{\rm f}}(z)|^2\frac{\rmd z}{\det z}\overset{N\gg1}{\approx}  \frac{(\lambda_+^{-1}-\lambda_-^{-1})^{2(N_{\rm f}^2-k_L^2)}\Delta^2_{N_{\rm f}+k_L}(\delta z_+)\Delta^2_{N_{\rm f}-k_L}(\delta z_-)}{N^{(N_{\rm f}+k_L)(N_{\rm f}+k_L-1)/2+(N_{\rm f}-k_L)(N_{\rm f}-k_L-1)/2}}\frac{(-1)^{k_L}\lambda_+^{N_{\rm f}+k_L}\lambda_-^{N_{\rm f}-k_L}\rmd \delta z_+\rmd \delta z_-}{N^{N_{\rm f}}}.
\end{equation}
Exploiting the identity $\lambda_+\lambda_-=-1$, one can explicitly write
\begin{equation}\label{a:1.10}
\begin{split}
 |\Delta_{2N_{\rm f}}(z)|^2\frac{\rmd z}{\det z}\overset{N\gg1}{\approx}&  \frac{(-1)^{N_{\rm f}}}{N^{N_{\rm f}^2+k_L^2}}\left(\frac{\sqrt{(N_{\rm f}+1)^2+k^2-k_L^2}+k}{\sqrt{(N_{\rm f}+1)^2+k^2-k_L^2}-k}\;\frac{N_{\rm f}+1-k_L}{N_{\rm f}+1+k_L}\right)^{k_L}\\
 &\times\left(2\frac{(N_{\rm f}+1)\sqrt{(N_{\rm f}+1)^2+k^2-k_L^2}-kk_L}{(N_{\rm f}+1)^2-k_L^2}\right)^{2(N_{\rm f}^2-k_L^2)}\Delta^2_{N_{\rm f}+k_L}(\delta z_+)\Delta^2_{N_{\rm f}-k_L}(\delta z_-)\rmd \delta z_+\rmd \delta z_-.
 \end{split}
\end{equation}
In the next step, we expand the determinant
\begin{equation}\label{a:1.11}
\begin{split}
{\det}^{-N} z \overset{N\gg1}{\approx}&\lambda_+^{N(N_{\rm f}+k_L)}\lambda_-^{N(N_{\rm f}-k_L)}\exp\left[-\sqrt{N} i(\lambda_+\Tr\delta z_+-\lambda_-\Tr\delta z_-)-\frac{1}{2}(\lambda_+^2\Tr\delta z_+^2+\lambda_-^2\Tr\delta z_-^2)\right]\\
=~&(-1)^{(N_{\rm f}+k_L)N}\left(\frac{\sqrt{(N_{\rm f}+1)^2+k^2-k_L^2}+k}{\sqrt{(N_{\rm f}+1)^2+k^2-k_L^2}-k}\;\frac{N_{\rm f}+1-k_L}{N_{\rm f}+1+k_L}\right)^{k_LN}\\
&\times\exp\left[-\sqrt{N} i(\lambda_+\Tr\delta z_+-\lambda_-\Tr\delta z_-)-\frac{1}{2}(\lambda_+^2\Tr\delta z_+^2+\lambda_-^2\Tr\delta z_-^2)\right]
\end{split}
\end{equation}
and the exponent
\begin{equation}\label{a:1.12}
\begin{split}
\frac{N}{2}\Tr z^2 -\frac{N(2k-\Tr z)^2}{2(2N_{\rm f}+1)}\overset{N\gg1}{\approx}& \frac{N}{2}\bigl((N_{\rm f}+k_L)\lambda_-^2+(N_{\rm f}-k_L)\lambda_+^2-(2N_{\rm f}+1)(\lambda_+^2+\lambda_-^2-2)\bigl)\\
&+\sqrt{N}i\bigl(\lambda_+\Tr\delta z_+-\lambda_-\Tr\delta z_-\bigl)-\frac{1}{2}\left(\Tr \delta z_+^2+\Tr \delta z_-^2\right)+\frac{(\Tr \delta z_+-\Tr \delta z_-)^2}{2(2N_{\rm f}+1)},
\end{split}
\end{equation}
where we employed the saddle point equation. The mass dependent term is independent of the massive modes $\delta z_\pm$. As it should be, the exponents proportional to $\sqrt{N}$ cancel each other so that we are left with the $\delta z_\pm$ integrals that can be computed as follows
\begin{equation}\label{a:1.13}
\begin{split}
&\int \rmd \delta z_+\rmd \delta z_- \Delta^2_{N_{\rm f}+k_L}(\delta z_+)\Delta^2_{N_{\rm f}-k_L}(\delta z_-)\exp\left[-\frac{1}{2}\bigl((\lambda_+^2+1)\Tr\delta z_+^2+(\lambda_-^2+1)\Tr\delta z_-^2\bigl)+\frac{(\Tr \delta z_+-\Tr \delta z_-)^2}{2(2N_{\rm f}+1)}\right]\\
=& \int_{-\infty}^\infty \frac{\sqrt{2N_{\rm f}+1}\;\rmd x}{\sqrt{2\pi}}\int \rmd \delta z_+\rmd \delta z_- \Delta^2_{N_{\rm f}+k_L}(\delta z_+)\Delta^2_{N_{\rm f}-k_L}(\delta z_-)\rme^{-\frac{2N_{\rm f}+1}{2}x^2-\frac{1}{2}((\lambda_+^2+1)\Tr\delta z_+^2+(\lambda_-^2+1)\Tr\delta z_-^2)+x(\Tr \delta z_+-\Tr \delta z_-)}\\
=&~\frac{\sqrt{2N_{\rm f}+1}}{(\lambda_+^2+1)^{(N_{\rm f}+k_L)^2/2}(\lambda_-^2+1)^{(N_{\rm f}-k_L)^2/2}}\int_{-\infty}^\infty \frac{\rmd x}{\sqrt{2\pi}}\exp\left[-\left(2N_{\rm f}+1-\frac{N_{\rm f}+k_L}{\lambda_+^2+1}-\frac{N_{\rm f}-k_L}{\lambda_-^2+1}\right)\frac{x^2}{2}\right]\\
&\quad\times\int \rmd \delta z_+ \Delta^2_{N_{\rm f}+k_L}(\delta z_+) \rme^{-\Tr\delta z_+^2/2}\int \rmd \delta z_- \Delta^2_{N_{\rm f}-k_L}(\delta z_-)\rme^{-\Tr\delta z_-^2/2}\\
=&~\frac{\sqrt{N_{\rm f}+1/2}}{\bigl((N_{\rm f}+1)^2+k^2-k_L^2\bigl)^{1/4}}(N_{\rm f}+k_L)!\left(\prod_{j=0}^{N_{\rm f}+k_L-1}\sqrt{2\pi} j!\right)(N_{\rm f}-k_L)!\left(\prod_{j=0}^{N_{\rm f}-k_L-1}\sqrt{2\pi} j!\right)\\
&\times\left(\frac{\sqrt{(N_{\rm f}+1)^2+k^2-k_L^2}+k}{\sqrt{(N_{\rm f}+1)^2+k^2-k_L^2}-k}\;\frac{N_{\rm f}+1-k_L}{N_{\rm f}+1+k_L}\right)^{-N_{\rm f}k_L}\left(2\frac{(N_{\rm f}+1)\sqrt{(N_{\rm f}+1)^2+k^2-k_L^2}-kk_L}{(N_{\rm f}+1)^2-k_L^2}\right)^{\frac{1}{2}-(N_{\rm f}^2+k_L^2)}.
\end{split}
\end{equation}

Now we are ready to put everything together and apply Sterling's formula. We eventually arrive at
\begin{equation}\label{a:1.14}
\begin{split}
  Z_{N_{\rm f}}\overset{N\gg1}{\approx}&\frac{(-1)^{(N_{\rm f}+k_L)N}
    2^{(N-1)/2}\pi^{N^2/2}\rme^{\frac{N}{2}\bigl(k_L(\lambda_-^2-\lambda_+^2)
      -(N_{\rm f}+1)(\lambda_+^2+\lambda_-^2)+2(2N_{\rm f}+1)\bigl)}}{N^{\frac{N^2}{2}+k_L^2-N_{\rm f}^2-\frac{1}{2}}
    \bigl((N_{\rm f}+1)^2+k^2-k_L^2\bigl)^{1/4}}\frac{\left(\prod_{j=0}^{N_{\rm f}+k_L-1}j!\right)\left(\prod_{j=0}^{N_{\rm f}-k_L-1}j!\right)}{\prod_{j=0}^{2N_{\rm f}-1}j!}\\
&\times \left(\frac{\sqrt{(N_{\rm f}+1)^2+k^2-k_L^2}+k}{\sqrt{(N_{\rm f}+1)^2+k^2-k_L^2}-k}\;\frac{N_{\rm f}+1-k_L}{N_{\rm f}+1+k_L}\right)^{k_L(N-N_{\rm f}+1)}\left(2\frac{(N_{\rm f}+1)\sqrt{(N_{\rm f}+1)^2+k^2-k_L^2}-kk_L}{(N_{\rm f}+1)^2-k_L^2}\right)^{N_{\rm f}^2-3k_L^2+\frac{1}{2}}\\
&\times \int \rmd\mu(U)\rme^{-\Tr U^\dagger \diag(\lambda_-\1_{N_{\rm f}+k_L},\lambda_+\1_{N_{\rm f}-k_L}) U\widehat{M}}.
\end{split}
\end{equation}
In the case when $k$ is an integer, meaning $k=k_L$ and $\lambda_\pm=\pm1$, this result simplifies drastically  to
\begin{equation}\label{a:1.15}
\begin{split}
Z_{N_{\rm f}}\overset{N\gg1}{\approx}&\frac{(-1)^{(N_{\rm f}+k)N}2^{\frac{N}{2}+N_{\rm f}^2-3k^2}\pi^{N^2/2}\rme^{N_{\rm f}N}}{N^{\frac{N^2}{2}+k^2-N_{\rm f}^2-\frac{1}{2}} \sqrt{N_{\rm f}+1}}\frac{\left(\prod_{j=0}^{N_{\rm f}+k-1}j!\right)\left(\prod_{j=0}^{N_{\rm f}-k-1}j!\right)}{\prod_{j=0}^{2N_{\rm f}-1}j!}\int \rmd\mu(U)\rme^{\Tr U^\dagger \diag(\1_{N_{\rm f}+k},-\1_{N_{\rm f}-k}) U\widehat{M}}.
\end{split}
\end{equation}
Note that the prefactors that only depend on $N$ and $N_{\rm f}$ could have been
absorbed in the definition of the partition function.
We would like to emphasize that the partition function for integer as well as non-integer $k$ is real, as can be already seen from its definition, and even positive when omitting the factor $(-1)^{(N_{\rm f}+k)N}$. The latter can be readily achieved by choosing an even matrix dimension $N$.

\subsection{\boldmath The $B$ integral\label{sec:Bint}}
\label{sec:a3}

In this section we evaluate the integral
\begin{equation}\label{a:2.1}
Y_{N_{\rm f}}=\int \rmd B\exp\left[\frac{\alpha_2}{2}(\Tr B-2i k)^2-\frac{N}{2}\Tr B^2\right]\prod_{f=1}^{2N_{\rm f}+2}\det[i B+(m_f+i\lambda')\1_{N-1}]
\end{equation}
in the large-$N$ limit. The computation proceeds along the same lines as for $Z_{N_{\rm f}}$. However, we cannot easily carry over the entire result for $N\to N-1$ and $N_{\rm f}\to N_{\rm f}+1$ since the standard deviations do not change in the same way.

Again, we introduce the auxiliary real variable $x$ to linearize the squared trace term and the complex Grassmann valued matrix $V$, albeit now it has the dimension $(N-1)\times (2N_{\rm f}+2)$. Collecting the masses in the diagonal matrix $\wt{M}=\diag(m_1,\ldots,m_{2N_{\rm f}+2})+i\lambda'\1_{2N_{\rm f}+2}$, we find the integral
\begin{equation}\label{a:2.2}
Y_{N_{\rm f}}=2^{(N-1)/2}\left(\frac{\pi}{N}\right)^{(N-1)^2/2}\int_{-\infty}^\infty\frac{\rmd x}{\sqrt{2\alpha_2 \pi}} \rme^{-\frac{N_{\rm f}+1}{2N}x^2+2ikx}\frac{\int \rmd V \exp[\frac{1}{2N}\Tr (V^\dagger V)^2 +\Tr V^\dagger V (\wt{M}-i\frac{x}{N}\1_{2N_{\rm f}+2})]}{\int \rmd V \exp[\Tr V^\dagger V]},
\end{equation}
 after the integration over $B$, which is the counterpart of~\eqref{a:1.3}. Next, we apply the bosonization formula~\cite{Som,LSZ,KSG} and integrate over the variable $x$  so that we obtain
\begin{equation}\label{a:2.3}
\begin{split}
Y_{N_{\rm f}}=&~\frac{2^{(N-2)/2}\left(\pi/N\right)^{(N-1)^2/2}\sqrt{N+2N_{\rm f}+1}}{\sqrt{N_{\rm f}+1}}\frac{\int \rmd\mu(\wt{U}) \det^{-N+1}\wt{U} \exp[\frac{N}{2}\Tr \wt{U}^2 +\Tr \wt{U}\widehat{\wt{M}} -\frac{N}{4(N_{\rm f}+1)}(2k-\Tr \wt{U})^2]}{\int \rmd\mu(\wt{U}) \det^{-N+1}\wt{U}\exp[N\Tr \wt{U}]}
\end{split}
\end{equation}
with $\widehat{\wt{M}}=N\wt{M}$. When diagonalizing $\wt{U}=U^\dagger \wt{z}U$ with $\wt{z}$ a $2N_{\rm f}+2$ dimensional diagonal matrix of complex phases we obtain
\begin{equation}\label{a:2.4}
\begin{split}
Y_{N_{\rm f}}=&~\frac{2^{(N-2)/2}\left(\pi/N\right)^{(N-1)^2/2}\sqrt{N+2N_{\rm f}+1}}{(2\pi i)^{2N_{\rm f}+2}N^{2(N_{\rm f}+1)(N-1)}(2N_{\rm f}+2)!\sqrt{N_{\rm f}+1}}\prod_{j=0}^{2N_{\rm f}+1}\frac{(N-1+j)!}{j!}\\
&\times \int \frac{\rmd \wt{z}}{\det^{N}\wt{z}} |\Delta_{2N_{\rm f}+2}(\wt{z})|^2 \exp\left[\frac{N}{2}\Tr \wt{z}^2-\frac{N}{4(N_{\rm f}+1)}(2k-\Tr \wt{z})^2\right] \int \rmd\mu(U)\rme^{\Tr U^\dagger \wt{z}U\widehat{\wt{M}} }
\end{split}
\end{equation}

We are ready for a  saddle point analysis of the $\tilde z$ integral
whose saddle point equation is
\begin{equation}\label{a:2.5}
\wt{z}-\wt{z}^{-1}+\frac{1}{2(N_{\rm f}+1)}(2k-\Tr \wt{z})\1_{2N_{\rm f}+2}=0.
\end{equation}
By plugging in the choice $\wt{z}_k=\diag(\lambda_+^{-1}\1_{N_{\rm f}+1+k_L},\lambda_-^{-1}\1_{N_{\rm f}+1-k_L})$ with exactly the same $\lambda_\pm$ and $k_L$ as in section~\ref{sec:Aunquenched}, one can easily verify that this
is a solution. The real part of the two corresponding actions is apart from an $N$ also the same, especially
\begin{equation}\label{a:2.6}
{\rm Re}\left(\frac{N}{2}\Tr \wt{z}_k^2-\frac{N(2k-\Tr \wt{z}_k)^2}{4(N_{\rm f}+1)}-N\Tr \ln \wt{z}_k\right)={\rm Re}\left(\frac{N}{2}\Tr z_k^2-\frac{N(2k-\Tr z_k)^2}{2(2N_{\rm f}+1)}-N\Tr \ln z_k\right)+N.
\end{equation}
Thus, the phase transition points from the original $A$ integral and from the original $B$ integral are the same as they should. If they were not equal, the
spectral density would be either exponentially small or exponentially large in $N$.
Yet, $Y_{N_{\rm f}}$, with two more flavors than $Z_{N_{\rm f}}$, 
has one additional phase transition compared to $Z_{N_{\rm f}}$ at about $k_c\approx\pm(N_{\rm f}+1/2)$. This suggest the presence of an additional phase transition
in the spectral density  as compared to 
the original partition function. Indeed, we have found in section~\ref{sec:micro} that the limit of the microscopic level density does not exist when
$|k|$ is larger than a critical value $k_c$. Hence we have to stay always in the regime where the two phases of the partition functions $Y_{N_{\rm f}}$ and $Z_{N_{\rm f}}$ agree. When this is not the case the oscillation will become dominant and a microscopic limit does not exist, see the discussions in sections~\ref{sec:micro} and~\ref{sec:unquenched}.

The expansion works exactly the same as before, i.e., $\wt{z}=\wt{z}_k+i\diag(\delta \wt{z}_+,-\delta \wt{z}_-)/\sqrt{N}$ with $\delta \wt{z}_+\in\mathbb{R}^{N_{\rm f}+1+k_L}$ and  $\delta \wt{z}_-\in\mathbb{R}^{N_{\rm f}+1-k_L}$. Thence, we have for the measure
\begin{equation}\label{a:2.7}
\begin{split}
 |\Delta_{2N_{\rm f}+2}(\wt{z})|^2\rmd \wt{z}\overset{N\gg1}{\approx}&  \frac{(-1)^{k_L}}{N^{(N_{\rm f}+1)^2+k_L^2}}\left(2\frac{(N_{\rm f}+1)\sqrt{(N_{\rm f}+1)^2+k^2-k_L^2}-kk_L}{(N_{\rm f}+1)^2-k_L^2}\right)^{2((N_{\rm f}+1)^2-k_L^2)}\\
 &\times\Delta^2_{N_{\rm f}+1+k_L}(\delta \wt{z}_+)\Delta^2_{N_{\rm f}+1-k_L}(\delta \wt{z}_-)\rmd \delta \wt{z}_+\rmd \delta \wt{z}_-,
 \end{split}
\end{equation}
for the determinant
\begin{equation}\label{a:2.8}
\begin{split}
{\det}^{-N} \wt{z} \overset{N\gg1}{\approx}&(-1)^{(N_{\rm f}+k_L+1)N}\left(\frac{\sqrt{(N_{\rm f}+1)^2+k^2-k_L^2}+k}{\sqrt{(N_{\rm f}+1)^2+k^2-k_L^2}-k}\;\frac{N_{\rm f}+1-k_L}{N_{\rm f}+1+k_L}\right)^{k_LN}\\
&\times\exp\left[-\sqrt{N} i(\lambda_+\Tr\delta \wt{z}_+-\lambda_-\Tr\delta \wt{z}_-)-\frac{1}{2}(\lambda_+^2\Tr\delta \wt{z}_+^2+\lambda_-^2\Tr\delta \wt{z}_-^2)\right],
\end{split}
\end{equation}
and for the exponent
\begin{equation}\label{a:2.9}
\begin{split}
\frac{N}{2}\Tr \wt{z}^2 -\frac{N(2k-\Tr \wt{z})^2}{4(N_{\rm f}+1)}\overset{N\gg1}{\approx}& \frac{N}{2}\bigl((N_{\rm f}+k_L)\lambda_-^2+(N_{\rm f}-k_L)\lambda_+^2-(2N_{\rm f}+1)(\lambda_+^2+\lambda_-^2-2)+2\bigl)\\
&+\sqrt{N}i\bigl(\lambda_+\Tr\delta \wt{z}_+-\lambda_-\Tr\delta \wt{z}_-\bigl)-\frac{1}{2}\left(\Tr \delta\wt{z}_+^2+\Tr \delta \wt{z}_-^2\right)+\frac{(\Tr \delta \wt{z}_+-\Tr \delta \wt{z}_-)^2}{4(N_{\rm f}+1)},
\end{split}
\end{equation}
where we again have used the saddle point equation to simplify the result.
The integral over the Gaussian fluctuation $\delta \wt{z}_\pm$ about the
saddle point is given by
\begin{equation}\label{a:2.10}
\begin{split}
&\int \rmd \delta \wt{z}_+\rmd \delta \wt{z}_- \Delta^2_{N_{\rm f}+1+k_L}(\delta \wt{z}_+)\Delta^2_{N_{\rm f}+1-k_L}(\delta \wt{z}_-)\exp\left[-\frac{1}{2}\bigl((\lambda_+^2+1)\Tr\delta \wt{z}_+^2+(\lambda_-^2+1)\Tr\delta \wt{z}_-^2\bigl)+\frac{(\Tr \delta \wt{z}_+-\Tr \delta \wt{z}_-)^2}{4(N_{\rm f}+1)}\right]\\
=&~\frac{\sqrt{N_{\rm f}+1}}{\bigl((N_{\rm f}+1)^2+k^2-k_L^2\bigl)^{1/4}}(N_{\rm f}+1+k_L)!\left(\prod_{j=0}^{N_{\rm f}+k_L}\sqrt{2\pi} j!\right)(N_{\rm f}+1-k_L)!\left(\prod_{j=0}^{N_{\rm f}-k_L}\sqrt{2\pi} j!\right)\\
&\times\left(\frac{\sqrt{(N_{\rm f}+1)^2+k^2-k_L^2}+k}{\sqrt{(N_{\rm f}+1)^2+k^2-k_L^2}-k}\;\frac{N_{\rm f}+1-k_L}{N_{\rm f}+1+k_L}\right)^{-(N_{\rm f}+1)k_L}\left(2\frac{(N_{\rm f}+1)\sqrt{(N_{\rm f}+1)^2+k^2-k_L^2}-kk_L}{(N_{\rm f}+1)^2-k_L^2}\right)^{\frac{1}{2}-((N_{\rm f}+1)^2+k_L^2)}.
\end{split}
\end{equation}
The degeneracy of the saddle points is $(2N_{\rm f}+2)!/[(N_{\rm f}+1+k_L)!(N_{\rm f}+1-k_L)!]$. Combining all contributions we arrive at the main result of this subsection,
\begin{equation}\label{a:2.11}
\begin{split}
  Y_{N_{\rm f}}\overset{N\gg1}{\approx}&\frac{(-1)^{(N_{\rm f}+k_L+1)(N-1)}2^{(N-2)/2}\pi^{(N-1)^2/2}\rme^{\frac{N}{2}\bigl(k_L(\lambda_-^2-\lambda_+^2)
      -(N_{\rm f}+1)(\lambda_+^2+\lambda_-^2-4)\bigl)}}
  {N^{\frac{(N-1)^2}{2}+k_L^2-(N_{\rm f}+1)^2-\frac{1}{2}} \bigl((N_{\rm f}+1)^2+k^2-k_L^2\bigl)^{1/4}}\frac{\left(\prod_{j=0}^{N_{\rm f}+k_L}j!\right)\left(\prod_{j=0}^{N_{\rm f}-k_L}j!\right)}{\prod_{j=0}^{2N_{\rm f}+1}j!}\\
&\hspace*{-1.2cm}\times \left(\frac{\sqrt{(N_{\rm f}+1)^2+k^2-k_L^2}+k}{\sqrt{(N_{\rm f}+1)^2+k^2-k_L^2}-k}\;\frac{N_{\rm f}+1-k_L}{N_{\rm f}+1+k_L}\right)^{k_L(N-N_{\rm f}-1)}\left(2\frac{(N_{\rm f}+1)\sqrt{(N_{\rm f}+1)^2+k^2-k_L^2}-kk_L}{(N_{\rm f}+1)^2-k_L^2}\right)^{(N_{\rm f}+1)^2-3k_L^2+\frac{1}{2}}\\
&\times \int \rmd\mu(U)\rme^{-\Tr U^\dagger \diag(\lambda_-\1_{N_{\rm f}+1+k_L},\lambda_+\1_{N_{\rm f}+1-k_L}) U\widehat{\wt{M}}}.
\end{split}
\end{equation}
and for integer $k$, implying $k=k_L$ and $\lambda_\pm=\pm1$, it reduces to
\begin{equation}\label{a:2.12}
\begin{split}
Y_{N_{\rm f}}\overset{N\gg1}{\approx}&\frac{(-1)^{(N_{\rm f}+k+1)(N-1)}2^{\frac{N-1}{2}+(N_{\rm f}+1)^2-3k^2}\pi^{(N-1)^2/2}\rme^{N(N_{\rm f}+1)}}{N^{\frac{(N-1)^2}{2}+k^2-(N_{\rm f}+1)^2-\frac{1}{2}} \sqrt{N_{\rm f}+1}}\frac{\left(\prod_{j=0}^{N_{\rm f}+k}j!\right)\left(\prod_{j=0}^{N_{\rm f}-k}j!\right)}{\prod_{j=0}^{2N_{\rm f}+1}j!}\\
&\times\int \rmd\mu(U)\rme^{\Tr U^\dagger \diag(\1_{N_{\rm f}+1+k},-\1_{N_{\rm f}+1-k}) U\widehat{\wt{M}}}.
\end{split}
\end{equation}

\bibliographystyle{apsrev4-1}
\bibliography{ref_unitary}

\begin{thebibliography}{93}%
\makeatletter
\providecommand \@ifxundefined [1]{%
 \@ifx{#1\undefined}
}%
\providecommand \@ifnum [1]{%
 \ifnum #1\expandafter \@firstoftwo
 \else \expandafter \@secondoftwo
 \fi
}%
\providecommand \@ifx [1]{%
 \ifx #1\expandafter \@firstoftwo
 \else \expandafter \@secondoftwo
 \fi
}%
\providecommand \natexlab [1]{#1}%
\providecommand \enquote  [1]{``#1''}%
\providecommand \bibnamefont  [1]{#1}%
\providecommand \bibfnamefont [1]{#1}%
\providecommand \citenamefont [1]{#1}%
\providecommand \href@noop [0]{\@secondoftwo}%
\providecommand \href [0]{\begingroup \@sanitize@url \@href}%
\providecommand \@href[1]{\@@startlink{#1}\@@href}%
\providecommand \@@href[1]{\endgroup#1\@@endlink}%
\providecommand \@sanitize@url [0]{\catcode `\\12\catcode `\$12\catcode
  `\&12\catcode `\#12\catcode `\^12\catcode `\_12\catcode `\%12\relax}%
\providecommand \@@startlink[1]{}%
\providecommand \@@endlink[0]{}%
\providecommand \url  [0]{\begingroup\@sanitize@url \@url }%
\providecommand \@url [1]{\endgroup\@href {#1}{\urlprefix }}%
\providecommand \urlprefix  [0]{URL }%
\providecommand \Eprint [0]{\href }%
\providecommand \doibase [0]{http://dx.doi.org/}%
\providecommand \selectlanguage [0]{\@gobble}%
\providecommand \bibinfo  [0]{\@secondoftwo}%
\providecommand \bibfield  [0]{\@secondoftwo}%
\providecommand \translation [1]{[#1]}%
\providecommand \BibitemOpen [0]{}%
\providecommand \bibitemStop [0]{}%
\providecommand \bibitemNoStop [0]{.\EOS\space}%
\providecommand \EOS [0]{\spacefactor3000\relax}%
\providecommand \BibitemShut  [1]{\csname bibitem#1\endcsname}%
\let\auto@bib@innerbib\@empty
%</preamble>
\bibitem [{\citenamefont {Leutwyler}\ and\ \citenamefont
  {Smilga}(1992)}]{Leutwyler:1992yt}%
  \BibitemOpen
  \bibfield  {author} {\bibinfo {author} {\bibfnamefont {H.}~\bibnamefont
  {Leutwyler}}\ and\ \bibinfo {author} {\bibfnamefont {A.~V.}\ \bibnamefont
  {Smilga}},\ }\href {\doibase 10.1103/PhysRevD.46.5607} {\bibfield  {journal}
  {\bibinfo  {journal} {Phys. Rev.}\ }\textbf {\bibinfo {volume} {D46}},\
  \bibinfo {pages} {5607} (\bibinfo {year} {1992})}\BibitemShut {NoStop}%
%%CITATION = PHRVA,D46,5607;%%
\bibitem [{\citenamefont {Shuryak}\ and\ \citenamefont
  {Verbaarschot}(1993)}]{Shuryak:1992pi}%
  \BibitemOpen
  \bibfield  {author} {\bibinfo {author} {\bibfnamefont {E.~V.}\ \bibnamefont
  {Shuryak}}\ and\ \bibinfo {author} {\bibfnamefont {J.~J.~M.}\ \bibnamefont
  {Verbaarschot}},\ }\href {\doibase 10.1016/0375-9474(93)90098-I} {\bibfield
  {journal} {\bibinfo  {journal} {Nucl. Phys.}\ }\textbf {\bibinfo {volume}
  {A560}},\ \bibinfo {pages} {306} (\bibinfo {year} {1993})},\ \Eprint
  {http://arxiv.org/abs/hep-th/9212088} {arXiv:hep-th/9212088 [hep-th]}
  \BibitemShut {NoStop}%
%%CITATION = HEP-TH/9212088;%%
\bibitem [{\citenamefont {Verbaarschot}\ and\ \citenamefont
  {Zahed}(1993)}]{Verbaarschot:1993pm}%
  \BibitemOpen
  \bibfield  {author} {\bibinfo {author} {\bibfnamefont {J.~J.~M.}\
  \bibnamefont {Verbaarschot}}\ and\ \bibinfo {author} {\bibfnamefont
  {I.}~\bibnamefont {Zahed}},\ }\href {\doibase 10.1103/PhysRevLett.70.3852}
  {\bibfield  {journal} {\bibinfo  {journal} {Phys. Rev. Lett.}\ }\textbf
  {\bibinfo {volume} {70}},\ \bibinfo {pages} {3852} (\bibinfo {year}
  {1993})},\ \Eprint {http://arxiv.org/abs/hep-th/9303012}
  {arXiv:hep-th/9303012 [hep-th]} \BibitemShut {NoStop}%
%%CITATION = HEP-TH/9303012;%%
\bibitem [{\citenamefont {Verbaarschot}(1994)}]{Verbaarschot:1994qf}%
  \BibitemOpen
  \bibfield  {author} {\bibinfo {author} {\bibfnamefont {J.~J.~M.}\
  \bibnamefont {Verbaarschot}},\ }\href {\doibase 10.1103/PhysRevLett.72.2531}
  {\bibfield  {journal} {\bibinfo  {journal} {Phys. Rev. Lett.}\ }\textbf
  {\bibinfo {volume} {72}},\ \bibinfo {pages} {2531} (\bibinfo {year}
  {1994})},\ \Eprint {http://arxiv.org/abs/hep-th/9401059}
  {arXiv:hep-th/9401059 [hep-th]} \BibitemShut {NoStop}%
%%CITATION = HEP-TH/9401059;%%
\bibitem [{\citenamefont {Verbaarschot}(1997)}]{Verbaarschot:1997bf}%
  \BibitemOpen
  \bibfield  {author} {\bibinfo {author} {\bibfnamefont {J.~J.~M.}\
  \bibnamefont {Verbaarschot}},\ }in\ \href@noop {} {\emph {\bibinfo
  {booktitle} {{Confinement, duality, and nonperturbative aspects of QCD.
  Proceedings, NATO Advanced Study Institute, Newton Institute Workshop,
  Cambridge, UK, June 23-July 4, 1997}}}}\ (\bibinfo {year} {1997})\ pp.\
  \bibinfo {pages} {343--378},\ \Eprint {http://arxiv.org/abs/hep-th/9710114}
  {arXiv:hep-th/9710114 [hep-th]} \BibitemShut {NoStop}%
%%CITATION = HEP-TH/9710114;%%
\bibitem [{\citenamefont {Verbaarschot}\ and\ \citenamefont
  {Wettig}(2000)}]{Verbaarschot:2000dy}%
  \BibitemOpen
  \bibfield  {author} {\bibinfo {author} {\bibfnamefont {J.~J.~M.}\
  \bibnamefont {Verbaarschot}}\ and\ \bibinfo {author} {\bibfnamefont
  {T.}~\bibnamefont {Wettig}},\ }\href {\doibase 10.1146/annurev.nucl.50.1.343}
  {\bibfield  {journal} {\bibinfo  {journal} {Ann. Rev. Nucl. Part. Sci.}\
  }\textbf {\bibinfo {volume} {50}},\ \bibinfo {pages} {343} (\bibinfo {year}
  {2000})},\ \Eprint {http://arxiv.org/abs/hep-ph/0003017}
  {arXiv:hep-ph/0003017 [hep-ph]} \BibitemShut {NoStop}%
%%CITATION = HEP-PH/0003017;%%
\bibitem [{\citenamefont {Verbaarschot}(2005)}]{Verbaarschot:2005rj}%
  \BibitemOpen
  \bibfield  {author} {\bibinfo {author} {\bibfnamefont {J.~J.~M.}\
  \bibnamefont {Verbaarschot}},\ }in\ \href@noop {} {\emph {\bibinfo
  {booktitle} {{Application of random matrices in physics. Proceedings, NATO
  Advanced Study Institute, Les Houches, France, June 6-25, 2004}}}}\ (\bibinfo
  {year} {2005})\ pp.\ \bibinfo {pages} {163--217},\ \Eprint
  {http://arxiv.org/abs/hep-th/0502029} {arXiv:hep-th/0502029 [hep-th]}
  \BibitemShut {NoStop}%
%%CITATION = HEP-TH/0502029;%%
\bibitem [{\citenamefont {Kanazawa}(2013)}]{Kanazawa:2012zzr}%
  \BibitemOpen
  \bibfield  {author} {\bibinfo {author} {\bibfnamefont {T.}~\bibnamefont
  {Kanazawa}},\ }\href {\doibase 10.1007/978-4-431-54165-3_3} {\emph {\bibinfo
  {title} {{Dirac Spectra in Dense QCD}}}},\ Springer theses Vol. 124\
  (\bibinfo  {publisher} {Springer Japan},\ \bibinfo {year} {2013})\BibitemShut
  {NoStop}%
\bibitem [{\citenamefont {Akemann}(2016)}]{Akemann:2016keq}%
  \BibitemOpen
  \bibfield  {author} {\bibinfo {author} {\bibfnamefont {G.}~\bibnamefont
  {Akemann}}\ }(\bibinfo {year} {2016})\ \bibinfo {note} {{Les Houches lecture
  notes}},\ \Eprint {http://arxiv.org/abs/1603.06011} {arXiv:1603.06011
  [math-ph]} \BibitemShut {NoStop}%
%%CITATION = ARXIV:1603.06011;%%
\bibitem [{\citenamefont {Pisarski}(1984)}]{Pisarski:1984dj}%
  \BibitemOpen
  \bibfield  {author} {\bibinfo {author} {\bibfnamefont {R.~D.}\ \bibnamefont
  {Pisarski}},\ }\href {\doibase 10.1103/PhysRevD.29.2423} {\bibfield
  {journal} {\bibinfo  {journal} {Phys.Rev.}\ }\textbf {\bibinfo {volume}
  {D29}},\ \bibinfo {pages} {2423} (\bibinfo {year} {1984})}\BibitemShut
  {NoStop}%
%%CITATION = PHRVA,D29,2423;%%
\bibitem [{\citenamefont {Appelquist}\ \emph
  {et~al.}(1986{\natexlab{a}})\citenamefont {Appelquist}, \citenamefont
  {Bowick}, \citenamefont {Karabali},\ and\ \citenamefont
  {Wijewardhana}}]{Appelquist:1986fd}%
  \BibitemOpen
  \bibfield  {author} {\bibinfo {author} {\bibfnamefont {T.~W.}\ \bibnamefont
  {Appelquist}}, \bibinfo {author} {\bibfnamefont {M.~J.}\ \bibnamefont
  {Bowick}}, \bibinfo {author} {\bibfnamefont {D.}~\bibnamefont {Karabali}}, \
  and\ \bibinfo {author} {\bibfnamefont {L.}~\bibnamefont {Wijewardhana}},\
  }\href {\doibase 10.1103/PhysRevD.33.3704} {\bibfield  {journal} {\bibinfo
  {journal} {Phys.Rev.}\ }\textbf {\bibinfo {volume} {D33}},\ \bibinfo {pages}
  {3704} (\bibinfo {year} {1986}{\natexlab{a}})}\BibitemShut {NoStop}%
%%CITATION = PHRVA,D33,3704;%%
\bibitem [{\citenamefont {Appelquist}\ \emph
  {et~al.}(1986{\natexlab{b}})\citenamefont {Appelquist}, \citenamefont
  {Bowick}, \citenamefont {Karabali},\ and\ \citenamefont
  {Wijewardhana}}]{Appelquist:1986qw}%
  \BibitemOpen
  \bibfield  {author} {\bibinfo {author} {\bibfnamefont {T.}~\bibnamefont
  {Appelquist}}, \bibinfo {author} {\bibfnamefont {M.~J.}\ \bibnamefont
  {Bowick}}, \bibinfo {author} {\bibfnamefont {D.}~\bibnamefont {Karabali}}, \
  and\ \bibinfo {author} {\bibfnamefont {L.}~\bibnamefont {Wijewardhana}},\
  }\href {\doibase 10.1103/PhysRevD.33.3774} {\bibfield  {journal} {\bibinfo
  {journal} {Phys.Rev.}\ }\textbf {\bibinfo {volume} {D33}},\ \bibinfo {pages}
  {3774} (\bibinfo {year} {1986}{\natexlab{b}})}\BibitemShut {NoStop}%
%%CITATION = PHRVA,D33,3774;%%
\bibitem [{\citenamefont {Lee}\ \emph {et~al.}(2006)\citenamefont {Lee},
  \citenamefont {Nagaosa},\ and\ \citenamefont {Wen}}]{Lee2006doping}%
  \BibitemOpen
  \bibfield  {author} {\bibinfo {author} {\bibfnamefont {P.~A.}\ \bibnamefont
  {Lee}}, \bibinfo {author} {\bibfnamefont {N.}~\bibnamefont {Nagaosa}}, \ and\
  \bibinfo {author} {\bibfnamefont {X.-G.}\ \bibnamefont {Wen}},\ }\href
  {\doibase 10.1103/RevModPhys.78.17} {\bibfield  {journal} {\bibinfo
  {journal} {Rev. Mod. Phys.}\ }\textbf {\bibinfo {volume} {78}},\ \bibinfo
  {pages} {17} (\bibinfo {year} {2006})}\BibitemShut {NoStop}%
\bibitem [{\citenamefont {Nayak}\ \emph {et~al.}(2008)\citenamefont {Nayak},
  \citenamefont {Simon}, \citenamefont {Stern}, \citenamefont {Freedman},\ and\
  \citenamefont {Das~Sarma}}]{Nayak2008}%
  \BibitemOpen
  \bibfield  {author} {\bibinfo {author} {\bibfnamefont {C.}~\bibnamefont
  {Nayak}}, \bibinfo {author} {\bibfnamefont {S.~H.}\ \bibnamefont {Simon}},
  \bibinfo {author} {\bibfnamefont {A.}~\bibnamefont {Stern}}, \bibinfo
  {author} {\bibfnamefont {M.}~\bibnamefont {Freedman}}, \ and\ \bibinfo
  {author} {\bibfnamefont {S.}~\bibnamefont {Das~Sarma}},\ }\href {\doibase
  10.1103/RevModPhys.80.1083} {\bibfield  {journal} {\bibinfo  {journal} {Rev.
  Mod. Phys.}\ }\textbf {\bibinfo {volume} {80}},\ \bibinfo {pages} {1083}
  (\bibinfo {year} {2008})}\BibitemShut {NoStop}%
\bibitem [{\citenamefont {Balents}(2010)}]{Balents2010}%
  \BibitemOpen
  \bibfield  {author} {\bibinfo {author} {\bibfnamefont {L.}~\bibnamefont
  {Balents}},\ }\href {\doibase 10.1038/nature08917} {\bibfield  {journal}
  {\bibinfo  {journal} {Nature}\ }\textbf {\bibinfo {volume} {464}},\ \bibinfo
  {pages} {199} (\bibinfo {year} {2010})}\BibitemShut {NoStop}%
\bibitem [{\citenamefont {Qi}\ and\ \citenamefont {Zhang}(2011)}]{QiZhang2011}%
  \BibitemOpen
  \bibfield  {author} {\bibinfo {author} {\bibfnamefont {X.-L.}\ \bibnamefont
  {Qi}}\ and\ \bibinfo {author} {\bibfnamefont {S.-C.}\ \bibnamefont {Zhang}},\
  }\href {\doibase 10.1103/RevModPhys.83.1057} {\bibfield  {journal} {\bibinfo
  {journal} {Rev. Mod. Phys.}\ }\textbf {\bibinfo {volume} {83}},\ \bibinfo
  {pages} {1057} (\bibinfo {year} {2011})}\BibitemShut {NoStop}%
\bibitem [{\citenamefont {Hansson}\ \emph {et~al.}(2017)\citenamefont
  {Hansson}, \citenamefont {Hermanns}, \citenamefont {Simon},\ and\
  \citenamefont {Viefers}}]{Hansson2017}%
  \BibitemOpen
  \bibfield  {author} {\bibinfo {author} {\bibfnamefont {T.~H.}\ \bibnamefont
  {Hansson}}, \bibinfo {author} {\bibfnamefont {M.}~\bibnamefont {Hermanns}},
  \bibinfo {author} {\bibfnamefont {S.~H.}\ \bibnamefont {Simon}}, \ and\
  \bibinfo {author} {\bibfnamefont {S.~F.}\ \bibnamefont {Viefers}},\ }\href
  {\doibase 10.1103/RevModPhys.89.025005} {\bibfield  {journal} {\bibinfo
  {journal} {Rev. Mod. Phys.}\ }\textbf {\bibinfo {volume} {89}},\ \bibinfo
  {pages} {025005} (\bibinfo {year} {2017})}\BibitemShut {NoStop}%
\bibitem [{\citenamefont {Hong}\ and\ \citenamefont
  {Park}(1993)}]{Hong:1992ww}%
  \BibitemOpen
  \bibfield  {author} {\bibinfo {author} {\bibfnamefont {D.~K.}\ \bibnamefont
  {Hong}}\ and\ \bibinfo {author} {\bibfnamefont {S.~H.}\ \bibnamefont
  {Park}},\ }\href {\doibase 10.1103/PhysRevD.47.3651} {\bibfield  {journal}
  {\bibinfo  {journal} {Phys. Rev.}\ }\textbf {\bibinfo {volume} {D47}},\
  \bibinfo {pages} {3651} (\bibinfo {year} {1993})}\BibitemShut {NoStop}%
%%CITATION = PHRVA,D47,3651;%%
\bibitem [{\citenamefont {Kondo}\ and\ \citenamefont
  {Maris}(1995)}]{Kondo:1994bt}%
  \BibitemOpen
  \bibfield  {author} {\bibinfo {author} {\bibfnamefont {K.~I.}\ \bibnamefont
  {Kondo}}\ and\ \bibinfo {author} {\bibfnamefont {P.}~\bibnamefont {Maris}},\
  }\href {\doibase 10.1103/PhysRevLett.74.18} {\bibfield  {journal} {\bibinfo
  {journal} {Phys. Rev. Lett.}\ }\textbf {\bibinfo {volume} {74}},\ \bibinfo
  {pages} {18} (\bibinfo {year} {1995})},\ \Eprint
  {http://arxiv.org/abs/hep-ph/9408210} {arXiv:hep-ph/9408210 [hep-ph]}
  \BibitemShut {NoStop}%
%%CITATION = HEP-PH/9408210;%%
\bibitem [{\citenamefont {Hong}(1998)}]{Hong:1997ws}%
  \BibitemOpen
  \bibfield  {author} {\bibinfo {author} {\bibfnamefont {D.~K.}\ \bibnamefont
  {Hong}},\ }\href {\doibase 10.1103/PhysRevD.57.1313} {\bibfield  {journal}
  {\bibinfo  {journal} {Phys. Rev.}\ }\textbf {\bibinfo {volume} {D57}},\
  \bibinfo {pages} {1313} (\bibinfo {year} {1998})},\ \Eprint
  {http://arxiv.org/abs/hep-th/9708027} {arXiv:hep-th/9708027 [hep-th]}
  \BibitemShut {NoStop}%
%%CITATION = HEP-TH/9708027;%%
\bibitem [{\citenamefont {Itoh}\ and\ \citenamefont
  {Kato}(1998)}]{Itoh:1998xk}%
  \BibitemOpen
  \bibfield  {author} {\bibinfo {author} {\bibfnamefont {T.}~\bibnamefont
  {Itoh}}\ and\ \bibinfo {author} {\bibfnamefont {H.}~\bibnamefont {Kato}},\
  }\href {\doibase 10.1103/PhysRevLett.81.30} {\bibfield  {journal} {\bibinfo
  {journal} {Phys. Rev. Lett.}\ }\textbf {\bibinfo {volume} {81}},\ \bibinfo
  {pages} {30} (\bibinfo {year} {1998})},\ \Eprint
  {http://arxiv.org/abs/hep-th/9802101} {arXiv:hep-th/9802101 [hep-th]}
  \BibitemShut {NoStop}%
%%CITATION = HEP-TH/9802101;%%
\bibitem [{\citenamefont {Matsuyama}\ \emph {et~al.}(1999)\citenamefont
  {Matsuyama}, \citenamefont {Nagahiro},\ and\ \citenamefont
  {Uchida}}]{Matsuyama:1999br}%
  \BibitemOpen
  \bibfield  {author} {\bibinfo {author} {\bibfnamefont {T.}~\bibnamefont
  {Matsuyama}}, \bibinfo {author} {\bibfnamefont {H.}~\bibnamefont {Nagahiro}},
  \ and\ \bibinfo {author} {\bibfnamefont {S.}~\bibnamefont {Uchida}},\ }\href
  {\doibase 10.1103/PhysRevD.60.105020} {\bibfield  {journal} {\bibinfo
  {journal} {Phys. Rev.}\ }\textbf {\bibinfo {volume} {D60}},\ \bibinfo {pages}
  {105020} (\bibinfo {year} {1999})},\ \Eprint
  {http://arxiv.org/abs/hep-th/9901049} {arXiv:hep-th/9901049 [hep-th]}
  \BibitemShut {NoStop}%
%%CITATION = HEP-TH/9901049;%%
\bibitem [{\citenamefont {Liu}\ and\ \citenamefont {Cheng}(2003)}]{Liu:2002yc}%
  \BibitemOpen
  \bibfield  {author} {\bibinfo {author} {\bibfnamefont {G.-Z.}\ \bibnamefont
  {Liu}}\ and\ \bibinfo {author} {\bibfnamefont {G.}~\bibnamefont {Cheng}},\
  }\href {\doibase 10.1103/PhysRevD.67.065010} {\bibfield  {journal} {\bibinfo
  {journal} {Phys. Rev.}\ }\textbf {\bibinfo {volume} {D67}},\ \bibinfo {pages}
  {065010} (\bibinfo {year} {2003})},\ \Eprint
  {http://arxiv.org/abs/hep-th/0211231} {arXiv:hep-th/0211231 [hep-th]}
  \BibitemShut {NoStop}%
%%CITATION = HEP-TH/0211231;%%
\bibitem [{\citenamefont {Hofmann}\ \emph {et~al.}(2010)\citenamefont
  {Hofmann}, \citenamefont {Raya},\ and\ \citenamefont
  {Madrigal}}]{Hofmann:2010zy}%
  \BibitemOpen
  \bibfield  {author} {\bibinfo {author} {\bibfnamefont {C.~P.}\ \bibnamefont
  {Hofmann}}, \bibinfo {author} {\bibfnamefont {A.}~\bibnamefont {Raya}}, \
  and\ \bibinfo {author} {\bibfnamefont {S.~S.}\ \bibnamefont {Madrigal}},\
  }\href {\doibase 10.1103/PhysRevD.82.096011} {\bibfield  {journal} {\bibinfo
  {journal} {Phys. Rev.}\ }\textbf {\bibinfo {volume} {D82}},\ \bibinfo {pages}
  {096011} (\bibinfo {year} {2010})},\ \Eprint {http://arxiv.org/abs/1010.3466}
  {arXiv:1010.3466 [hep-ph]} \BibitemShut {NoStop}%
%%CITATION = ARXIV:1010.3466;%%
\bibitem [{\citenamefont {Karthik}\ and\ \citenamefont
  {Narayanan}(2016{\natexlab{a}})}]{Karthik:2015sgq}%
  \BibitemOpen
  \bibfield  {author} {\bibinfo {author} {\bibfnamefont {N.}~\bibnamefont
  {Karthik}}\ and\ \bibinfo {author} {\bibfnamefont {R.}~\bibnamefont
  {Narayanan}},\ }\href {\doibase 10.1103/PhysRevD.93.045020} {\bibfield
  {journal} {\bibinfo  {journal} {Phys. Rev.}\ }\textbf {\bibinfo {volume}
  {D93}},\ \bibinfo {pages} {045020} (\bibinfo {year} {2016}{\natexlab{a}})},\
  \Eprint {http://arxiv.org/abs/1512.02993} {arXiv:1512.02993 [hep-lat]}
  \BibitemShut {NoStop}%
%%CITATION = ARXIV:1512.02993;%%
\bibitem [{\citenamefont {Karthik}\ and\ \citenamefont
  {Narayanan}(2016{\natexlab{b}})}]{Karthik:2016ppr}%
  \BibitemOpen
  \bibfield  {author} {\bibinfo {author} {\bibfnamefont {N.}~\bibnamefont
  {Karthik}}\ and\ \bibinfo {author} {\bibfnamefont {R.}~\bibnamefont
  {Narayanan}},\ }\href {\doibase 10.1103/PhysRevD.94.065026} {\bibfield
  {journal} {\bibinfo  {journal} {Phys. Rev.}\ }\textbf {\bibinfo {volume}
  {D94}},\ \bibinfo {pages} {065026} (\bibinfo {year} {2016}{\natexlab{b}})},\
  \Eprint {http://arxiv.org/abs/1606.04109} {arXiv:1606.04109 [hep-th]}
  \BibitemShut {NoStop}%
%%CITATION = ARXIV:1606.04109;%%
\bibitem [{\citenamefont {Karthik}\ and\ \citenamefont
  {Narayanan}(2017)}]{Karthik:2017hol}%
  \BibitemOpen
  \bibfield  {author} {\bibinfo {author} {\bibfnamefont {N.}~\bibnamefont
  {Karthik}}\ and\ \bibinfo {author} {\bibfnamefont {R.}~\bibnamefont
  {Narayanan}},\ }\href {\doibase 10.1103/PhysRevD.96.054509} {\bibfield
  {journal} {\bibinfo  {journal} {Phys. Rev.}\ }\textbf {\bibinfo {volume}
  {D96}},\ \bibinfo {pages} {054509} (\bibinfo {year} {2017})},\ \Eprint
  {http://arxiv.org/abs/1705.11143} {arXiv:1705.11143 [hep-th]} \BibitemShut
  {NoStop}%
%%CITATION = ARXIV:1705.11143;%%
\bibitem [{\citenamefont {Roscher}\ \emph {et~al.}(2016)\citenamefont
  {Roscher}, \citenamefont {Torres},\ and\ \citenamefont
  {Strack}}]{Roscher:2016wox}%
  \BibitemOpen
  \bibfield  {author} {\bibinfo {author} {\bibfnamefont {D.}~\bibnamefont
  {Roscher}}, \bibinfo {author} {\bibfnamefont {E.}~\bibnamefont {Torres}}, \
  and\ \bibinfo {author} {\bibfnamefont {P.}~\bibnamefont {Strack}},\ }\href
  {\doibase 10.1007/JHEP11(2016)017} {\bibfield  {journal} {\bibinfo  {journal}
  {JHEP}\ }\textbf {\bibinfo {volume} {11}},\ \bibinfo {pages} {017} (\bibinfo
  {year} {2016})},\ \Eprint {http://arxiv.org/abs/1605.05347} {arXiv:1605.05347
  [cond-mat.str-el]} \BibitemShut {NoStop}%
%%CITATION = ARXIV:1605.05347;%%
\bibitem [{\citenamefont {Vafa}\ and\ \citenamefont
  {Witten}(1984)}]{Vafa:1983tf}%
  \BibitemOpen
  \bibfield  {author} {\bibinfo {author} {\bibfnamefont {C.}~\bibnamefont
  {Vafa}}\ and\ \bibinfo {author} {\bibfnamefont {E.}~\bibnamefont {Witten}},\
  }\href {\doibase 10.1016/0550-3213(84)90230-X} {\bibfield  {journal}
  {\bibinfo  {journal} {Nucl. Phys.}\ }\textbf {\bibinfo {volume} {B234}},\
  \bibinfo {pages} {173} (\bibinfo {year} {1984})}\BibitemShut {NoStop}%
%%CITATION = NUPHA,B234,173;%%
\bibitem [{\citenamefont {Appelquist}\ and\ \citenamefont
  {Nash}(1990)}]{Appelquist:1989tc}%
  \BibitemOpen
  \bibfield  {author} {\bibinfo {author} {\bibfnamefont {T.}~\bibnamefont
  {Appelquist}}\ and\ \bibinfo {author} {\bibfnamefont {D.}~\bibnamefont
  {Nash}},\ }\href {\doibase 10.1103/PhysRevLett.64.721} {\bibfield  {journal}
  {\bibinfo  {journal} {Phys.Rev.Lett.}\ }\textbf {\bibinfo {volume} {64}},\
  \bibinfo {pages} {721} (\bibinfo {year} {1990})}\BibitemShut {NoStop}%
%%CITATION = PRLTA,64,721;%%
\bibitem [{\citenamefont {Ferretti}\ \emph {et~al.}(1992)\citenamefont
  {Ferretti}, \citenamefont {Rajeev},\ and\ \citenamefont
  {Yang}}]{Ferretti:1992fga}%
  \BibitemOpen
  \bibfield  {author} {\bibinfo {author} {\bibfnamefont {G.}~\bibnamefont
  {Ferretti}}, \bibinfo {author} {\bibfnamefont {S.~G.}\ \bibnamefont
  {Rajeev}}, \ and\ \bibinfo {author} {\bibfnamefont {Z.}~\bibnamefont
  {Yang}},\ }\href {\doibase 10.1142/S0217751X92003616} {\bibfield  {journal}
  {\bibinfo  {journal} {Int. J. Mod. Phys.}\ }\textbf {\bibinfo {volume}
  {A7}},\ \bibinfo {pages} {7989} (\bibinfo {year} {1992})},\ \Eprint
  {http://arxiv.org/abs/hep-th/9204075} {arXiv:hep-th/9204075 [hep-th]}
  \BibitemShut {NoStop}%
%%CITATION = HEP-TH/9204075;%%
\bibitem [{\citenamefont {Diamantini}\ \emph {et~al.}(1993)\citenamefont
  {Diamantini}, \citenamefont {Sodano},\ and\ \citenamefont
  {Semenoff}}]{Diamantini:1993vn}%
  \BibitemOpen
  \bibfield  {author} {\bibinfo {author} {\bibfnamefont {M.}~\bibnamefont
  {Diamantini}}, \bibinfo {author} {\bibfnamefont {P.}~\bibnamefont {Sodano}},
  \ and\ \bibinfo {author} {\bibfnamefont {G.}~\bibnamefont {Semenoff}},\
  }\href {\doibase 10.1103/PhysRevLett.70.3848} {\bibfield  {journal} {\bibinfo
   {journal} {Phys.Rev.Lett.}\ }\textbf {\bibinfo {volume} {70}},\ \bibinfo
  {pages} {3848} (\bibinfo {year} {1993})},\ \Eprint
  {http://arxiv.org/abs/hep-ph/9301256} {arXiv:hep-ph/9301256 [hep-ph]}
  \BibitemShut {NoStop}%
%%CITATION = HEP-PH/9301256;%%
\bibitem [{\citenamefont {Damgaard}\ \emph {et~al.}(1998)\citenamefont
  {Damgaard}, \citenamefont {Heller}, \citenamefont {Krasnitz},\ and\
  \citenamefont {Madsen}}]{Damgaard:1998yv}%
  \BibitemOpen
  \bibfield  {author} {\bibinfo {author} {\bibfnamefont {P.~H.}\ \bibnamefont
  {Damgaard}}, \bibinfo {author} {\bibfnamefont {U.~M.}\ \bibnamefont
  {Heller}}, \bibinfo {author} {\bibfnamefont {A.}~\bibnamefont {Krasnitz}}, \
  and\ \bibinfo {author} {\bibfnamefont {T.}~\bibnamefont {Madsen}},\ }\href
  {\doibase 10.1016/S0370-2693(98)01073-9} {\bibfield  {journal} {\bibinfo
  {journal} {Phys. Lett.}\ }\textbf {\bibinfo {volume} {B440}},\ \bibinfo
  {pages} {129} (\bibinfo {year} {1998})},\ \Eprint
  {http://arxiv.org/abs/hep-lat/9803012} {arXiv:hep-lat/9803012 [hep-lat]}
  \BibitemShut {NoStop}%
%%CITATION = HEP-LAT/9803012;%%
\bibitem [{\citenamefont {Karthik}\ and\ \citenamefont
  {Narayanan}(2016{\natexlab{c}})}]{Karthik:2016bmf}%
  \BibitemOpen
  \bibfield  {author} {\bibinfo {author} {\bibfnamefont {N.}~\bibnamefont
  {Karthik}}\ and\ \bibinfo {author} {\bibfnamefont {R.}~\bibnamefont
  {Narayanan}},\ }\href {\doibase 10.1103/PhysRevD.94.045020} {\bibfield
  {journal} {\bibinfo  {journal} {Phys. Rev.}\ }\textbf {\bibinfo {volume}
  {D94}},\ \bibinfo {pages} {045020} (\bibinfo {year} {2016}{\natexlab{c}})},\
  \Eprint {http://arxiv.org/abs/1607.03905} {arXiv:1607.03905 [hep-lat]}
  \BibitemShut {NoStop}%
%%CITATION = ARXIV:1607.03905;%%
\bibitem [{\citenamefont {Verbaarschot}\ and\ \citenamefont
  {Zahed}(1994)}]{Verbaarschot:1994ip}%
  \BibitemOpen
  \bibfield  {author} {\bibinfo {author} {\bibfnamefont {J.~J.~M.}\
  \bibnamefont {Verbaarschot}}\ and\ \bibinfo {author} {\bibfnamefont
  {I.}~\bibnamefont {Zahed}},\ }\href {\doibase 10.1103/PhysRevLett.73.2288}
  {\bibfield  {journal} {\bibinfo  {journal} {Phys. Rev. Lett.}\ }\textbf
  {\bibinfo {volume} {73}},\ \bibinfo {pages} {2288} (\bibinfo {year}
  {1994})},\ \Eprint {http://arxiv.org/abs/hep-th/9405005}
  {arXiv:hep-th/9405005 [hep-th]} \BibitemShut {NoStop}%
%%CITATION = HEP-TH/9405005;%%
\bibitem [{\citenamefont {Nagao}\ and\ \citenamefont
  {Slevin}(1993)}]{NagaoSlevin93}%
  \BibitemOpen
  \bibfield  {author} {\bibinfo {author} {\bibfnamefont {T.}~\bibnamefont
  {Nagao}}\ and\ \bibinfo {author} {\bibfnamefont {K.}~\bibnamefont {Slevin}},\
  }\href {\doibase 10.1063/1.530157} {\bibfield  {journal} {\bibinfo  {journal}
  {J. Math. Phys.}\ }\textbf {\bibinfo {volume} {34}},\ \bibinfo {pages} {2075}
  (\bibinfo {year} {1993})}\BibitemShut {NoStop}%
\bibitem [{\citenamefont {Damgaard}\ and\ \citenamefont
  {Nishigaki}(1998)}]{Damgaard:1997pw}%
  \BibitemOpen
  \bibfield  {author} {\bibinfo {author} {\bibfnamefont {P.~H.}\ \bibnamefont
  {Damgaard}}\ and\ \bibinfo {author} {\bibfnamefont {S.~M.}\ \bibnamefont
  {Nishigaki}},\ }\href {\doibase 10.1103/PhysRevD.57.5299} {\bibfield
  {journal} {\bibinfo  {journal} {Phys. Rev.}\ }\textbf {\bibinfo {volume}
  {D57}},\ \bibinfo {pages} {5299} (\bibinfo {year} {1998})},\ \Eprint
  {http://arxiv.org/abs/hep-th/9711096} {arXiv:hep-th/9711096 [hep-th]}
  \BibitemShut {NoStop}%
%%CITATION = HEP-TH/9711096;%%
\bibitem [{\citenamefont {Akemann}\ and\ \citenamefont
  {Damgaard}(1998)}]{Akemann:1998ta}%
  \BibitemOpen
  \bibfield  {author} {\bibinfo {author} {\bibfnamefont {G.}~\bibnamefont
  {Akemann}}\ and\ \bibinfo {author} {\bibfnamefont {P.~H.}\ \bibnamefont
  {Damgaard}},\ }\href {\doibase 10.1016/S0550-3213(98)00338-1} {\bibfield
  {journal} {\bibinfo  {journal} {Nucl. Phys.}\ }\textbf {\bibinfo {volume}
  {B528}},\ \bibinfo {pages} {411} (\bibinfo {year} {1998})},\ \Eprint
  {http://arxiv.org/abs/hep-th/9801133} {arXiv:hep-th/9801133 [hep-th]}
  \BibitemShut {NoStop}%
%%CITATION = HEP-TH/9801133;%%
\bibitem [{\citenamefont {Christiansen}(1999)}]{Christiansen:1998mu}%
  \BibitemOpen
  \bibfield  {author} {\bibinfo {author} {\bibfnamefont {J.}~\bibnamefont
  {Christiansen}},\ }\href {\doibase 10.1016/S0550-3213(99)00091-7} {\bibfield
  {journal} {\bibinfo  {journal} {Nucl.Phys.}\ }\textbf {\bibinfo {volume}
  {B547}},\ \bibinfo {pages} {329} (\bibinfo {year} {1999})},\ \Eprint
  {http://arxiv.org/abs/hep-th/9809194} {arXiv:hep-th/9809194 [hep-th]}
  \BibitemShut {NoStop}%
%%CITATION = HEP-TH/9809194;%%
\bibitem [{\citenamefont {Magnea}(2000{\natexlab{a}})}]{Magnea:1999iv}%
  \BibitemOpen
  \bibfield  {author} {\bibinfo {author} {\bibfnamefont {U.}~\bibnamefont
  {Magnea}},\ }\href {\doibase 10.1103/PhysRevD.61.056005} {\bibfield
  {journal} {\bibinfo  {journal} {Phys.Rev.}\ }\textbf {\bibinfo {volume}
  {D61}},\ \bibinfo {pages} {056005} (\bibinfo {year} {2000}{\natexlab{a}})},\
  \Eprint {http://arxiv.org/abs/hep-th/9907096} {arXiv:hep-th/9907096 [hep-th]}
  \BibitemShut {NoStop}%
%%CITATION = HEP-TH/9907096;%%
\bibitem [{\citenamefont {Magnea}(2000{\natexlab{b}})}]{Magnea:1999ey}%
  \BibitemOpen
  \bibfield  {author} {\bibinfo {author} {\bibfnamefont {U.}~\bibnamefont
  {Magnea}},\ }\href {\doibase 10.1103/PhysRevD.62.016005} {\bibfield
  {journal} {\bibinfo  {journal} {Phys.Rev.}\ }\textbf {\bibinfo {volume}
  {D62}},\ \bibinfo {pages} {016005} (\bibinfo {year} {2000}{\natexlab{b}})},\
  \Eprint {http://arxiv.org/abs/hep-th/9912207} {arXiv:hep-th/9912207 [hep-th]}
  \BibitemShut {NoStop}%
%%CITATION = HEP-TH/9912207;%%
\bibitem [{\citenamefont {Hilmoine}\ and\ \citenamefont
  {Niclasen}(2000)}]{Hilmoine:2000ca}%
  \BibitemOpen
  \bibfield  {author} {\bibinfo {author} {\bibfnamefont {C.}~\bibnamefont
  {Hilmoine}}\ and\ \bibinfo {author} {\bibfnamefont {R.}~\bibnamefont
  {Niclasen}},\ }\href {\doibase 10.1103/PhysRevD.62.096013} {\bibfield
  {journal} {\bibinfo  {journal} {Phys.Rev.}\ }\textbf {\bibinfo {volume}
  {D62}},\ \bibinfo {pages} {096013} (\bibinfo {year} {2000})},\ \Eprint
  {http://arxiv.org/abs/hep-th/0004081} {arXiv:hep-th/0004081 [hep-th]}
  \BibitemShut {NoStop}%
%%CITATION = HEP-TH/0004081;%%
\bibitem [{\citenamefont {Nagao}\ and\ \citenamefont
  {Nishigaki}(2001)}]{Nagao:2000um}%
  \BibitemOpen
  \bibfield  {author} {\bibinfo {author} {\bibfnamefont {T.}~\bibnamefont
  {Nagao}}\ and\ \bibinfo {author} {\bibfnamefont {S.~M.}\ \bibnamefont
  {Nishigaki}},\ }\href {\doibase 10.1103/PhysRevD.63.045011} {\bibfield
  {journal} {\bibinfo  {journal} {Phys.Rev.}\ }\textbf {\bibinfo {volume}
  {D63}},\ \bibinfo {pages} {045011} (\bibinfo {year} {2001})},\ \Eprint
  {http://arxiv.org/abs/hep-th/0005077} {arXiv:hep-th/0005077 [hep-th]}
  \BibitemShut {NoStop}%
%%CITATION = HEP-TH/0005077;%%
\bibitem [{\citenamefont {Szabo}(2001)}]{Szabo:2000qq}%
  \BibitemOpen
  \bibfield  {author} {\bibinfo {author} {\bibfnamefont {R.~J.}\ \bibnamefont
  {Szabo}},\ }\href {\doibase 10.1016/S0550-3213(00)00775-6} {\bibfield
  {journal} {\bibinfo  {journal} {Nucl. Phys.}\ }\textbf {\bibinfo {volume}
  {B598}},\ \bibinfo {pages} {309} (\bibinfo {year} {2001})},\ \Eprint
  {http://arxiv.org/abs/hep-th/0009237} {arXiv:hep-th/0009237} \BibitemShut
  {NoStop}%
%%CITATION = HEP-TH/0009237;%%
\bibitem [{\citenamefont {Szabo}(2005)}]{Szabo:2005gi}%
  \BibitemOpen
  \bibfield  {author} {\bibinfo {author} {\bibfnamefont {R.~J.}\ \bibnamefont
  {Szabo}},\ }\href {\doibase 10.1016/j.nuclphysb.2005.06.028} {\bibfield
  {journal} {\bibinfo  {journal} {Nucl. Phys.}\ }\textbf {\bibinfo {volume}
  {B723}},\ \bibinfo {pages} {163} (\bibinfo {year} {2005})},\ \Eprint
  {http://arxiv.org/abs/hep-th/0504202} {arXiv:hep-th/0504202 [hep-th]}
  \BibitemShut {NoStop}%
%%CITATION = HEP-TH/0504202;%%
\bibitem [{\citenamefont {Deser}\ \emph {et~al.}(1982)\citenamefont {Deser},
  \citenamefont {Jackiw},\ and\ \citenamefont {Templeton}}]{Deser:1981wh}%
  \BibitemOpen
  \bibfield  {author} {\bibinfo {author} {\bibfnamefont {S.}~\bibnamefont
  {Deser}}, \bibinfo {author} {\bibfnamefont {R.}~\bibnamefont {Jackiw}}, \
  and\ \bibinfo {author} {\bibfnamefont {S.}~\bibnamefont {Templeton}},\ }\href
  {\doibase 10.1016/0003-4916(82)90164-6} {\bibfield  {journal} {\bibinfo
  {journal} {Annals Phys.}\ }\textbf {\bibinfo {volume} {140}},\ \bibinfo
  {pages} {372} (\bibinfo {year} {1982})},\ \bibinfo {note} {[Annals
  Phys.281,409(2000)]}\BibitemShut {NoStop}%
%%CITATION = APNYA,140,372;%%
\bibitem [{\citenamefont {Komargodski}\ and\ \citenamefont
  {Seiberg}(2018)}]{Komargodski:2017keh}%
  \BibitemOpen
  \bibfield  {author} {\bibinfo {author} {\bibfnamefont {Z.}~\bibnamefont
  {Komargodski}}\ and\ \bibinfo {author} {\bibfnamefont {N.}~\bibnamefont
  {Seiberg}},\ }\href {\doibase 10.1007/JHEP01(2018)109} {\bibfield  {journal}
  {\bibinfo  {journal} {JHEP}\ }\textbf {\bibinfo {volume} {01}},\ \bibinfo
  {pages} {109} (\bibinfo {year} {2018})},\ \Eprint
  {http://arxiv.org/abs/1706.08755} {arXiv:1706.08755 [hep-th]} \BibitemShut
  {NoStop}%
%%CITATION = ARXIV:1706.08755;%%
\bibitem [{\citenamefont {Gaiotto}\ \emph {et~al.}(2018)\citenamefont
  {Gaiotto}, \citenamefont {Komargodski},\ and\ \citenamefont
  {Seiberg}}]{Gaiotto:2017tne}%
  \BibitemOpen
  \bibfield  {author} {\bibinfo {author} {\bibfnamefont {D.}~\bibnamefont
  {Gaiotto}}, \bibinfo {author} {\bibfnamefont {Z.}~\bibnamefont
  {Komargodski}}, \ and\ \bibinfo {author} {\bibfnamefont {N.}~\bibnamefont
  {Seiberg}},\ }\href {\doibase 10.1007/JHEP01(2018)110} {\bibfield  {journal}
  {\bibinfo  {journal} {JHEP}\ }\textbf {\bibinfo {volume} {01}},\ \bibinfo
  {pages} {110} (\bibinfo {year} {2018})},\ \Eprint
  {http://arxiv.org/abs/1708.06806} {arXiv:1708.06806 [hep-th]} \BibitemShut
  {NoStop}%
%%CITATION = ARXIV:1708.06806;%%
\bibitem [{\citenamefont {Armoni}\ and\ \citenamefont
  {Niarchos}(2018)}]{Armoni:2017jkl}%
  \BibitemOpen
  \bibfield  {author} {\bibinfo {author} {\bibfnamefont {A.}~\bibnamefont
  {Armoni}}\ and\ \bibinfo {author} {\bibfnamefont {V.}~\bibnamefont
  {Niarchos}},\ }\href {\doibase 10.1103/PhysRevD.97.106001} {\bibfield
  {journal} {\bibinfo  {journal} {Phys. Rev.}\ }\textbf {\bibinfo {volume}
  {D97}},\ \bibinfo {pages} {106001} (\bibinfo {year} {2018})},\ \Eprint
  {http://arxiv.org/abs/1711.04832} {arXiv:1711.04832 [hep-th]} \BibitemShut
  {NoStop}%
%%CITATION = ARXIV:1711.04832;%%
\bibitem [{\citenamefont {Marino}(2004)}]{Marino:2002fk}%
  \BibitemOpen
  \bibfield  {author} {\bibinfo {author} {\bibfnamefont {M.}~\bibnamefont
  {Marino}},\ }\href {\doibase 10.1007/s00220-004-1194-4} {\bibfield  {journal}
  {\bibinfo  {journal} {Commun. Math. Phys.}\ }\textbf {\bibinfo {volume}
  {253}},\ \bibinfo {pages} {25} (\bibinfo {year} {2004})},\ \Eprint
  {http://arxiv.org/abs/hep-th/0207096} {arXiv:hep-th/0207096 [hep-th]}
  \BibitemShut {NoStop}%
%%CITATION = HEP-TH/0207096;%%
\bibitem [{\citenamefont {Aganagic}\ \emph {et~al.}(2004)\citenamefont
  {Aganagic}, \citenamefont {Klemm}, \citenamefont {Marino},\ and\
  \citenamefont {Vafa}}]{Aganagic:2002wv}%
  \BibitemOpen
  \bibfield  {author} {\bibinfo {author} {\bibfnamefont {M.}~\bibnamefont
  {Aganagic}}, \bibinfo {author} {\bibfnamefont {A.}~\bibnamefont {Klemm}},
  \bibinfo {author} {\bibfnamefont {M.}~\bibnamefont {Marino}}, \ and\ \bibinfo
  {author} {\bibfnamefont {C.}~\bibnamefont {Vafa}},\ }\href {\doibase
  10.1088/1126-6708/2004/02/010} {\bibfield  {journal} {\bibinfo  {journal}
  {JHEP}\ }\textbf {\bibinfo {volume} {02}},\ \bibinfo {pages} {010} (\bibinfo
  {year} {2004})},\ \Eprint {http://arxiv.org/abs/hep-th/0211098}
  {arXiv:hep-th/0211098 [hep-th]} \BibitemShut {NoStop}%
%%CITATION = HEP-TH/0211098;%%
\bibitem [{\citenamefont {Tierz}(2004)}]{Tierz:2002jj}%
  \BibitemOpen
  \bibfield  {author} {\bibinfo {author} {\bibfnamefont {M.}~\bibnamefont
  {Tierz}},\ }\href {\doibase 10.1142/S0217732304014100} {\bibfield  {journal}
  {\bibinfo  {journal} {Mod. Phys. Lett.}\ }\textbf {\bibinfo {volume} {A19}},\
  \bibinfo {pages} {1365} (\bibinfo {year} {2004})},\ \Eprint
  {http://arxiv.org/abs/hep-th/0212128} {arXiv:hep-th/0212128 [hep-th]}
  \BibitemShut {NoStop}%
%%CITATION = HEP-TH/0212128;%%
\bibitem [{\citenamefont {Kapustin}\ \emph {et~al.}(2010)\citenamefont
  {Kapustin}, \citenamefont {Willett},\ and\ \citenamefont
  {Yaakov}}]{Kapustin:2009kz}%
  \BibitemOpen
  \bibfield  {author} {\bibinfo {author} {\bibfnamefont {A.}~\bibnamefont
  {Kapustin}}, \bibinfo {author} {\bibfnamefont {B.}~\bibnamefont {Willett}}, \
  and\ \bibinfo {author} {\bibfnamefont {I.}~\bibnamefont {Yaakov}},\ }\href
  {\doibase 10.1007/JHEP03(2010)089} {\bibfield  {journal} {\bibinfo  {journal}
  {JHEP}\ }\textbf {\bibinfo {volume} {03}},\ \bibinfo {pages} {089} (\bibinfo
  {year} {2010})},\ \Eprint {http://arxiv.org/abs/0909.4559} {arXiv:0909.4559
  [hep-th]} \BibitemShut {NoStop}%
%%CITATION = ARXIV:0909.4559;%%
\bibitem [{\citenamefont {Drukker}\ \emph {et~al.}(2011)\citenamefont
  {Drukker}, \citenamefont {Marino},\ and\ \citenamefont
  {Putrov}}]{Drukker:2010nc}%
  \BibitemOpen
  \bibfield  {author} {\bibinfo {author} {\bibfnamefont {N.}~\bibnamefont
  {Drukker}}, \bibinfo {author} {\bibfnamefont {M.}~\bibnamefont {Marino}}, \
  and\ \bibinfo {author} {\bibfnamefont {P.}~\bibnamefont {Putrov}},\ }\href
  {\doibase 10.1007/s00220-011-1253-6} {\bibfield  {journal} {\bibinfo
  {journal} {Commun. Math. Phys.}\ }\textbf {\bibinfo {volume} {306}},\
  \bibinfo {pages} {511} (\bibinfo {year} {2011})},\ \Eprint
  {http://arxiv.org/abs/1007.3837} {arXiv:1007.3837 [hep-th]} \BibitemShut
  {NoStop}%
%%CITATION = ARXIV:1007.3837;%%
\bibitem [{\citenamefont {Das}\ \emph {et~al.}(1990)\citenamefont {Das},
  \citenamefont {Dhar}, \citenamefont {Sengupta},\ and\ \citenamefont
  {Wadia}}]{Das:1989fq}%
  \BibitemOpen
  \bibfield  {author} {\bibinfo {author} {\bibfnamefont {S.~R.}\ \bibnamefont
  {Das}}, \bibinfo {author} {\bibfnamefont {A.}~\bibnamefont {Dhar}}, \bibinfo
  {author} {\bibfnamefont {A.~M.}\ \bibnamefont {Sengupta}}, \ and\ \bibinfo
  {author} {\bibfnamefont {S.~R.}\ \bibnamefont {Wadia}},\ }\href {\doibase
  10.1142/S0217732390001165} {\bibfield  {journal} {\bibinfo  {journal} {Mod.
  Phys. Lett.}\ }\textbf {\bibinfo {volume} {A5}},\ \bibinfo {pages} {1041}
  (\bibinfo {year} {1990})}\BibitemShut {NoStop}%
%%CITATION = MPLAE,A5,1041;%%
\bibitem [{\citenamefont {Cicuta}\ and\ \citenamefont
  {Montaldi}(1990)}]{Cicuta:1990uc}%
  \BibitemOpen
  \bibfield  {author} {\bibinfo {author} {\bibfnamefont {G.~M.}\ \bibnamefont
  {Cicuta}}\ and\ \bibinfo {author} {\bibfnamefont {E.}~\bibnamefont
  {Montaldi}},\ }\href {\doibase 10.1142/S0217732390002183} {\bibfield
  {journal} {\bibinfo  {journal} {Mod. Phys. Lett.}\ }\textbf {\bibinfo
  {volume} {A5}},\ \bibinfo {pages} {1927} (\bibinfo {year}
  {1990})}\BibitemShut {NoStop}%
%%CITATION = MPLAE,A5,1927;%%
\bibitem [{\citenamefont {Ueda}(1991)}]{Ueda:1991xa}%
  \BibitemOpen
  \bibfield  {author} {\bibinfo {author} {\bibfnamefont {H.}~\bibnamefont
  {Ueda}},\ }\href {\doibase 10.1143/PTP.86.23} {\bibfield  {journal} {\bibinfo
   {journal} {Prog. Theor. Phys.}\ }\textbf {\bibinfo {volume} {86}},\ \bibinfo
  {pages} {23} (\bibinfo {year} {1991})}\BibitemShut {NoStop}%
%%CITATION = PTPKA,86,23;%%
\bibitem [{\citenamefont {Sawada}\ and\ \citenamefont
  {Ueda}(1991)}]{Sawada:1991uh}%
  \BibitemOpen
  \bibfield  {author} {\bibinfo {author} {\bibfnamefont {S.}~\bibnamefont
  {Sawada}}\ and\ \bibinfo {author} {\bibfnamefont {H.}~\bibnamefont {Ueda}},\
  }\href {\doibase 10.1142/S0217732391004310} {\bibfield  {journal} {\bibinfo
  {journal} {Mod. Phys. Lett.}\ }\textbf {\bibinfo {volume} {A6}},\ \bibinfo
  {pages} {3717} (\bibinfo {year} {1991})}\BibitemShut {NoStop}%
%%CITATION = MPLAE,A6,3717;%%
\bibitem [{\citenamefont {Korchemsky}(1992)}]{Korchemsky:1992tt}%
  \BibitemOpen
  \bibfield  {author} {\bibinfo {author} {\bibfnamefont {G.~P.}\ \bibnamefont
  {Korchemsky}},\ }\href {\doibase 10.1142/S0217732392002470} {\bibfield
  {journal} {\bibinfo  {journal} {Mod. Phys. Lett.}\ }\textbf {\bibinfo
  {volume} {A7}},\ \bibinfo {pages} {3081} (\bibinfo {year} {1992})},\ \Eprint
  {http://arxiv.org/abs/hep-th/9205014} {arXiv:hep-th/9205014 [hep-th]}
  \BibitemShut {NoStop}%
%%CITATION = HEP-TH/9205014;%%
\bibitem [{\citenamefont {David}(1997)}]{David:1996vp}%
  \BibitemOpen
  \bibfield  {author} {\bibinfo {author} {\bibfnamefont {F.}~\bibnamefont
  {David}},\ }\href {\doibase 10.1016/S0550-3213(96)00716-X} {\bibfield
  {journal} {\bibinfo  {journal} {Nucl. Phys.}\ }\textbf {\bibinfo {volume}
  {B487}},\ \bibinfo {pages} {633} (\bibinfo {year} {1997})},\ \Eprint
  {http://arxiv.org/abs/hep-th/9610037} {arXiv:hep-th/9610037 [hep-th]}
  \BibitemShut {NoStop}%
%%CITATION = HEP-TH/9610037;%%
\bibitem [{\citenamefont {Kieburg}\ \emph {et~al.}(2013)\citenamefont
  {Kieburg}, \citenamefont {Verbaarschot},\ and\ \citenamefont
  {Zafeiropoulos}}]{Kieburg:2013xta}%
  \BibitemOpen
  \bibfield  {author} {\bibinfo {author} {\bibfnamefont {M.}~\bibnamefont
  {Kieburg}}, \bibinfo {author} {\bibfnamefont {J.~J.~M.}\ \bibnamefont
  {Verbaarschot}}, \ and\ \bibinfo {author} {\bibfnamefont {S.}~\bibnamefont
  {Zafeiropoulos}},\ }\href {\doibase 10.1103/PhysRevD.88.094502} {\bibfield
  {journal} {\bibinfo  {journal} {Phys. Rev.}\ }\textbf {\bibinfo {volume}
  {D88}},\ \bibinfo {pages} {094502} (\bibinfo {year} {2013})},\ \Eprint
  {http://arxiv.org/abs/1307.7251} {arXiv:1307.7251 [hep-lat]} \BibitemShut
  {NoStop}%
%%CITATION = ARXIV:1307.7251;%%
\bibitem [{\citenamefont {Gasser}\ and\ \citenamefont
  {Leutwyler}(1987)}]{Gasser:1987ah}%
  \BibitemOpen
  \bibfield  {author} {\bibinfo {author} {\bibfnamefont {J.}~\bibnamefont
  {Gasser}}\ and\ \bibinfo {author} {\bibfnamefont {H.}~\bibnamefont
  {Leutwyler}},\ }\href {\doibase 10.1016/0370-2693(87)91652-2} {\bibfield
  {journal} {\bibinfo  {journal} {Phys. Lett.}\ }\textbf {\bibinfo {volume}
  {B188}},\ \bibinfo {pages} {477} (\bibinfo {year} {1987})}\BibitemShut
  {NoStop}%
%%CITATION = PHLTA,B188,477;%%
\bibitem [{\citenamefont {Alvarez-Gaume}\ \emph {et~al.}(1985)\citenamefont
  {Alvarez-Gaume}, \citenamefont {Della~Pietra},\ and\ \citenamefont
  {Moore}}]{AlvarezGaume:1984nf}%
  \BibitemOpen
  \bibfield  {author} {\bibinfo {author} {\bibfnamefont {L.}~\bibnamefont
  {Alvarez-Gaume}}, \bibinfo {author} {\bibfnamefont {S.}~\bibnamefont
  {Della~Pietra}}, \ and\ \bibinfo {author} {\bibfnamefont {G.~W.}\
  \bibnamefont {Moore}},\ }\href {\doibase 10.1016/0003-4916(85)90383-5}
  {\bibfield  {journal} {\bibinfo  {journal} {Annals Phys.}\ }\textbf {\bibinfo
  {volume} {163}},\ \bibinfo {pages} {288} (\bibinfo {year}
  {1985})}\BibitemShut {NoStop}%
%%CITATION = APNYA,163,288;%%
\bibitem [{\citenamefont {Leutwyler}(1990)}]{Leutwyler:1990an}%
  \BibitemOpen
  \bibfield  {author} {\bibinfo {author} {\bibfnamefont {H.}~\bibnamefont
  {Leutwyler}},\ }\href@noop {} {\bibfield  {journal} {\bibinfo  {journal}
  {Helv. Phys. Acta}\ }\textbf {\bibinfo {volume} {63}},\ \bibinfo {pages}
  {660} (\bibinfo {year} {1990})}\BibitemShut {NoStop}%
%%CITATION = HPACA,63,660;%%
\bibitem [{\citenamefont {Redlich}(1984)}]{Redlich:1983dv}%
  \BibitemOpen
  \bibfield  {author} {\bibinfo {author} {\bibfnamefont {A.~N.}\ \bibnamefont
  {Redlich}},\ }\href {\doibase 10.1103/PhysRevD.29.2366} {\bibfield  {journal}
  {\bibinfo  {journal} {Phys. Rev.}\ }\textbf {\bibinfo {volume} {D29}},\
  \bibinfo {pages} {2366} (\bibinfo {year} {1984})}\BibitemShut {NoStop}%
%%CITATION = PHRVA,D29,2366;%%
\bibitem [{\citenamefont {Deser}\ \emph {et~al.}(1998)\citenamefont {Deser},
  \citenamefont {Griguolo},\ and\ \citenamefont {Seminara}}]{Deser:1997gp}%
  \BibitemOpen
  \bibfield  {author} {\bibinfo {author} {\bibfnamefont {S.}~\bibnamefont
  {Deser}}, \bibinfo {author} {\bibfnamefont {L.}~\bibnamefont {Griguolo}}, \
  and\ \bibinfo {author} {\bibfnamefont {D.}~\bibnamefont {Seminara}},\ }\href
  {\doibase 10.1103/PhysRevD.57.7444} {\bibfield  {journal} {\bibinfo
  {journal} {Phys. Rev.}\ }\textbf {\bibinfo {volume} {D57}},\ \bibinfo {pages}
  {7444} (\bibinfo {year} {1998})},\ \Eprint
  {http://arxiv.org/abs/hep-th/9712066} {arXiv:hep-th/9712066 [hep-th]}
  \BibitemShut {NoStop}%
%%CITATION = HEP-TH/9712066;%%
\bibitem [{\citenamefont {Halasz}\ and\ \citenamefont
  {Verbaarschot}(1995)}]{Halasz:1995qb}%
  \BibitemOpen
  \bibfield  {author} {\bibinfo {author} {\bibfnamefont {A.~M.}\ \bibnamefont
  {Halasz}}\ and\ \bibinfo {author} {\bibfnamefont {J.~J.~M.}\ \bibnamefont
  {Verbaarschot}},\ }\href {\doibase 10.1103/PhysRevD.52.2563} {\bibfield
  {journal} {\bibinfo  {journal} {Phys. Rev.}\ }\textbf {\bibinfo {volume}
  {D52}},\ \bibinfo {pages} {2563} (\bibinfo {year} {1995})},\ \Eprint
  {http://arxiv.org/abs/hep-th/9502096} {arXiv:hep-th/9502096 [hep-th]}
  \BibitemShut {NoStop}%
%%CITATION = HEP-TH/9502096;%%
\bibitem [{\citenamefont {Bertuola}\ \emph {et~al.}(2004)\citenamefont
  {Bertuola}, \citenamefont {Bohigas},\ and\ \citenamefont
  {Pato}}]{Bertuola2004}%
  \BibitemOpen
  \bibfield  {author} {\bibinfo {author} {\bibfnamefont {A.~C.}\ \bibnamefont
  {Bertuola}}, \bibinfo {author} {\bibfnamefont {O.}~\bibnamefont {Bohigas}}, \
  and\ \bibinfo {author} {\bibfnamefont {M.~P.}\ \bibnamefont {Pato}},\ }\href
  {\doibase 10.1103/PhysRevE.70.065102} {\bibfield  {journal} {\bibinfo
  {journal} {Phys. Rev. E}\ }\textbf {\bibinfo {volume} {70}},\ \bibinfo
  {pages} {065102(R)} (\bibinfo {year} {2004})},\ \Eprint
  {http://arxiv.org/abs/math-ph/0411033} {math-ph/0411033} \BibitemShut
  {NoStop}%
\bibitem [{\citenamefont {Abul-Magd}(2005)}]{AbulMagd2005}%
  \BibitemOpen
  \bibfield  {author} {\bibinfo {author} {\bibfnamefont {A.~Y.}\ \bibnamefont
  {Abul-Magd}},\ }\href {\doibase 10.1103/PhysRevE.72.066114} {\bibfield
  {journal} {\bibinfo  {journal} {Phys. Rev.}\ }\textbf {\bibinfo {volume}
  {E72}},\ \bibinfo {pages} {066114} (\bibinfo {year} {2005})},\ \Eprint
  {http://arxiv.org/abs/cond-mat/0510494} {cond-mat/0510494} \BibitemShut
  {NoStop}%
\bibitem [{\citenamefont {Akemann}\ and\ \citenamefont
  {Vivo}(2008)}]{AkeVivo2008}%
  \BibitemOpen
  \bibfield  {author} {\bibinfo {author} {\bibfnamefont {G.}~\bibnamefont
  {Akemann}}\ and\ \bibinfo {author} {\bibfnamefont {P.}~\bibnamefont {Vivo}},\
  }\href {\doibase 10.1088/1742-5468/2008/09/p09002} {\bibfield  {journal}
  {\bibinfo  {journal} {J. Stat. Mech.}\ }\textbf {\bibinfo {volume} {2008}},\
  \bibinfo {pages} {P09002} (\bibinfo {year} {2008})},\ \Eprint
  {http://arxiv.org/abs/0806.1861} {arXiv:0806.1861 [math-ph]} \BibitemShut
  {NoStop}%
\bibitem [{\citenamefont {Kanazawa}(2016)}]{Kanazawa:2016nlh}%
  \BibitemOpen
  \bibfield  {author} {\bibinfo {author} {\bibfnamefont {T.}~\bibnamefont
  {Kanazawa}},\ }\href {\doibase 10.1007/JHEP05(2016)166} {\bibfield  {journal}
  {\bibinfo  {journal} {JHEP}\ }\textbf {\bibinfo {volume} {05}},\ \bibinfo
  {pages} {166} (\bibinfo {year} {2016})},\ \Eprint
  {http://arxiv.org/abs/1602.05631} {arXiv:1602.05631 [hep-th]} \BibitemShut
  {NoStop}%
%%CITATION = ARXIV:1602.05631;%%
\bibitem [{\citenamefont {Pisarski}(2006)}]{Pisarski:2006hz}%
  \BibitemOpen
  \bibfield  {author} {\bibinfo {author} {\bibfnamefont {R.~D.}\ \bibnamefont
  {Pisarski}},\ }\href {\doibase 10.1103/PhysRevD.74.121703} {\bibfield
  {journal} {\bibinfo  {journal} {Phys. Rev.}\ }\textbf {\bibinfo {volume}
  {D74}},\ \bibinfo {pages} {121703} (\bibinfo {year} {2006})},\ \Eprint
  {http://arxiv.org/abs/hep-ph/0608242} {arXiv:hep-ph/0608242 [hep-ph]}
  \BibitemShut {NoStop}%
%%CITATION = HEP-PH/0608242;%%
\bibitem [{\citenamefont {Myers}\ and\ \citenamefont
  {Ogilvie}(2008)}]{Myers:2007vc}%
  \BibitemOpen
  \bibfield  {author} {\bibinfo {author} {\bibfnamefont {J.~C.}\ \bibnamefont
  {Myers}}\ and\ \bibinfo {author} {\bibfnamefont {M.~C.}\ \bibnamefont
  {Ogilvie}},\ }\href {\doibase 10.1103/PhysRevD.77.125030} {\bibfield
  {journal} {\bibinfo  {journal} {Phys. Rev.}\ }\textbf {\bibinfo {volume}
  {D77}},\ \bibinfo {pages} {125030} (\bibinfo {year} {2008})},\ \Eprint
  {http://arxiv.org/abs/0707.1869} {arXiv:0707.1869 [hep-lat]} \BibitemShut
  {NoStop}%
%%CITATION = ARXIV:0707.1869;%%
\bibitem [{\citenamefont {Unsal}\ and\ \citenamefont
  {Yaffe}(2008)}]{Unsal:2008ch}%
  \BibitemOpen
  \bibfield  {author} {\bibinfo {author} {\bibfnamefont {M.}~\bibnamefont
  {Unsal}}\ and\ \bibinfo {author} {\bibfnamefont {L.~G.}\ \bibnamefont
  {Yaffe}},\ }\href {\doibase 10.1103/PhysRevD.78.065035} {\bibfield  {journal}
  {\bibinfo  {journal} {Phys. Rev.}\ }\textbf {\bibinfo {volume} {D78}},\
  \bibinfo {pages} {065035} (\bibinfo {year} {2008})},\ \Eprint
  {http://arxiv.org/abs/0803.0344} {arXiv:0803.0344 [hep-th]} \BibitemShut
  {NoStop}%
%%CITATION = ARXIV:0803.0344;%%
\bibitem [{\citenamefont {Akemann}\ \emph {et~al.}(2001)\citenamefont
  {Akemann}, \citenamefont {Dalmazi}, \citenamefont {Damgaard},\ and\
  \citenamefont {Verbaarschot}}]{Akemann:2000df}%
  \BibitemOpen
  \bibfield  {author} {\bibinfo {author} {\bibfnamefont {G.}~\bibnamefont
  {Akemann}}, \bibinfo {author} {\bibfnamefont {D.}~\bibnamefont {Dalmazi}},
  \bibinfo {author} {\bibfnamefont {P.}~\bibnamefont {Damgaard}}, \ and\
  \bibinfo {author} {\bibfnamefont {J.}~\bibnamefont {Verbaarschot}},\ }\href
  {\doibase 10.1016/S0550-3213(01)00066-9} {\bibfield  {journal} {\bibinfo
  {journal} {Nucl.Phys.}\ }\textbf {\bibinfo {volume} {B601}},\ \bibinfo
  {pages} {77} (\bibinfo {year} {2001})},\ \Eprint
  {http://arxiv.org/abs/hep-th/0011072} {arXiv:hep-th/0011072 [hep-th]}
  \BibitemShut {NoStop}%
%%CITATION = HEP-TH/0011072;%%
\bibitem [{\citenamefont {Akemann}\ and\ \citenamefont
  {Vernizzi}(2003)}]{Akemann:2002vy}%
  \BibitemOpen
  \bibfield  {author} {\bibinfo {author} {\bibfnamefont {G.}~\bibnamefont
  {Akemann}}\ and\ \bibinfo {author} {\bibfnamefont {G.}~\bibnamefont
  {Vernizzi}},\ }\href {\doibase 10.1016/S0550-3213(03)00221-9} {\bibfield
  {journal} {\bibinfo  {journal} {Nucl. Phys.}\ }\textbf {\bibinfo {volume}
  {B660}},\ \bibinfo {pages} {532} (\bibinfo {year} {2003})},\ \Eprint
  {http://arxiv.org/abs/hep-th/0212051} {arXiv:hep-th/0212051 [hep-th]}
  \BibitemShut {NoStop}%
%%CITATION = HEP-TH/0212051;%%
\bibitem [{\citenamefont {Strahov}\ and\ \citenamefont
  {Fyodorov}(2003)}]{Strahov:2002zu}%
  \BibitemOpen
  \bibfield  {author} {\bibinfo {author} {\bibfnamefont {E.}~\bibnamefont
  {Strahov}}\ and\ \bibinfo {author} {\bibfnamefont {Y.~V.}\ \bibnamefont
  {Fyodorov}},\ }\href {\doibase 10.1007/s00220-003-0938-x} {\bibfield
  {journal} {\bibinfo  {journal} {Commun. Math. Phys.}\ }\textbf {\bibinfo
  {volume} {241}},\ \bibinfo {pages} {343} (\bibinfo {year} {2003})},\ \Eprint
  {http://arxiv.org/abs/math-ph/0210010} {arXiv:math-ph/0210010 [math-ph]}
  \BibitemShut {NoStop}%
%%CITATION = MATH-PH/0210010;%%
\bibitem [{\citenamefont {Splittorff}\ and\ \citenamefont
  {Verbaarschot}(2004)}]{Splittorff:2003cu}%
  \BibitemOpen
  \bibfield  {author} {\bibinfo {author} {\bibfnamefont {K.}~\bibnamefont
  {Splittorff}}\ and\ \bibinfo {author} {\bibfnamefont {J.~J.~M.}\ \bibnamefont
  {Verbaarschot}},\ }\href {\doibase 10.1016/j.nuclphysb.2004.01.031}
  {\bibfield  {journal} {\bibinfo  {journal} {Nucl. Phys.}\ }\textbf {\bibinfo
  {volume} {B683}},\ \bibinfo {pages} {467} (\bibinfo {year} {2004})},\ \Eprint
  {http://arxiv.org/abs/hep-th/0310271} {arXiv:hep-th/0310271 [hep-th]}
  \BibitemShut {NoStop}%
%%CITATION = HEP-TH/0310271;%%
\bibitem [{\citenamefont {Kanazawa}\ and\ \citenamefont
  {Kieburg}(2018)}]{KiebKana}%
  \BibitemOpen
  \bibfield  {author} {\bibinfo {author} {\bibfnamefont {T.}~\bibnamefont
  {Kanazawa}}\ and\ \bibinfo {author} {\bibfnamefont {M.}~\bibnamefont
  {Kieburg}},\ }\href {\doibase 10.1007/JHEP11(2018)205} {\bibfield  {journal}
  {\bibinfo  {journal} {J. High Energ. Phys.}\ }\textbf {\bibinfo {volume}
  {2018}},\ \bibinfo {pages} {205} (\bibinfo {year} {2018})},\ \Eprint
  {http://arxiv.org/abs/1809.10602} {arXiv:1809.10602 [hep-th]} \BibitemShut
  {NoStop}%
\bibitem [{\citenamefont {Harish-Chandra}(1956)}]{HC}%
  \BibitemOpen
  \bibfield  {author} {\bibinfo {author} {\bibnamefont {Harish-Chandra}},\
  }\href {\doibase 10.1073/pnas.42.5.252} {\bibfield  {journal} {\bibinfo
  {journal} {Proc. Natl. Acad. Sci. USA}\ }\textbf {\bibinfo {volume} {42}},\
  \bibinfo {pages} {252} (\bibinfo {year} {1956})}\BibitemShut {NoStop}%
\bibitem [{\citenamefont {Itzykson}\ and\ \citenamefont {Zuber}(1980)}]{IZ}%
  \BibitemOpen
  \bibfield  {author} {\bibinfo {author} {\bibfnamefont {C.}~\bibnamefont
  {Itzykson}}\ and\ \bibinfo {author} {\bibfnamefont {J.~B.}\ \bibnamefont
  {Zuber}},\ }\href {\doibase 10.1063/1.524438} {\bibfield  {journal} {\bibinfo
   {journal} {J. Math. Phys.}\ }\textbf {\bibinfo {volume} {21}},\ \bibinfo
  {pages} {411} (\bibinfo {year} {1980})}\BibitemShut {NoStop}%
\bibitem [{\citenamefont {Andr\'eief}(1883)}]{Andreief}%
  \BibitemOpen
  \bibfield  {author} {\bibinfo {author} {\bibfnamefont {K.~A.}\ \bibnamefont
  {Andr\'eief}},\ }\href@noop {} {\bibfield  {journal} {\bibinfo  {journal}
  {M\'em. de la Soc. Sci., Bordeaux}\ }\textbf {\bibinfo {volume} {2}},\
  \bibinfo {pages} {1} (\bibinfo {year} {1883})}\BibitemShut {NoStop}%
\bibitem [{\citenamefont {Gradshteyn}\ and\ \citenamefont
  {Ryzhik}(2007)}]{GRmathbook2007}%
  \BibitemOpen
  \bibfield  {author} {\bibinfo {author} {\bibfnamefont {I.~S.}\ \bibnamefont
  {Gradshteyn}}\ and\ \bibinfo {author} {\bibfnamefont {I.~M.}\ \bibnamefont
  {Ryzhik}},\ }\href@noop {} {\emph {\bibinfo {title} {{Table of Integrals,
  Series, and Products}}}},\ \bibinfo {edition} {{Seventh}}\ ed.\ (\bibinfo
  {publisher} {Elsevier},\ \bibinfo {year} {2007})\BibitemShut {NoStop}%
\bibitem [{\citenamefont {Akemann}\ \emph {et~al.}(2005)\citenamefont
  {Akemann}, \citenamefont {Osborn}, \citenamefont {Splittorff},\ and\
  \citenamefont {Verbaarschot}}]{Akemann:2004dr}%
  \BibitemOpen
  \bibfield  {author} {\bibinfo {author} {\bibfnamefont {G.}~\bibnamefont
  {Akemann}}, \bibinfo {author} {\bibfnamefont {J.~C.}\ \bibnamefont {Osborn}},
  \bibinfo {author} {\bibfnamefont {K.}~\bibnamefont {Splittorff}}, \ and\
  \bibinfo {author} {\bibfnamefont {J.~J.~M.}\ \bibnamefont {Verbaarschot}},\
  }\href {\doibase 10.1016/j.nuclphysb.2005.01.018} {\bibfield  {journal}
  {\bibinfo  {journal} {Nucl. Phys.}\ }\textbf {\bibinfo {volume} {B712}},\
  \bibinfo {pages} {287} (\bibinfo {year} {2005})},\ \Eprint
  {http://arxiv.org/abs/hep-th/0411030} {arXiv:hep-th/0411030 [hep-th]}
  \BibitemShut {NoStop}%
%%CITATION = HEP-TH/0411030;%%
\bibitem [{\citenamefont {Osborn}\ \emph {et~al.}(2005)\citenamefont {Osborn},
  \citenamefont {Splittorff},\ and\ \citenamefont
  {Verbaarschot}}]{Osborn:2005ss}%
  \BibitemOpen
  \bibfield  {author} {\bibinfo {author} {\bibfnamefont {J.~C.}\ \bibnamefont
  {Osborn}}, \bibinfo {author} {\bibfnamefont {K.}~\bibnamefont {Splittorff}},
  \ and\ \bibinfo {author} {\bibfnamefont {J.~J.~M.}\ \bibnamefont
  {Verbaarschot}},\ }\href {\doibase 10.1103/PhysRevLett.94.202001} {\bibfield
  {journal} {\bibinfo  {journal} {Phys. Rev. Lett.}\ }\textbf {\bibinfo
  {volume} {94}},\ \bibinfo {pages} {202001} (\bibinfo {year} {2005})},\
  \Eprint {http://arxiv.org/abs/hep-th/0501210} {arXiv:hep-th/0501210 [hep-th]}
  \BibitemShut {NoStop}%
%%CITATION = HEP-TH/0501210;%%
\bibitem [{\citenamefont {Akemann}\ \emph {et~al.}(2011)\citenamefont
  {Akemann}, \citenamefont {Kanazawa}, \citenamefont {Phillips},\ and\
  \citenamefont {Wettig}}]{Akemann:2010tv}%
  \BibitemOpen
  \bibfield  {author} {\bibinfo {author} {\bibfnamefont {G.}~\bibnamefont
  {Akemann}}, \bibinfo {author} {\bibfnamefont {T.}~\bibnamefont {Kanazawa}},
  \bibinfo {author} {\bibfnamefont {M.~J.}\ \bibnamefont {Phillips}}, \ and\
  \bibinfo {author} {\bibfnamefont {T.}~\bibnamefont {Wettig}},\ }\href
  {\doibase 10.1007/JHEP03(2011)066} {\bibfield  {journal} {\bibinfo  {journal}
  {JHEP}\ }\textbf {\bibinfo {volume} {03}},\ \bibinfo {pages} {066} (\bibinfo
  {year} {2011})},\ \Eprint {http://arxiv.org/abs/1012.4461} {arXiv:1012.4461
  [hep-lat]} \BibitemShut {NoStop}%
%%CITATION = ARXIV:1012.4461;%%
\bibitem [{\citenamefont {Kanazawa}\ \emph {et~al.}(2011)\citenamefont
  {Kanazawa}, \citenamefont {Wettig},\ and\ \citenamefont
  {Yamamoto}}]{Kanazawa:2011tt}%
  \BibitemOpen
  \bibfield  {author} {\bibinfo {author} {\bibfnamefont {T.}~\bibnamefont
  {Kanazawa}}, \bibinfo {author} {\bibfnamefont {T.}~\bibnamefont {Wettig}}, \
  and\ \bibinfo {author} {\bibfnamefont {N.}~\bibnamefont {Yamamoto}},\ }\href
  {\doibase 10.1007/JHEP12(2011)007} {\bibfield  {journal} {\bibinfo  {journal}
  {JHEP}\ }\textbf {\bibinfo {volume} {12}},\ \bibinfo {pages} {007} (\bibinfo
  {year} {2011})},\ \Eprint {http://arxiv.org/abs/1110.5858} {arXiv:1110.5858
  [hep-ph]} \BibitemShut {NoStop}%
%%CITATION = ARXIV:1110.5858;%%
\bibitem [{\citenamefont {Verbaarschot}\ and\ \citenamefont
  {Wettig}(2014)}]{Verbaarschot:2014upa}%
  \BibitemOpen
  \bibfield  {author} {\bibinfo {author} {\bibfnamefont {J.~J.~M.}\
  \bibnamefont {Verbaarschot}}\ and\ \bibinfo {author} {\bibfnamefont
  {T.}~\bibnamefont {Wettig}},\ }\href {\doibase 10.1103/PhysRevD.90.116004}
  {\bibfield  {journal} {\bibinfo  {journal} {Phys. Rev.}\ }\textbf {\bibinfo
  {volume} {D90}},\ \bibinfo {pages} {116004} (\bibinfo {year} {2014})},\
  \Eprint {http://arxiv.org/abs/1407.8393} {arXiv:1407.8393 [hep-th]}
  \BibitemShut {NoStop}%
%%CITATION = ARXIV:1407.8393;%%
\bibitem [{\citenamefont {Dominici}(2007)}]{Dominici2007}%
  \BibitemOpen
  \bibfield  {author} {\bibinfo {author} {\bibfnamefont {D.}~\bibnamefont
  {Dominici}},\ }\href {\doibase 10.1080/10236190701458824} {\bibfield
  {journal} {\bibinfo  {journal} {J. Difference Equ. Appl.}\ }\textbf {\bibinfo
  {volume} {13}},\ \bibinfo {pages} {1115} (\bibinfo {year} {2007})},\ \Eprint
  {http://arxiv.org/abs/math/0601078} {math/0601078} \BibitemShut {NoStop}%
\bibitem [{\citenamefont {Peskin}(1980)}]{Peskin:1980gc}%
  \BibitemOpen
  \bibfield  {author} {\bibinfo {author} {\bibfnamefont {M.~E.}\ \bibnamefont
  {Peskin}},\ }\href {\doibase 10.1016/0550-3213(80)90051-6} {\bibfield
  {journal} {\bibinfo  {journal} {Nucl. Phys.}\ }\textbf {\bibinfo {volume}
  {B175}},\ \bibinfo {pages} {197} (\bibinfo {year} {1980})}\BibitemShut
  {NoStop}%
%%CITATION = NUPHA,B175,197;%%
\bibitem [{\citenamefont {Sommers}(2007)}]{Som}%
  \BibitemOpen
  \bibfield  {author} {\bibinfo {author} {\bibfnamefont {H.-J.}\ \bibnamefont
  {Sommers}},\ }\href {\doibase
  http://www.actaphys.uj.edu.pl/fulltext?series=Reg\&vol=38\&page=4105}
  {\bibfield  {journal} {\bibinfo  {journal} {Acta Phys. Pol. B}\ }\textbf
  {\bibinfo {volume} {38}},\ \bibinfo {pages} {4105} (\bibinfo {year}
  {2007})},\ \Eprint {http://arxiv.org/abs/0710.5375} {arXiv:0710.5375
  [cond-mat.stat-mech]} \BibitemShut {NoStop}%
\bibitem [{\citenamefont {Littlemann}\ \emph {et~al.}(2008)\citenamefont
  {Littlemann}, \citenamefont {Sommers},\ and\ \citenamefont
  {Zirnbauer}}]{LSZ}%
  \BibitemOpen
  \bibfield  {author} {\bibinfo {author} {\bibfnamefont {P.}~\bibnamefont
  {Littlemann}}, \bibinfo {author} {\bibfnamefont {H.-J.}\ \bibnamefont
  {Sommers}}, \ and\ \bibinfo {author} {\bibfnamefont {M.~R.}\ \bibnamefont
  {Zirnbauer}},\ }\href {\doibase 10.1007/s00220-008-0535-0} {\bibfield
  {journal} {\bibinfo  {journal} {Commun. Math. Phys.}\ }\textbf {\bibinfo
  {volume} {283}},\ \bibinfo {pages} {343} (\bibinfo {year} {2008})},\ \Eprint
  {http://arxiv.org/abs/0707.2929} {arXiv:0707.2929 [math-ph]} \BibitemShut
  {NoStop}%
\bibitem [{\citenamefont {Kieburg}\ \emph {et~al.}(2009)\citenamefont
  {Kieburg}, \citenamefont {Sommers},\ and\ \citenamefont {Guhr}}]{KSG}%
  \BibitemOpen
  \bibfield  {author} {\bibinfo {author} {\bibfnamefont {M.}~\bibnamefont
  {Kieburg}}, \bibinfo {author} {\bibfnamefont {H.-J.}\ \bibnamefont
  {Sommers}}, \ and\ \bibinfo {author} {\bibfnamefont {T.}~\bibnamefont
  {Guhr}},\ }\href {\doibase 10.1088/1751-8113/42/27/275206} {\bibfield
  {journal} {\bibinfo  {journal} {J. Phys. A}\ }\textbf {\bibinfo {volume}
  {42}},\ \bibinfo {pages} {275206} (\bibinfo {year} {2009})},\ \Eprint
  {http://arxiv.org/abs/0905.3256} {arXiv:0905.3256 [math-ph]} \BibitemShut
  {NoStop}%
\end{thebibliography}%
\end{document}